\documentclass[reprint,preprintnumbers,amsmath,amssymb,aps,nofootinbib,showkeys, superscriptaddress]{revtex4-2}

\usepackage[utf8]{inputenc}  
\usepackage{graphicx}  
\usepackage{xcolor}  
\usepackage{amsmath}  
\usepackage{amsfonts}  
\usepackage{latexsym}  
\usepackage{amssymb,bbm}  
\usepackage{bm}  
\usepackage[colorlinks=true,allcolors=blue]{hyperref}  
\usepackage{lipsum}  
\usepackage{setspace}
\usepackage{braket}

\newcommand{\xa}{\alpha}

\newcommand{\xt}{\vartheta}

\newcommand{\xl}{\lambda}
\newcommand{\xr}{\rho}
\newcommand{\xo}{\omega}

\newcommand{\Cs}{{}^{13}\R{C}}

\newcommand{\rt}{\rightarrow}

\newcommand{\dg}{\dagger}

\newcommand{\beq}{\begin{equation}}
\newcommand{\eeq}{\end{equation}}
                  
\newcommand{\benum}{\begin{enumerate}}
\newcommand{\eenum}{\end{enumerate}}
                    
\newcommand{\bit}{\begin{itemize}}
\newcommand{\eit}{\end{itemize}}

\newcommand{\bea}{\begin{eqnarray}}
\newcommand{\eea}{\end{eqnarray}}

\newcommand{\bev}{\begin{verbatim}}
\newcommand{\eev}{\end{verbatim}}

\newcommand{\noi}{\noindent}

\newcommand{\T}[1]{\textbf{#1}}
\newcommand{\I}[1]{\textit{#1}}
\newcommand{\R}[1]{\textrm{#1}}

\newcommand{\zl}[1]{\label{eqn:#1}}
\newcommand{\zr}[1]{Eq.\,(\ref{eqn:#1})}
\newcommand{\zfl}[1]{\protect\label{fig:#1}}
\newcommand{\zfr}[1]{\figurename\,\ref{fig:#1}}

\newcommand{\expec}[1]{\left\langle #1\right\rangle}

\newcommand{\ba}{\left\{ \begin{array}{lr}}
\newcommand{\ea}{\end{array}\right.}

\newcommand{\blist}[1]{
 \begin{list}{#1}
 \begin{align}
	 arrow
 \end{align}
 $\checkmark\star
  { \setlength{\itemsep}{3pt}
     \setlength{\parsep}{2pt}
     \setlength{\topsep}{3pt}
     \setlength{\partopsep}{0pt}
     \setlength{\leftmargin}{1em}
     \setlength{\labelwidth}{1em}
     \setlength{\labelsep}{0.5em} } }
\newcommand{\elist}{
  \end{list}  }

\DeclareMathSymbol{\vartheta}{\mathalpha}{letters}{"12}
\DeclareMathSymbol{\theta}{\mathalpha}{letters}{"23}
\DeclareMathSymbol{\phi}{\mathalpha}{letters}{"27}
\DeclareMathSymbol{\varphi}{\mathalpha}{letters}{"1E}

\newcommand{\bef}
{
\begin{figure}[htbp]
\centering
}

\newcommand{\eef}{\end{figure}}

\renewcommand{\figurename}{Fig.} 

\begin{document}

\title{Emergent Decoherence Dynamics in Doubly Disordered Spin Networks}

\author{Cooper M. Selco}
\thanks{These authors contributed equally to this work.}
\affiliation{Department of Chemistry, University of California, Berkeley, Berkeley, CA 94720, USA}

\author{Christian Bengs}
\thanks{These authors contributed equally to this work.}
\affiliation{Department of Chemistry, University of California, Berkeley, Berkeley, CA 94720, USA}
\affiliation{Chemical Sciences Division, Lawrence Berkeley National Laboratory, Berkeley, CA 94720, USA}
\affiliation{School of Chemistry, University of Southampton, Southampton, SO17 1BJ, UK}

\author{Chaitali Shah}
\affiliation{Department of Chemistry, University of California, Berkeley, Berkeley, CA 94720, USA}

\author{Zhuorui Zhang}
\affiliation{Department of Chemistry, University of California, Berkeley, Berkeley, CA 94720, USA}

\author{Ashok Ajoy}
\email{ashokaj@berkeley.edu}
\affiliation{Department of Chemistry, University of California, Berkeley, Berkeley, CA 94720, USA}
\affiliation{Chemical Sciences Division, Lawrence Berkeley National Laboratory, Berkeley, CA 94720, USA}

\begin{abstract}

Elucidating the emergence of irreversible macroscopic laws from reversible quantum many-body dynamics is a question of broad importance across all quantum science~\cite{trotzkyProbingRelaxationEquilibrium2012, eisertQuantumManybodySystems2015, smithManybodyLocalizationQuantum2016b, choiExploringManybodyLocalization2016a, bernienProbingManybodyDynamics2017, miInformationScramblingQuantum2021, millington-hotzeNuclearSpinDiffusion2023}. Many-body decoherence
plays a key role in this transition~\cite{zurekDecoherenceEinselectionQuantum2003, schlosshauerDecoherenceQuantumToClassicalTransition2007}, yet connecting microscopic dynamics to emergent macroscopic behavior remains challenging. Here, in a doubly disordered electron-nuclear spin network, we uncover an emergent decoherence law for nuclear polarization, $e^{-\sqrt{R_{p}t}}e^{-R_{d}t}$, that is robust across broad parameter regimes. We trace its microscopic origins to two interdependent decoherence channels: long-range interactions mediated by the electron network and spin transport within the nuclear network exhibiting anomalous, sub-diffusive dynamics. We demonstrate the capacity to control—and even eliminate—either channel individually through a combination of Floquet engineering~\cite{choiRobustDynamicHamiltonian2020} and (optical) environment modulation. We find that disorder, typically viewed as detrimental, here proves protective, generating isolated electron-free clusters that localize polarization and prolong coherence lifetimes. These findings establish a microscopic framework for manipulating decoherence pathways and suggests engineered disorder as a new design principle for realizing long-lived quantum memories and sensors.

\end{abstract}

\maketitle


\pagebreak
\noi Quantum many-body systems obey complex yet time-reversible microscopic laws, while their macroscopic evolution is often described by comparatively simple but irreversible models~\cite{rauReversibleQuantumMicrodynamics1996, schlosshauerQuantumDecoherence2019}. Predicting emergent decoherence laws that bridge these regimes remains a central challenge.
In ordered spin lattices, translational symmetry enables accurate prediction of decoherence resulting in dissipative hydrodynamics~\cite{khemaniOperatorSpreadingEmergence2018a, gopalakrishnanHydrodynamicsOperatorSpreading2018, yeEmergentHydrodynamicsNonequilibrium2020, zuEmergentHydrodynamicsStrongly2021, joshiObservingEmergentHydrodynamics2022a}. Disordered lattices, however, lack such symmetry, yielding richer and less predictable decoherence behavior~\cite{abaninColloquiumManybodyLocalization2019, davisProbingManybodyDynamics2023}. Understanding these effects is crucial for extending coherence lifetimes across quantum platforms where disorder naturally arises~\cite{golubevQuantumDecoherenceDisordered1998, wuIntrinsicDecoherenceIsolated2017}.
\newline\indent
Here we study decoherence dynamics in a \I{doubly-disordered} spin network comprising nuclear and electronic spins in nitrogen-doped diamond. The system hosts a dilute network of $^{13}$C nuclei (1.1\%) coupled to paramagnetic defects (NV and P1 centers)~\cite{jelezkoSingleDefectCentres2006}. The $\Cs$ spins experience decoherence through a complex interplay of spin transport and random electron–nuclear couplings. This platform offers exceptional microscopic control: inter-nuclear couplings can be tuned through Floquet engineering~\cite{ryanRobustDecouplingTechniques2010, alvarezMeasuringSpectrumColored2011a} and nuclear coherence can be monitored continuously, providing simultaneous access to short- and long-time dynamics~\cite{beatrezFloquetPrethermalizationLifetime2021a}.
\newline\indent
Over hundreds of seconds (many decades of the nuclear $T_2^{\ast}$ period), we observe the $\Cs$ magnetization follows a universal emergent decoherence law, $M(t) = e^{-\sqrt{R_{p}t}}e^{-R_{d}t}$, consisting of a product of stretched- and mono-exponential components corresponding to two distinct relaxation channels. While stretched-exponential dynamics have been reported in restricted temporal regimes~\cite{tseNuclearSpinLatticeRelaxation1968a, davisProbingManybodyDynamics2023}, this composite form persists across a wide range of Hamiltonian parameters and spin concentrations, and subsumes prior models as limiting cases~\cite{bloembergenInteractionNuclearSpins1949a, blumbergNuclearSpinLatticeRelaxation1960b, tseNuclearSpinLatticeRelaxation1968a}.
\newline\indent
To uncover how the decoherence law emerges from the rich underlying microscopic dynamics, we construct a theoretical framework that reproduces experiment and reveals subdiffusive polarization transport within the nuclear network—behavior inconsistent with conventional hydrodynamic expectations~\cite{abragamPrinciplesNuclearMagnetism1962, boutisSpinDiffusionCorrelated2004, friedmanDiffusiveHydrodynamicsIntegrability2020}. We identify the two relaxation channels as distinct electron-mediated processes: $R_d$ reflects transport toward localized, electron-centered relaxation basins, whereas $R_p$ represents mean-field-like decoherence from the fluctuating electron bath (\zfr{fig1}a-b). We find that the the doubly disordered network produces rare electron-free regions that act as long-lived polarization traps, producing exceptionally slow relaxation. Finally, by combining Hamiltonian engineering with all-optical environment modulation, we show the ability to independently tune—and even eliminate—each channel individually, demonstrating that modifications of the microscopic Hamiltonian yield deterministic control over the emergent macroscopic behavior.

\begin{figure*}[t]
    \centering
    \includegraphics[width=1\textwidth]{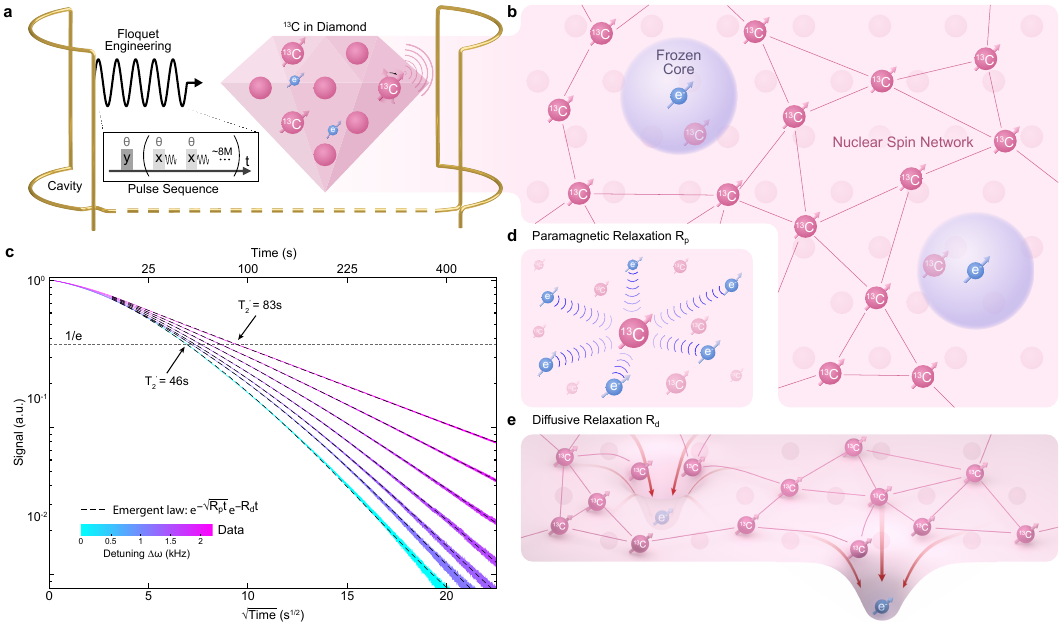}
    \caption{\textbf{Emergent decoherence law in a doubly disordered spin network.} (a) \emph{\textbf{Protocol and setup}}. Floquet engineering pulse sequence consists of one $\xt_{y}$ pulse followed by $\sim$8M $\xt_{x}$ pulses. $^{13}$C magnetization is monitored quasi-continuously after each $\xt_{x}$ pulse via inductive cavity readout. (b) \emph{\textbf{Doubly disordered spin network}} illustrating disordered $^{13}$C network embedded within disordered electron network (NV and P1 centers; concentrations not to scale). Pink lines indicate inter-spin couplings; transparent circles show spinless $^{12}$C; blue spheres mark electron frozen cores where inward diffusion is suppressed (see Methods). (c) \emph{\textbf{Experimentally measured decoherence}} traces vs. pulse detuning, $\Delta\omega$ (colorbar). Experimental data (smoothed with a 100-point moving average) plotted on log vs. $\sqrt{t}$ scale sampled every $\approx$80 $\mu$s over $>$500 s. Top $x$-axis shows $t$. Horizontal dashed line marks $1/e$ crossing with surrogate $T_{2}'$ values labeled. Black dashed lines are fits to emergent law, $e^{-\sqrt{R_{p}t}}e^{-R_{d}t}$, showing excellent agreement across entire dataset (for residuals see SI Sec.~\ref{section_fitting}). At $\Delta\omega \simeq 2.25$ kHz (top pink trace), decay reduces to $e^{-\sqrt{R_{p}t}}$. (d)-(e) \emph{\textbf{Microscopic decoherence mechanisms}}. (d) \emph{\textbf{Paramagnetic relaxation $R_{p}$}}: direct relaxation pathway via dipolar coupling to disordered electron environment (blue waves) (e) \emph{\textbf{Diffusive relaxation $R_{d}$}}: indirect pathway via polarization transport through nuclear spin network toward electron ``sinkholes'' (red arrows). Nuclei near electrons relax rapidly, generating polarization gradients that drive spin transport.}
    \label{Fig1}
    \zfl{fig1}
\end{figure*} 

\begin{figure*}[t]
    \centering
    \includegraphics[width=1\textwidth]{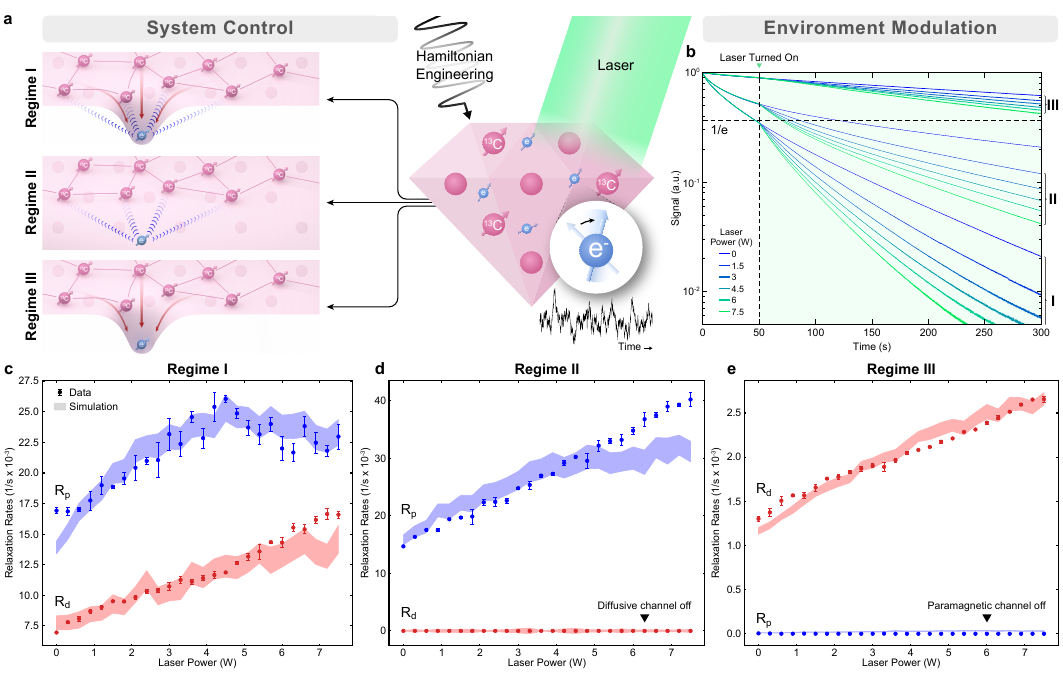}
    \caption{\textbf{Controlling emergent dynamics via simultaneous system and environment engineering.} (a) \emph{\textbf{System control}}: engineering microscopic Hamiltonians enables selective elimination of either decoherence channel. In Regime \T{I} ($\Delta\omega=0$ kHz, $\xt=90^{\circ}$), both paramagnetic (blue waves) and diffusive (red arrows) pathways are active. In Regime \T{II} ($\Delta\omega=2.25$ kHz, $\xt=90^{\circ}$), $H_{nn}^{(1)} \rightarrow 0$ effectively ``turns off" diffusive pathway, leaving only paramagnetic relaxation. In Regime \T{III} ($\Delta\omega=5$ kHz, $\xt=5^{\circ}$), energy mismatch suppresses paramagnetic channel, isolating diffusive pathway (see Methods).
    (b) \emph{\textbf{Environmental modulation}} via simultaneous laser illumination. Decay curves (log scale) for each regime are shown with laser turned on 50 s into Floquet sequence at various powers (colorbar). Increasing power causes curves to ``fan out.'' Laser-driven electron fluctuations reshape spectral distribution of magnetic noise. (c)-(e) \emph{\textbf{Probing relaxation rates $R_{p}$ and $R_{d}$}} via fits to \zr{emerge}. Laser remains on throughout; error bars denote standard error from three random trials. (c) \emph{\textbf{Regime \T{I}}}: $R_{p}$ increases then decreases ($>$4 W), while $R_{d}$ increases linearly. (d) \emph{\textbf{Regime \T{II}}}: diffusive channel ``turned off'' (flat line) yielding decay as $e^{-\sqrt{R_{p}t}}$. $R_{p}$ increases linearly with power. (e) \emph{\textbf{Regime \T{III}}}: paramagnetic channel suppressed (flat line) yielding decay as $e^{-R_{d}t}$. $R_{d}$ increases linearly with power, with overall smaller rates. Shaded regions indicate simulated relaxation rates from Markov chain Monte Carlo analysis, with thickness representing standard error (see SI Sec.~\ref{section_montecarlo}).}
    \label{Fig2}
    \zfl{fig2}
\end{figure*}

\vspace{0.5em}
\noi{\normalsize\bfseries Doubly disordered spin platform\par}

\noi Our platform is a diamond crystal containing a dilute, disordered network of $^{13}$C nuclear spins at 1.1\% natural abundance (\zfr{fig1}a-b). The nuclei interact via long-range magnetic dipolar couplings described by \mbox{$H_{\rm nn}=\sum_{i<j}d_{ij}(2I^{i}_{z}I^{j}_{z}-I^{i}_{x}I^{j}_{x}-I^{i}_{y}I^{j}_{y})$}, where $I^{j}_{\mu}$ are spin-$1/2$ Pauli operators. A sparse network of electron spins—NV centers ($\sim$1 ppm) and P1 centers ($\sim$30 ppm)—creates a strongly inhomogeneous relaxation landscape for the nuclei (\zfr{fig1}b).
\newline\indent
The measurement protocol (\zfr{fig1}a) was conducted at 100 K~\cite{dsouzaCryogenicFieldcyclingInstrument2025}. The $\Cs$ nuclei were optically hyperpolarized via NV centers following established procedures~\cite{ajoyOrientationindependentRoomTemperature2018a, sarkarRapidlyEnhancedSpinPolarization2022a} (see Methods). After spatial homogenization of the $\Cs$ polarization, the spins were subjected to a Floquet drive at 9.4 T (\zfr{fig1}a, inset). An initial state $\xr(0) \propto I_{x}$ evolves under a train of eight million pulses of flip angle $\xt_{x}$ and detuning $\Delta\omega$, while the resulting nuclear magnetization $M(t)=\R{Tr}\{\xr^{\dg}(t)I_x\}$ is quasi-continuously monitored through a resonant cavity during each of the interpulse windows~\cite{moonHighspeedHighmemoryNMR2025}.

\vspace{0.5em}
\noi {\normalsize\bfseries Emergent decoherence law\par}
\noi We first consider the case of resonant driving ($\Delta\omega=0$, cyan trace in \zfr{fig1}c). Data acquired over 600 s reveal a decay that deviates strongly from a simple exponential. After an initial 10 s transient—attributed to spins within the frozen core (see Methods)—the dynamics are well described by a universal emergent law,
\beq
M(t)=e^{-\sqrt{R_{p}t}}e^{-R_{d}t},
\zl{emerge}
\eeq
which combines stretched- and mono-exponential dependencies with only two free parameters, $R_p$ and $R_d$. The fit (dashed black line in \zfr{fig1}c) reproduces the data closely, with minimal residuals (SI Sec.~\ref{section_fitting}). Varying $\Delta\omega$ tunes the microscopic Hamiltonian (see Methods), including regimes where the leading-order nuclear dipolar term $H_{nn}^{(1)}{\to}0$ (top pink trace) effectively vanishes. Yet the same functional form of \zr{emerge} remains valid across all conditions (black dashed lines, \zfr{fig1}c). 
\newline\indent
\zfr{fig1}d–e illustrate microscopic interpretations of $R_p$ and $R_d$ individually; their combination in a product form of \zr{emerge} is nonetheless surprising. Thermal fluctuations of NV and P1 center electrons generate magnetic noise at the $^{13}$C sites, producing two distinct decoherence pathways: a \emph{direct} channel from dipolar coupling to the electronic impurities (\zfr{fig1}d; see SI Sec.~\ref{subsection_diffusionless} for derivation), and an \emph{indirect} channel mediated by polarization transport through the nuclear network (\zfr{fig1}e). Spins located near electronic impurities relax rapidly under strong local dipolar field fluctuations, establishing polarization gradients that drive transport from more distant nuclei.

\vspace{0.5em}
\noi{\normalsize\bfseries Microscopic control of emergent dynamics\par}
\noi To construct a microscopic picture of the emergent dynamics, we show that each decoherence channel can be independently tuned and even selectively eliminated. This control is achieved through a combination of system control via Hamiltonian engineering and laser-driven environmental modulation, enabling isolation of the individual pathways (\zfr{fig2}a-b).
\newline\indent
We focus on three representative regimes (schematized in \zfr{fig2}a). Regime \T{I}, corresponding to on-resonance driving ($\Delta\omega{=}0$, $\xt{=}90^\circ$), serves as the reference case where both decoherence channels are active (top panel, \zfr{fig2}a), corresponding to the cyan trace in \zfr{fig1}c. In Regime \T{II}, the drive parameters ($\Delta\omega{=}2.25$ kHz, $\xt{=}90^\circ$) suppress inter-nuclear interactions (see Methods), isolating the direct paramagnetic relaxation channel and yielding a so-called \I{``diffusion-limited''} regime (middle panel, \zfr{fig2}b). Conversely, Regime \T{III} employs short-angle, far-detuned pulses ($\Delta\omega{=}5$ kHz, $\xt{=}5^\circ$), creating an energy mismatch that suppresses the electron-mediated pathway (see Methods), producing a \I{``diffusion-dominated''} regime. In these limits, the dynamics simplify: Regime \T{II} approaches a pure stretched-exponential form $\propto e^{-\sqrt{R_{p}t}}$ (top pink line, \zfr{fig1}c), while Regime \T{III} follows a mono-exponential decay.

\vspace{0.5em}
\noi{\normalsize\bfseries All-optical environment engineering\par} 
\noi We demonstrate active control of the decoherence channels through \emph{all-optical} modulation of the electronic bath. This approach operates at arbitrary magnetic fields and is compatible with Hamiltonian engineering. Illumination of NV centers with 532 nm light drives intersystem crossing~\cite{dohertyTheoryGroundstateSpin2012}, producing fluctuating dipolar fields that reshape the magnetic noise experienced by $^{13}$C nuclei (see SI Sec.~\ref{subsection_model}). Unlike conventional spin-bath driving schemes~\cite{ernstNuclearMagneticDouble1966, levittBroadbandHeteronuclearDecoupling1982, joosProtectingQubitCoherence2022a}, the effect here is realized entirely optically. Although NV centers are the optically active sites, their fluctuations enhance noise in nearby P1 centers through nonsecular interactions, which propagate across the dense P1 network. Correlated NV–P1 clustering in diamond, observed previously~\cite{bussandriP1CenterElectron2024, nir-aradNitrogenSubstitutionsAggregation2024}, likely facilitates this process.
\newline\indent
\zfr{fig2}b shows decay curves measured with the laser activated at various powers (0–7.5 W) after 50 s of Floquet driving across all three regimes. Nuclear relaxation accelerates (\I{``fans out"}) systematically with optical power, confirming active environmental noise modulation. Measurements in SI Sec.~\ref{section_heating} confirm the effect is not trivially due to sample heating.
\newline\indent
\zfr{fig2}c-e shows the extracted relaxation rates following fits to \zr{emerge} as a function of laser power across the three regimes. In Regime \T{I} (\zfr{fig2}c), $R_{p}$ first increases and then begins to decline, suggesting the onset of all-optical decoupling at higher powers (see SI Sec.~\ref{subsection_allopticaldecoupling}), though very strong illumination may introduce additional effects such as charge-state conversion or heating~\cite{aslamPhotoinducedIonizationDynamics2013}. In Regimes \T{II} and \T{III}, one channel is eliminated through Hamiltonian engineering (flat lines in \zfr{fig2}d-e), while the other is continuously tunable via optical control. A nearly linear dependence is observed for $R_p$ and $R_d$, respectively, with the complementary channel suppressed. Notably, $R_p$ in Regime \T{II} exceeds its value in Regime \T{I} reflecting the suppression of nuclear diffusion: when transport is frozen, local paramagnetic relaxation is enhanced as nuclear spins can no longer redistribute polarization away from impurity sites.

\vspace{0.5em}
\noi{\normalsize\bfseries Minimal microscopic model\par}
\noi To microscopically interpret the emergent dynamics, we construct a Markov chain Monte Carlo model (see Methods and SI Sec.~\ref{subsection_model}) on a diamond lattice populated with $\Cs$ and electronic spins at the experimental concentrations. For simplicity, all electronic spins are treated as optically active. The polarization dynamics $p(t)$ are described by a semiclassical hopping model,
\beq
\dot{p}(t) = (W+R)p(t)
\zl{hopping_model}
\eeq
where $W$ and $R$ represent hopping and dissipation, respectively (\zfr{fig1}d-e). 
\newline\indent
\begin{figure}[htbp]
    \centering
    \includegraphics[width=\columnwidth]{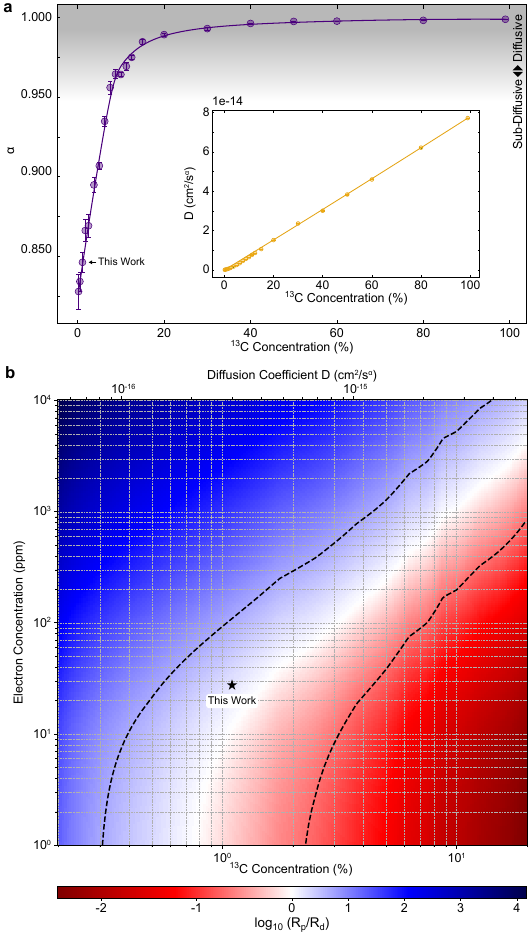}
    \caption{\textbf{Concentration-dependent transport and relaxation.} (a) \emph{\textbf{Polarization transport}}. Simulations track mean squared displacement and fit to $6Dt^{\alpha}$ (see Methods, SI Sec.~\ref{subsection_transport}). Main plot (purple) shows diffusion exponent $\alpha$ increases with $^{13}$C concentration; solid line is guide to eye. Each point is mean of 5 independent runs of 100 trajectories; error bars denote standard error. Inset (yellow) shows corresponding diffusion coefficient $D$. At 1.1\% $^{13}$C (this work), $D=3.86$\r{A}/s$^{0.85}$, and $\alpha=0.85$, indicating subdiffusive behavior that deviates from hydrodynamic expectations. (b) \emph{\textbf{Relative contributions}}, $\rm{log_{10}}(R_p/R_d)$, of the two relaxation channels shown as heat map (colorbar) across varying $^{13}$C ($x$-axis) and electron ($y$-axis) concentrations; top $x$-axis indicates corresponding diffusion coefficient $D$. Blue regions denote ``diffusion-limited'' regimes (${\rightarrow} e^{-\sqrt{R_{p}t}}$), while red indicates ``diffusion-dominated'' regimes (${\rightarrow} e^{-R_{d}t}$) (see SI Sec.~\ref{section_relaxationlandscape}). Dashed lines mark $R_p/R_d = 3$ or $1/3$. Star (this work) marks a newly accessed regime, (see comparison with literature in SI Sec.~\ref{section_context}).}
    \label{Fig3}
    \zfl{fig3}
\end{figure}
The hopping rates follow Fermi’s golden rule $W\propto \kappa^{2}\: d_{ij}^{2}\: T_{2}$, scaling with the squared internuclear coupling and the intrinsic nuclear coherence time $T_{2}$ (see SI Sec.~\ref{subsection_model}). A sequence-dependent prefactor $\kappa$ (Methods) incorporates Floquet driving; with $\kappa{\rt} 0$ in Regime \T{II}. Dissipation $R$ arises from electronic dipolar field fluctuations, with magnitude set by the spectral density $J_{\rm env}(\xo_{\R{eff}})$ at frequencies corresponding to the Floquet drive; $J_{\rm env}(\xo_{\R{eff}}){\rt} 0$ in Regime \T{III}. Laser illumination is modeled as reducing the electron correlation time inversely with optical power (see SI Sec.~\ref{subsection_lasereffects}).
\newline\indent
Simulated dynamics reproduce \zr{emerge} with high fidelity: best fits yield a stretched-exponential factor of $1/2$ combined with a monoexponential (SI Sec.~\ref{subsection_optimalstretchingfactor}), demonstrating that the minimal ingredients of \zfr{fig1}d-e suffice. Incorporating laser effects (see details in Methods and SI Sec.~\ref{subsection_model}), produces excellent agreement with experimental data across all three regimes, as shown by the shaded bands in \zfr{fig2}c-e, whose widths represents standard error over 100 random lattice realizations. Altogether, this framework illustrates how tuning the microscopic Hamiltonian—through system or environment engineering—can directly tailor emergent decoherence behavior. 

\begin{figure*}[t]
    \centering
    \includegraphics[width=1\textwidth]{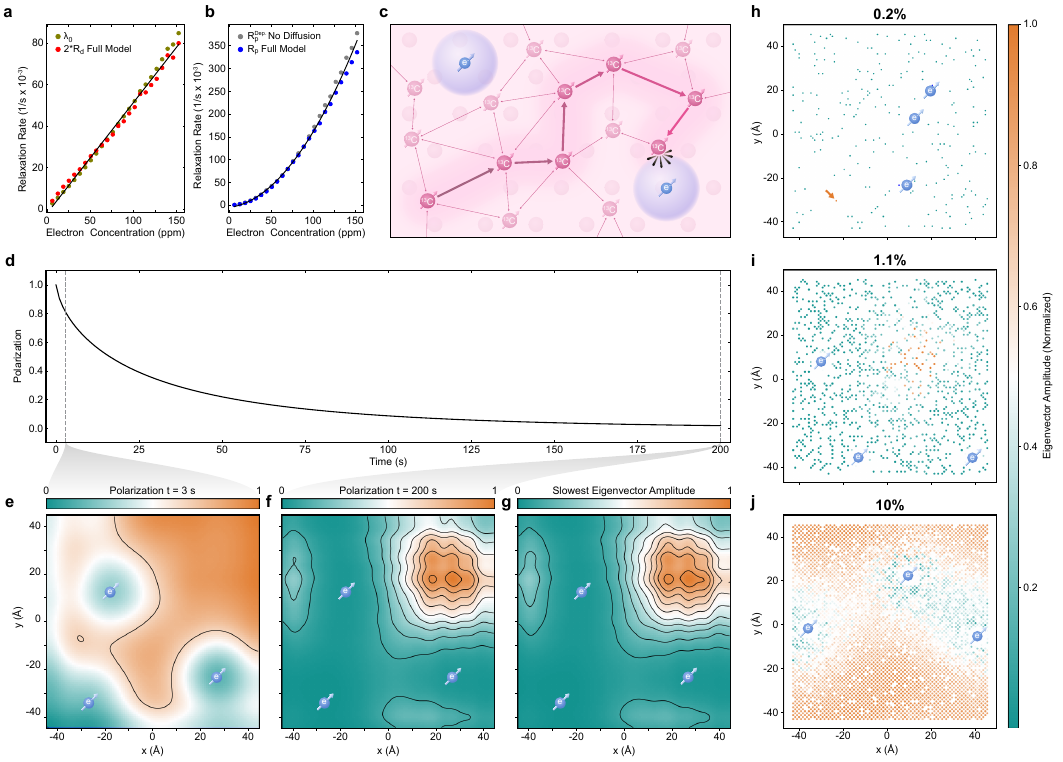}
    \caption{\textbf{Microscopic origins of emergent dynamics.} (a)-(b) \emph{\textbf{Eigenvalue and rate scaling}} with electron concentration. (a) Mean slowest eigenvalue $\langle \lambda_0 \rangle$ alongside $2R_d$; solid line is linear guide to eye. Scaling indicates $\langle \lambda_0 \rangle$ is origin of monoexponential component in \zr{emerge}. (b) $R_p$ from full Monte Carlo model with $R_p^{\text{dep}}$ (excluding diffusion); solid line is quadratic guide to eye. Results show $R_p$ is electron-mediated on-site decoherence channel independent of diffusion. (c) \emph{\textbf{Random walks in random environments}} conceptual schematic, drawing analogy of polarization diffusion to particle walks in media with randomly distributed static traps (electrons). (d) \emph{\textbf{Polarization decay}} averaged over 100 $^{13}$C configurations for fixed electron positions, decay follows \zr{emerge}. (e)-(f) \emph{\textbf{Polarization heatmaps}} showing polarization (colorbar) projected onto $xy$-plane at $t=3$ s and $t=200$ s (dashed lines in d). Black contours mark equal polarization levels. At late times polarization is confined to electron-free domains. (g) \emph{\textbf{Profile of slowest eigenmode}} displayed similarly, shows close match to the confinement pattern in (f). (h-j) \emph{\textbf{Slowest eigenmode profile versus $^{13}$C concentration}}. 2D projections as in (g) for single $^{13}$C configurations at 0.2\%, 1.1\%, and 10\% concentrations, respectively. Panels (h-i) show trapped regions (orange) similar to (f-g). At higher concentrations (j), increased network connectivity progressively eliminates trap-free domains, marking transition to diffusion dominated regime and hydrodynamic behavior (\zfr{fig3}a). For eigenvalue spectra, see SI Sec.~\ref{subsection_eigenvalues}.}
    \label{Fig4}
    \zfl{fig4}
\end{figure*}

\vspace{0.5em}
\noi {\normalsize\bfseries Subdiffusive polarization transport\par}
\noi Motivated by strong agreement with experiment, we use the model to probe the microscopic nature of polarization transport. Starting from an initially localized $^{13}$C polarization (see Methods and SI Sec.~\ref{subsection_transport}), the mean-squared displacement follows 
$\langle r^2(t) \rangle = 6Dt^{\alpha}$, with $\xa = 0.85< 1$, indicating anomalous, \emph{subdiffusive} transport, and with $D\simeq3.86$ \r{A}$^{2}$/s$^{0.85}$ (\zfr{fig3}a). This behavior challenges the standard assumption that polarization transport in solids is inherently diffusive~\cite{abragamPrinciplesNuclearMagnetism1962, khutsishviliSPINDIFFUSION1966, ramanathanDynamicNuclearPolarization2008a}, showing instead that dilution and disorder qualitatively alter dynamics. Increasing the $^{13}$C concentration drives a crossover to normal diffusion (shaded region, \zfr{fig3}a), with $\xa \rt 1$ near 20\% enrichment, yet the emergent law \zr{emerge} remains valid across the entire range.
\newline\indent
We next vary electronic and nuclear spin concentrations, plotting the resulting relaxation rates as a 2D map where color encodes the ratio $R_{p}/R_{d}$, capturing the relative strength of the two relaxation mechanisms (\zfr{fig3}b). High electron and low $\Cs$ concentrations (blue region), yield a diffusion-limited regime ($\rightarrow e^{-\sqrt{R_{p}t}}$), dynamically equivalent to Regime \T{II}. Conversely, high $\Cs$ and low electron concentrations (red region) produce a diffusion-dominated regime ($\rightarrow e^{-R_{d}t}$), dynamically equivalent to Regime \T{III}. These represent the two limiting cases of \zr{emerge}. Most prior studies~\cite{bloembergenInteractionNuclearSpins1949a, blumbergNuclearSpinLatticeRelaxation1960b, simmonsNuclearSpinLatticeRelaxation1962, tseNuclearSpinLatticeRelaxation1968a, furmanNuclearSpinlatticeRelaxation1997a}, primarily focused on ordered systems, fall within these two limiting cases (see SI Sec.~\ref{section_context} for a comparison). In contrast, our results (star in \zfr{fig3}b) in the doubly disordered system here occupy the central region where electron-induced noise and nuclear transport are comparably important.

\vspace{0.5em}
\noi
{\normalsize\bfseries Microscopic origins of the emergent behavior\par}
\noi To elucidate the microscopic origins of the observed decoherence, we consider the matrix $M=W+R$ introduced in \zr{hopping_model}. For each lattice realization in the Monte Carlo model, the evolution can be expanded in terms of eigenmodes,
\begin{equation}
\begin{aligned}
p(t)=\sum_{j=0}^{N-1}a_{j}e^{-\lambda_{j}t},
\zl{eigenmode}
\end{aligned}
\end{equation} 
where eigenvalues $\lambda_{j}$ are sorted in ascending order. Since $N\gg 1$, this can be approximated as (see SI Sec.~\ref{subsection_derivation}), 
\begin{equation}
\begin{aligned}
p(t)\sim a_{0}\exp(-\lambda_{0}t)\exp(a^{-1}_{0}\sum_{j=1}^{N-1}a_{j}e^{-(\lambda_{j}-\lambda_{0})t}),
\label{eigenmodes}
\end{aligned}
\end{equation}
The resulting expression mirrors \zr{emerge}, identifying $e^{-\xl_0t}$ with $e^{-R_{d}t}$, while the superposition of higher modes may yield the stretched-exponential component $\propto e^{-\sqrt{R_{p}t}}$.
\newline\indent
To verify this interpretation, we computed the mean of the slowest eigenvalue, $\expec{\xl_0}$, across 400 random lattice realizations. The same configurations were then used to generate decay trajectories, which were averaged and fit to \zr{emerge} to extract $R_{d}$. Across a range of electron concentrations, we find $\expec{\xl_0}\simeq 2R_{d}$ (\zfr{fig4}a), identifying $\expec{\xl_0}$ as the microscopic origin of the mono-exponential component. Comparing Eq.~(\ref{eigenmodes}) and \zr{emerge}, this then identifies the stretched exponential component as arising from the remaining modes.
\newline\indent
The connection between the eigenmodes of the microscopic model and the macroscopic rates $R_{p}$ and $R_{d}$ offers deeper insight into the underlying physics. To probe the origin of $R_{p}$, we performed simulations with the hopping term in \zr{hopping_model} eliminated ($W=0$), denoting $R_p$ here as $R_{p}^{\rm dep}$, and compared the resulting dynamics to those from the full matrix $M$ (\zfr{fig4}b). The identical values of $R_{p}$ in both cases demonstrate that the stretched component arises from collective, long-range electron-induced relaxation, acting as an on-site decoherence channel independent of diffusion, as schematized in \zfr{fig1}d. Consequently, $R_{d}$ (or $\expec{\xl_0}$) encodes all information about diffusive contributions to the decoherence process (\zfr{fig1}e). This interpretation parallels hard-sphere trapping models~\cite{kirkpatrickTimeDependentTransport1982, grassbergerLongTimeProperties1982}, where polarization takes on the role of the random walker and randomly distributed electrons serve as static traps which terminate the walker upon contact (\zfr{fig4}c).
\newline\indent
To refine this connection, \zfr{fig4}d tracks the net $^{13}$C polarization for 100 random $^{13}$C configurations with fixed electron positions. \zfr{fig4}e-f visualize the evolving polarization in the $^{13}$C network at short ($t{=}3$s) and long ($t{=}200$s) times. Over time, polarization is drawn into electron-centered relaxation basins (\zfr{fig1}e), generating pronounced spatial inhomogeneities. At late times, polarization becomes confined within \emph{isolated clusters} of nuclear spins remote from any electrons and weakly coupled to the rest of the network. Remarkably, these clusters closely match the spatial profile of the eigenmode associated with $\lambda_{0}$ (\zfr{fig4}g), enabling a priori prediction of the polarization confinement pattern. The emergent mono-exponential contribution thus reflects the gradual depletion of these isolated polarization reservoirs via spin diffusion toward electron sinks. A single realization is shown in SI Sec.~\ref{subsection_polarization}. 
\newline\indent
Acting as long-lived reservoirs of polarization, these isolated clusters are important to the long lifetimes observed in this work. Their formation is a direct consequence of the doubly disordered network—random in both electron and nuclear configurations—which creates trap-free regions that are otherwise suppressed in ordered systems. As shown in SI Sec.~\ref{section_disorder}, arranging the electrons instead in an ordered configuration reduces these regions and accelerates relaxation. Similarly, \zfr{fig4}h-j show that increasing $\Cs$ enrichment expands the spatial profile of the slowest eigenmode across the network, erasing trap-free domains, driving the system into the diffusion-dominated regime, and shortening lifetimes (see SI Sec.~\ref{section_relaxationlandscape}). 

\vspace{0.5em}
\noi {\normalsize\bfseries Outlook\par}

\noi Our results illustrate how complex many-body dynamics can yield simple emergent laws at the macroscopic scale. In the doubly disordered electron–nuclear spin network studied here, decoherence obeys a factorized law arising from two elementary channels driven by diffusion and on-site fluctuations. Because electron—or two-level-fluctuator—mediated decoherence is pervasive across quantum systems~\cite{sangtawesinOriginsDiamondSurface2019, siddiqiEngineeringHighcoherenceSuperconducting2021}, such emergent behavior likely extends to a wide class of solid-state and molecular platforms~\cite{lunghiHowPhononsRelax2019, onizhukColloquiumDecoherenceSolidstate2025}. The results identify practical routes to suppress nuclear decoherence, for instance through all-optical electron decoupling or the design of materials in which electronic states can be dynamically shelved~\cite{capozziThermalAnnihilationPhotoinduced2017} to prolong nuclear spin lifetimes.
\newline\indent
\zfr{fig4} shows that disorder—typically viewed as detrimental~\cite{gaoEffectsDisorderSuperconducting2025}—here acts protectively by producing localized clusters that trap polarization and retard relaxation. This suggests that \I{engineered disorder} can serve as a resource for coherence preservation. The observation that long-time decay is dominated by a single slow eigenmode localized within isolated network clusters links this behavior to percolation and random-network physics~\cite{chakrabartiQuantumSemiclassicalPercolation2009, wiersmaRandomQuantumNetworks2010}, providing new conceptual tools for controlling decoherence in complex quantum media.
\newline\indent
Targeted polarization transfer into electron-free clusters could create long-lived spin domains for quantum memories and sensing~\cite{abobeihAtomicscaleImaging27nuclearspin2019, sahinHighFieldMagnetometry2022c}. More broadly, the combined use of Hamiltonian engineering and (optical) environment control demonstrated here offers a deterministic means to manipulate individual decoherence pathways, establishing a framework for steering emergent dynamics across diverse open quantum systems.


\section*{Methods}

\noi{\normalsize\bfseries Sample -} Experiments are carried out using a $3.0\times 3.0\times 0.3$ mm$^{3}$, (100)-cut, type-Ib single-crystal diamond (Element6). The diamond hosts NV centers at a concentration of $\approx$1 ppm and P1 centers at $\approx30$ ppm. The $^{13}$C isotope is present at its natural abundance of $1.1$\%. At this concentration, the average nearest-neighbor distance between $^{13}$C nuclei is $\approx4.5$ \r{A} (with a standard deviation of 1.63 \r{A}), corresponding to a typical dipolar coupling strength of $\approx80$ Hz. The nuclear free induction decay (FID) time, $T_{2}^{\star}$, is $\approx1.5$ ms, and the nuclear $T_{1}$ time at 100 K and 9.4 T is $\approx3100$ s~\cite{dsouzaCryogenicFieldcyclingInstrument2025}.
\newline
\noi{\normalsize\bfseries Experiment Apparatus -} The experimental setup utilizes a custom-built cryogenic field-cycling system, as described in~\cite{dsouzaCryogenicFieldcyclingInstrument2025}. The diamond sample is housed in a glass sample tube, which is mounted on a home-built NMR probe. The probe includes a planar loop coil situated beneath the sample for microwave (MW) excitation during hyperpolarization and a copper saddle coil for RF control and detection of $^{13}$C nuclear spins at high magnetic field ($\sim100$ MHz). The NMR probe is positioned inside a continuous-flow cryostat (Oxford Instruments), maintained at $100$ K via continuous liquid nitrogen flow. The cryostat features vacuum-sealed glass windows at the bottom of each chamber, enabling optical access for laser illumination. Optical pumping is achieved using a $532$ nm continuous-wave laser (Verdi G8, Coherent), modulated by a mechanical shutter (Thorlabs) with millisecond-scale timing resolution. For magnetic field cycling, the cryostat is mounted on a belt-driven actuator that transports the entire system from low field ($27$ mT) to high field ($9.4$ T) over a $90$ second interval at a translation speed of $7$ mm/s. 
\newline
\noi{\normalsize\bfseries Experiment Protocol -} The experimental sequence begins at low magnetic field (27 mT), where NV centers are optically polarized using 7 W of laser power for 120 seconds. During this period, polarization is transferred from the electron spins of the NV centers to nearby $^{13}$C nuclear spins via chirped microwave excitation, as described in \cite{ajoyOrientationindependentRoomTemperature2018a, sarkarRapidlyEnhancedSpinPolarization2022a}. Following hyperpolarization, the cryostat is mechanically shuttled from low to high magnetic field (9.4 T) over a duration of 90 s. Upon reaching high field, a Floquet driving protocol is applied to the $^{13}$C nuclei, consisting of an initial $\xt_{y}$ pulse followed by a train of $\xt_{x}$ pulses. In experiments with $\xt = 90^{\circ}$, the pulse duration is 38 $\mu$s, and the full sequence consists of approximately 8 million pulses. While the nominal flip angle $\xt$ varies across the three regimes, the delay between pulses remains fixed at 40 $\mu$s. Following each $\xt_{x}$ pulse, the free induction decay (FID) of the nuclear spin ensemble is recorded and digitized at a 600 MHz sampling rate over a 2 $\mu$s acquisition window. The signal is then Fourier transformed, and the amplitude of the resulting spectrum is tracked over successive pulses to assess the decay dynamics. More details of this procedure have been described previously, see Ref.~\cite{moonHighspeedHighmemoryNMR2025}.
\begin{figure}[b]
    \centering
    \includegraphics[width=0.9\columnwidth]{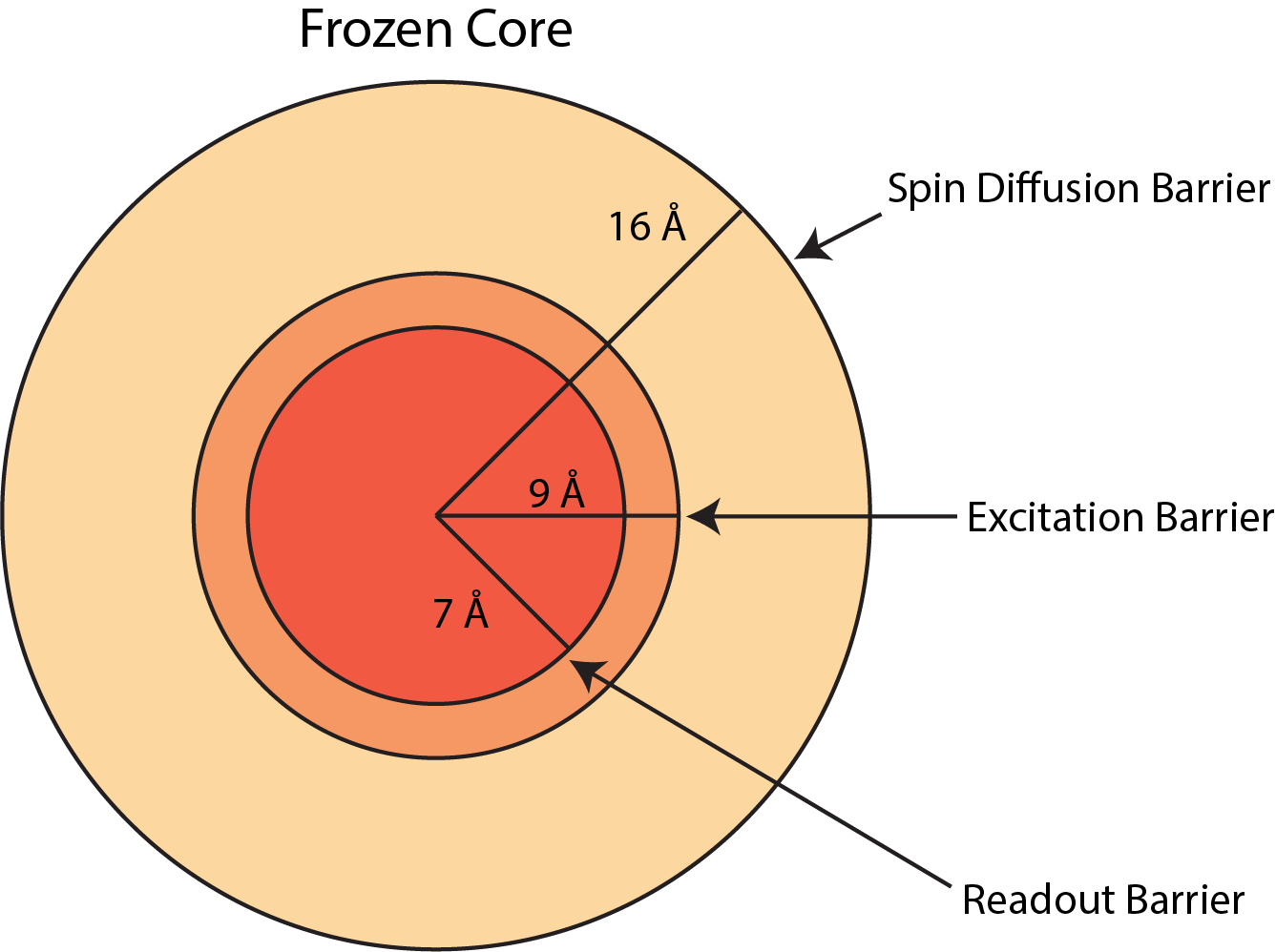}
    \caption{A detailed breakdown of the frozen core, into regions that fall within the excitation bandwidth and detection window, relative to their distance from the electron. Frozen-core spins within the red inner-most region are neither excited nor detected. A small shell of frozen-core spins (orange, middle region) may in principle be observed, but are not excited. The outer layer of frozen-core spins (light orange) are excited and observed, but decay rapidly due to their proximity to the electron.}
    \label{FrozenCore}
\end{figure}
\newline
\noi{\normalsize\bfseries Readout and Excitation Windows -} Each electron spin creates a ``frozen core'' region in which nearby $^{13}$C nuclei experience large hyperfine shifts and are approximately decoupled from the broader nuclear network~\cite{wolfeDirectObservationNuclear1973, ramanathanDynamicNuclearPolarization2008a, priscoScalingAnalysesHyperpolarization2021, sternDirectObservationHyperpolarization2021a}. The radius of this frozen core can be estimated using the expression from Khutsishvili~\cite{khutsishviliSPINDIFFUSIONNUCLEAR1969},
\begin{equation}
r_{c} = a\left(P_{e}\frac{\gamma_{e}}{\gamma_{C}}\right)^{1/4},
\end{equation}
where $P_{e}$ is the electron spin polarization at 100 K, $\gamma_{e}$ and $\gamma_{C}$ are the gyromagnetic ratios of the electron and $^{13}$C, respectively, and $a$ is the average nearest-neighbor distance between $^{13}$C nuclei ($4.5$ \r{A}). This yields a spin diffusion barrier radius of approximately 16 \r{A}. The frozen core can be further divided into three regions as illustrated in Fig.~\ref{FrozenCore}: a readout window, defined by the finite acquisition time of the FID; an excitation window, determined by the finite RF pulse bandwidth; and the spin diffusion barrier, set by the electron–nuclear coupling strength. These regions together determine which nuclei contribute to the measured signal. 

Each FID is acquired for 2 $\mu$s at a 600 MHz sampling rate, resulting in Fourier transform bins of 500 kHz. Because the decay of the ``central" Fourier amplitude is used to infer decoherence, this amplitude effectively captures magnetization from spins within $\pm250$ kHz of the $\sim100$ MHz Larmor frequency of the bulk $^{13}$C nuclei. A $^{13}$C–electron dipolar coupling of 250 kHz corresponds to a distance of 4.3 \r{A}. However, spins within a 7 \r{A} radius also experience strong Fermi contact interactions, and are therefore shifted outside the detection window~\cite{galiInitioSupercellCalculations2008}.
Additionally, the excitation bandwidth for our pulse parameters is approximately 25 kHz, corresponding to a nuclear–electron distance of 9.3 \r{A}. Spins located at shorter distances are thus weakly excited and do not contribute to the observable signal. Fig.~\ref{FrozenCore} illustrates the combined excitation and readout windows together with the 16 \r{A} frozen core. We conclude that nuclei situated in the light-orange region of the frozen core ($\approx$9.3 \r{A} to $\approx$16 \r{A}) are excited and observed. However, these spins relax rapidly due to their proximity to the electron, providing a possible explanation for the transient period and the deviations from the model observed in the first 10 seconds of decay. 
\newline
\indent
We also note that the experimental observations beyond the transient period in \zfr{fig1} can be fitted using a Kohlrausch–Williams–Watts (KWW) form, $e^{-(t/\tau)^{\gamma}}$ . While such fits reproduce the data, they require varying power-law exponents ($\gamma$) across the dataset. As a result, the KWW ansatz offers no additional insight into the underlying physical decoherence dynamics compared to our factorized decoherence law.
\newline
\noi{\normalsize\bfseries Floquet Engineering -} Below, we outline the derivation of the scaled nuclear-nuclear ($d_{ij}$) and electron-nuclear ($h_{i\mu})$ dipolar coupling constants under periodic driving.
For a complete analysis we refer to SI Sec.~\ref{subsection_model}. For a general periodic pulse sequence the propagator takes the form $U_{\rm rf}(t)=P(t)e^{-i \overline{H}_{\rm rf}t}$, where $P(t+T)=P(t)$ represents the periodic part of the motion, and $\overline{H}_{\rm rf}=\omega_{\rm eff}\overline{{\bf n}}_{z}\cdot \bf{I}$ sets the new quantization ``$z$-axis''. We analyze the dynamics in the combined interaction frame of the micromotion and effective quantization axis: $\tilde{Q}(t)=V(t)QV^{\dagger}(t)$, where $V(t)=e^{+i \overline{H}_{\rm rf}t}P^{\dagger}(t)$. This leads to a bimodal Floquet expansion of the interaction frame operators: $\tilde{Q}(t)=\sum_{nk}Q^{(nk)}e^{+i (n \omega_{\rm eff}+k\omega_{d})t}$, where $\omega_{d}=2\pi/T$ is the driving frequency.
\newline
\indent
The effective hopping rates are related to the static part $H^{(00)}_{\rm nn}$ of the nuclear-nuclear interaction frame Hamiltonian
\begin{equation}
\label{eq:coupling_scaling}
\begin{aligned}
&H^{(00)}_{\rm nn}=\kappa\sum_{i<j}\;d_{ij}\{3 ({\bf I}_{i}\cdot\overline{{\bf n}}_{z})({\bf I}_{j}\cdot\overline{{\bf n}}_{z})-{\bf I}_{i}\cdot {\bf I}_{j}\},
\\
&\kappa= \sum_{m=-2}^{+2}T^{-1}\int_{0}^{T}d^{2}_{0m}(\theta_{\rm eff})e^{+i m \phi_{\rm eff}}D^{2}_{m0}{[}P^{\dagger}(t){]}dt,
\end{aligned}
\end{equation}
Here, $\theta_{\rm eff}$ and $\phi_{\rm eff}$ parameterize the effective quantization axis, $d^{l}_{mn}$ are reduced Wigner matrix elements, and $D^{l}_{m0}{[}P^{\dagger}(t){]}$ are Wigner matrix elements parametrized by the micromotion $P(t)\in {\rm SU}(2)$. The driving protocol dependent scaling factor $\kappa$ may be tuned to access different dipolar hopping regimes. In particular, $\kappa$ approximately vanishes in Regime \T{II}. The bimodal Floquet expansion of the hyperfine interactions is given by
\begin{equation}
\begin{aligned}
&\tilde{H}_{\rm ne}(t)=\sum_{i,\mu}h_{i\mu}\sum_{n=-1}^{+1}\sum_{k=-\infty}^{+\infty}c^{q}_{k}({\bf I}_{i}\cdot\overline{{\bf n}}_{q})S^{z}_{\mu}e^{+i (n \omega_{\rm eff}+k\omega_{d})t},
\\
&c^{q}_{k}=T^{-1}\sum_{m=-1}^{+1}e^{+i m \phi_{\rm eff}}d^{1}_{qm}(\theta_{\rm eff})\int_{0}^{T}D^{1}_{m0}{[}P^{\dagger}(t){]}e^{-i k \omega_{d}}dt,
\end{aligned}
\end{equation}
where $h_{i\mu}$ is the dipolar coupling constant between nuclear spin $i$ and paramagnetic impurity $\mu$. For far off-resonant driving in Regime \T{III} ($\theta_{\rm eff}\rightarrow 0$), the $n\neq 0$ components responsible for electron-mediated on-site relaxation are strongly non-secular and become increasingly suppressed. For the reference case, Regime \T{I}, neither contribution vanishes in general, and both channels remain active. 
\newline
\noi{\normalsize\bfseries Monte Carlo Simulations -} The decoherence dynamics are described using a semi-classical random hopping model. For a complete treatment, we refer to SI Sec.~\ref{subsection_model}. The model incorporates the exact diamond lattice geometry, with lattice sites randomly populated by electrons (30 ppm) and $^{13}$C nuclei (1.1\%). Paramagnetic decoherence is modeled as a spatially inhomogeneous fluctuating hyperfine field, where each electron contributes to on-site depolarization with a strength that scales as $1/r^{6}$. Polarization migration is implemented via a dipolar hopping process governed by Fermi’s golden rule~\cite{abragamNuclearMagnetismOrder1982}. Within the model, both the decoherence and hopping rates are random in nature. To capture the disorder-induced effects, we perform a configurational average over different lattice realizations. For any particular configuration the polarization dynamics are determined by: 
\begin{equation}
\begin{aligned}
\dot{p}(t) = (W+R)p(t),
\end{aligned}
\end{equation}
where $p(t)$ contains the time-dependent polarization of each nuclear spin, and the matrices $W$ and $R$ account for dipolar polarization transport and electron-induced decoherence, respectively. The hopping rates describing the diffusive dynamics can be expressed as follows
\begin{equation}
\label{eq:coupling_scaling}
\begin{aligned}
W_{ij}=\kappa^{2}d_{ij}^{2}T_{2},
\end{aligned}
\end{equation}
where $d_{ij}$ is the dipolar coupling constant between nuclear spins $i$ and $j$, $T_{2}$ is the intrinsic nuclear coherence time which we treat as a model parameter, and $0<\kappa<1$ is the driving protocol dependent dipolar scaling factor based on a bimodal high-frequency expansion of the Floquet Hamiltonian (see SI Sec.~\ref{subsection_model}).
\newline
\indent
For a fluctuating spin model, the relaxation rates due to electronic dipolar field fluctuations are approximately given by:
\begin{equation}
\begin{aligned}
&R_{ij} = -\eta J_{\rm env}(\omega_{\rm eff})\delta_{ij},
\\
&J_{\rm env}(\omega)= \left\{\sum_{k=-\infty}^{+\infty}J_{\rm e}(\omega+k \omega_{d})c^{+1}_{k}c^{-1}_{k}\right\}\sum_{\mu=1}^{N_{e}}h^{2}_{i\mu}.
\end{aligned}
\end{equation}
Here, $\delta_{ij}$ is the Kronecker delta and $\eta$ is a free parameter accounting for the limited number of electrons in the simulation volume and possible deviations in the exact experimental conditions. $J_{\rm env}(\omega)$ represents the filtered electron noise spectral density, derived from a high-frequency expansion of the Floquet Hamiltonian (see SI Sec.~\ref{subsection_model}), and $J_{e}(\omega)$ represents the bare electron spectral density assumed to be Lorentzian.
\newline
\indent
The effects of laser illumination are incorporated into the model by extending the NV center optical pumping framework described in Ref.~\cite{hincksStatisticalInferenceQuantum2018}. Within this approach, the NV center spin states are described within an effective Hilbert space, and their population dynamics are governed by a master equation formulated using Lindblad dissipators. We augment this model by including $T_{1}$ relaxation processes within the ground-state manifold (see SI Sec.~\ref{subsection_lasereffects} for more details).
\newline
\indent
To compare with the experimental observations in \zfr{fig2}c-e across all three dynamical regimes of the Floquet driving, we proceed as follows. We allow $\eta$ to vary across each of the regimes to compensate for pulse imperfections, resonance frequency distributions, and possible deviations in the exact experimental conditions, which cannot be captured by the idealized filtered density $J_{\rm env}(\omega)$. The values of $\eta$ used are $1.5\times10^{-3}$, $2.0\times10^{-3}$ and $3.4\times10^{-5}$ in Regimes \T{I}, \T{II}, and \T{III}, respectively. Additionally, laser illumination preferentially populates the $m_{s}=0$ state of the NV center and enhances fluctuations in the surrounding electron spin bath. These effects reduce the effective hyperfine field experienced by the nuclei, thereby enhancing diffusive behavior. To capture these effects, we allow $T_{2}$ to vary linearly with laser power from $2.5\times10^{-5}$ to $5.0\times10^{-5}$, using the same linear dependence across all three regimes. 

To investigate the spin diffusion process, we ignore relaxation effects ($R = 0$), and initialize the polarization on a single nuclear spin located at the origin. As a measure of the diffusive character, we compute the mean squared displacement of polarization, $\langle r^2(t) \rangle$, as a function of time. For a general diffusive process, the mean squared displacement typically follows a power-law dependence on time
\begin{equation}
\label{r^2}
\begin{aligned}
\langle r^2(t) \rangle = 6Dt^{\alpha}.
\end{aligned}
\end{equation}
To extract the diffusion exponent $\alpha$ and diffusion coefficient $D$, we calculate the configurational averaged $\langle r^2(t) \rangle$, and fit the resulting data to Eq.~\ref{r^2}. For more details including a finite-size scaling analysis, see SI Sec.~\ref{subsection_transport}.

\vspace{0.5em}
\noi {\normalsize\bfseries Acknowledgements\par}
\noi We gratefully thank D. Suter, C. Ramanathan, and L. J. I. Moon for insightful discussions. This work was supported in part by the U.S. Department of Energy National Nuclear Security Administration through the NNSA Office of Defense Nuclear Nonproliferation R\&D through the LB24-NV center $^{13}$C quantum sensor-PD3Ta project and the Nonproliferation Stewardship Program (NSP). We additionally acknowledge funding from ONR (N00014-20-1-2806), AFOSR YIP (FA9550-23-1-0106), and instrumentation support from AFOSR DURIP (FA9550-
22-1-0156) and NSF MRI (2320520). CMS acknowledges the NDSEG fellowship. 

\vspace{0.5em}
\noi{\normalsize\bfseries Author Contributions\par}
\noi CMS performed measurements with assistance from CS and ZZ, and analysed the data. CMS, CB, ZZ and CS built instrumentation. CMS and CB performed Monte-Carlo analysis, CB performed theoretical analysis. AA conceived the research and supervised the project. CMS, CB and AA wrote the paper with input from all authors.

\noi{\normalsize\bfseries Competing Interests\par}
\noi The authors declare no competing interests

\clearpage
\renewcommand{\tocname}{Supplementary Information}
\tableofcontents

\section{Summary}
\label{section_summary}

In this Supplementary Information, we provide additional context for our results by comparing them with previous studies and presenting further details of the Monte Carlo simulations and eigenmode analysis of the relaxation dynamics. Section~\ref{section_context} reviews key prior experimental studies on nuclear relaxation driven by paramagnetic impurities and highlights the novel contributions of our work in this area. In Sec.~\ref{section_montecarlo}, we describe the random hopping model used to simulate the relaxation dynamics, including a detailed discussion of the model’s assumptions, free parameters, and the procedure used to compute the process matrix $M$ (sub-section~\ref{subsection_model}). We also discuss the interplay between the noise filter function of the pulsed spin-locking sequence and the electron noise spectral density (see Fig.~\ref{SpectralOverlap}). Sub-sections~\ref{subsection_lasereffects} and~\ref{subsection_transport} detail the incorporation of laser illumination effects into the model and the analysis of polarization transport used to extract $\alpha$ and $D$, as well as a finite-size scaling analysis. Additional subsections examine the implications of the model for all-optical decoupling (sub-section~\ref{subsection_allopticaldecoupling}) and the temperature dependence of the dynamics (sub-section~\ref{subsection_temperature}). Section~\ref{section_relaxationlandscape} provides additional insight into the relaxation landscape shown in Fig.~3b of the main text by presenting a one-dimensional slice as a function of $^{13}$C concentration at fixed electron concentration (see Fig.~\ref{Concentrations}). Section~\ref{section_eigenmode} expands on the eigenmode decomposition of the relaxation dynamics, including a derivation of Eq.~4 from the main text (sub-section~\ref{subsection_derivation}). In sub-section~\ref{subsection_polarization}, we show that in the long-time limit, the nuclear polarization aligns with the spatial profile of the slowest decaying eigenmode for a fixed nuclear and electron configuration (see Fig.~\ref{PolarizationVsEigenvector}), complementing Figs.~4d–g of the main text. Sub-section~\ref{subsection_eigenvalues} presents the eigenvalue spectrum of the process matrix $M$ for various $^{13}$C concentrations (Fig.~\ref{RelaxationEigenvalues}), complementing Figs.~4h–j of the main text. In Sec.~\ref{section_disorder}, we show that disorder in the electron network leads to prolonged relaxation lifetimes. Section~\ref{section_fitting} presents a goodness-of-fit analysis based on residuals between experimental data and the emergent relaxation law. Finally, in Sec.~\ref{section_heating}, we demonstrate that the effects of laser illumination on the relaxation rates cannot be trivially attributed to sample heating.

\section{Comparison with previous Literature} \label{section_context}

The relaxation of nuclear spins in solids containing dilute paramagnetic centers has a storied history and has been investigated for over 75 years. Beginning with Bloembergen (1949)~\cite{bloembergenInteractionNuclearSpins1949a}, it was recognized that nuclear $T_{1}$ reflects an interplay between local electron–nuclear coupling and transport of nuclear Zeeman energy by spin diffusion. In that framework, the nuclear spin-diffusion constant is given by the seminal estimate $D=a^{2}/(50T_{2})$ where $a$ is the lattice spacing. Subsequent work by Blumberg (1960)~\cite{blumbergNuclearSpinLatticeRelaxation1960b} found that immediately after saturation, before polarization gradients are established, the magnetization recovery exhibits a $\sqrt{t}$ behavior, followed by an exponential dependence at longer times. Closely related experiments by Simmons, et al. (1962)~\cite{simmonsNuclearSpinLatticeRelaxation1962} on $^{27}$Al in sapphire showed that even in quadrupolar systems, the recovery of Zeeman magnetization is effectively monoexponential, consistent with rapid diffusion homogenizing the nuclear spin temperature. A critical advance came with the work of Tse and Hartmann (1968)~\cite{tseNuclearSpinLatticeRelaxation1968a}, who introduced magic-angle spin locking experiments to suppress flip–flops, isolating the diffusionless regime in which each nucleus relaxes independently via direct coupling to surrounding paramagnetic centers. In this limit, they observed a stretched-exponential decay of the form $e^{-\sqrt{t}}$. 

\begin{figure*}[t]
    \centering
    \includegraphics[width=0.9\textwidth]{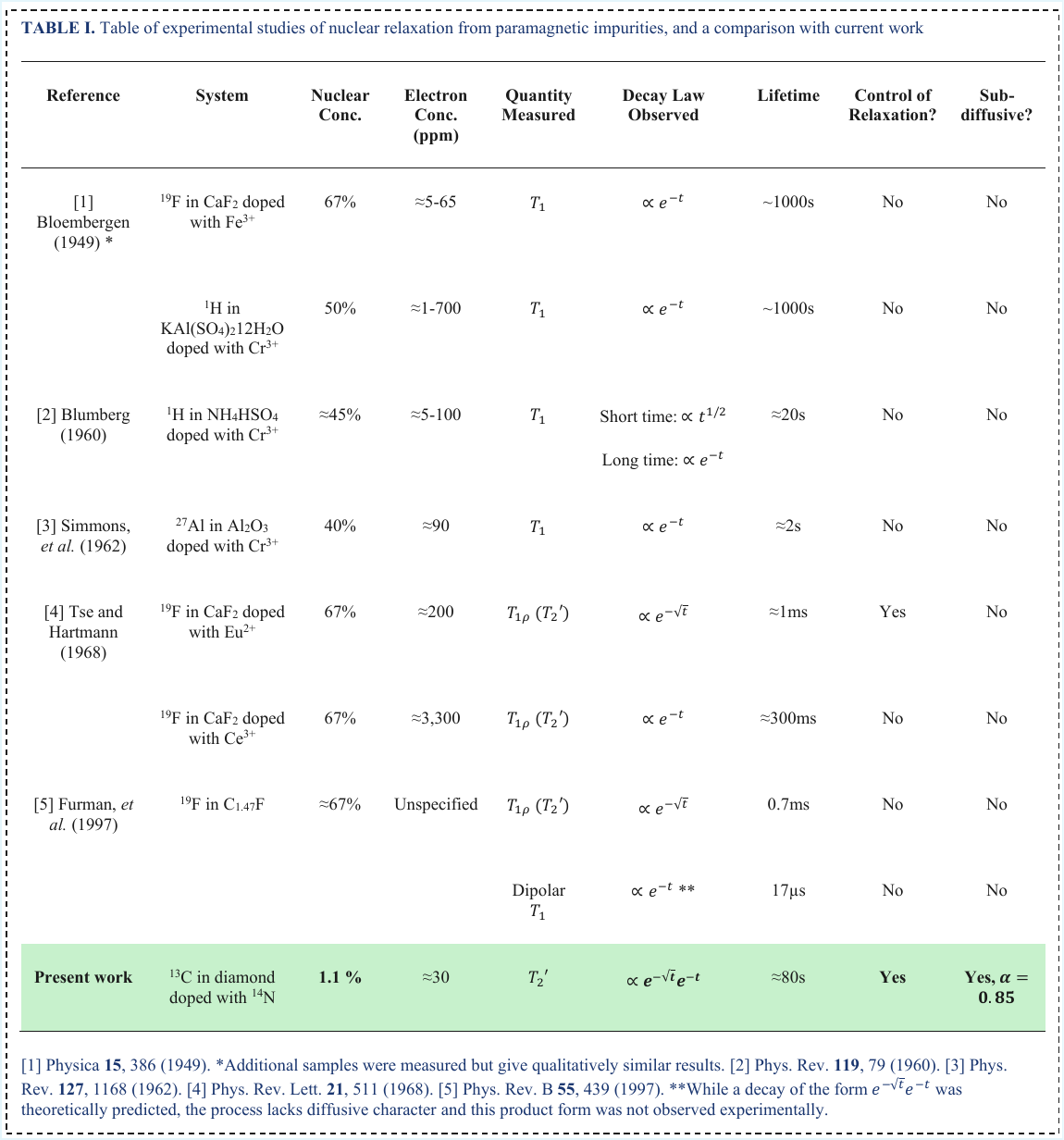}
    \caption{\textbf{Comparison with previous literature.} Table shows a comparison of key quantities from experimental studies on the relaxation of nuclear spins in solids containing paramagnetic impurities with that of the current work. }
    \label{Table}
\end{figure*}

All of these pioneering studies focused  on systems with highly abundant nuclear spins—such as $^{19}$F in CaF$_{2}$ or $^{27}$Al in sapphire—guaranteeing strong average nearest-neighbor dipolar coupling and rapid diffusion in the absence of any externally applied radio-frequency fields. By contrast, our work provides the first experimental investigation of nuclear relaxation in a dilute spin network, using natural abundance $^{13}$C (1.1\%)—a regime made accessible only through recent breakthroughs in optical hyperpolarization that enhance signal-to-noise by over three orders of magnitude~\cite{ajoyOrientationindependentRoomTemperature2018a, sarkarRapidlyEnhancedSpinPolarization2022a}.
 
This low connectivity fundamentally alters the transport: we find that nuclear spin diffusion in this regime is sub-diffusive, a behavior that has been largely overlooked in previous studies due to the well-percolated, diffusive nature of high-abundance spin networks. The closest antecedent on diffusion in rare nuclear species is the treatment of Goldman and Jacquinot (1982)~\cite{goldmanNuclearSpinDiffusion1982a} for $^{43}$Ca in CaF$_{2}$ (0.045\% concentration), in which they predict an extremely small diffusion coefficient but did not report on sub-diffusion. While modern reciprocal-space measurements by Boutis et al. (2004)~\cite{boutisSpinDiffusionCorrelated2004} provided the first direct determination of the spin diffusion coefficient in abundant-spin CaF$_{2}$, comparable insight into dilute spin systems has remained limited. Our work helps close this gap by presenting some of the first evidence of anomalous, sub-diffusive spin transport in a sparse $^{13}$C network. More recently, Zu et al. (2021)~\cite{zuEmergentHydrodynamicsStrongly2021} reported ``Fickian yet non-Gaussian'' hydrodynamics in a dilute P1-center network—qualitatively deviating from normal diffusion—echoing our finding that sparse, disordered spin graphs exhibit anomalous transport statistics. 

Within this landscape, our central experimental result is a unifying framework captured by the product decay law for nuclear relaxation, $e^{-\sqrt{R_{p}t}}e^{-R_{d}t}$, which holds over hundreds of seconds. This factorized form unifies, in a single expression, the two canonical ingredients of impurity-driven relaxation: a paramagnetic channel from direct dipolar coupling to surrounding electron spins and a diffusive channel from polarization transport toward impurities. This result cleanly bridges the gap between two well-established limits: it reduces to the diffusionless form of Tse and Hartmann $e^{-\sqrt{R_{p}t}}$, when the diffusive pathway is quenched ($R_{d}\rightarrow0$), and to the rapid diffusion limit (e.g. Bloembergen/Blumberg) $e^{-R_{d}t}$ when diffusion dominates, as in a strongly-connected spin network. We emphasize that although Furman et al. (1997)~\cite{furmanNuclearSpinlatticeRelaxation1997a} derived a similar product form for the relaxation of dipolar-order, to the best of our knowledge, that prediction neither involved diffusion nor was it experimentally verified. By contrast, our results demonstrate the first experimental observation of such a product decay law. We trace its microscopic origins to two independently controllable channels, and show that it persists even when the nuclear transport itself is sub-diffusive. A further advance of our work is the development of Hamiltonian engineering and all-optical control techniques that allow us to independently modulate each relaxation pathway. For the first time, we demonstrate the ability to selectively suppress, enhance, or decouple either channel—establishing a new level of experimental control over nuclear spin relaxation. Additionally, we introduce a novel all-optical method to dynamically modulate paramagnetic impurities, laying the groundwork for new strategies to manipulate nuclear spin dynamics in solid-state systems.

\section{Monte Carlo Analysis}
\label{section_montecarlo}

\subsection{Relaxation in the Diffusionless Limit}
\label{subsection_diffusionless}

The relaxation of a single nuclear spin induced by a single nearby electron can be modeled by an exponential of the form ${\rm exp}(-\frac{At}{r^6})$, where $r$ is the distance between the nuclear spin and the electron, and $A$ is a coupling constant. The $1/r^6$ scaling reflects the fact that the relaxation rate is proportional to the square of the dipolar coupling, consistent with second-order perturbation theory (i.e., Fermi’s golden rule). We ignore the angular dependence here for simplicity ($A(\theta_{i},\phi_{i})=A)$. When a nuclear spin is surrounded by many paramagnetic impurities, each at a distance $r_{i}$, the total relaxation is given by:
\begin{equation}
\begin{aligned}
{\rm exp}\left(\sum_{i}-\frac{At}{r_{i}^{6}}\right) = \prod_{i}{\rm exp}\left(-\frac{At}{r_{i}^{6}}\right)
\end{aligned}
\end{equation}
In the experiments, the observed signal arises from an ensemble of nuclear spins, each experiencing a different local electronic environment due to the random spatial distribution of electron spins. This is accounted for by a configurational average
\begin{equation}
\begin{aligned}
S(t)=\left\langle \prod_{i} \exp\left(-\frac{At}{r_{i}^{6}}\right) \right\rangle_{\rm conf}
\end{aligned}
\end{equation}
Assuming the electron spins are distributed uniformly in 
$\mathbb{R}^{3}$ according to a homogeneous Poisson point process of density $\rho$, the ensemble average over spin configurations can be computed via the Laplace functional of the process~\cite{resnickAdventuresStochasticProcesses2002}. The resulting spatial integral yields a stretched exponential of the form 
\begin{equation}
\begin{aligned}
S(t)={\rm exp}\left[-(Bt)^{\frac{1}{2}}\right].
\end{aligned}
\end{equation}
More generally, for interactions that scale as $1/r^{\alpha}$, the stretching exponent becomes $d/\alpha$, where $d$ is the spatial dimensionality of the system~\cite{furmanNuclearSpinlatticeRelaxation1995a}.

\subsection{Random Hopping Model}
\label{subsection_model}

Semiclassical approximations to the polarization dynamics are generated by treating polarization transport as a Markovian hopping process. Starting from a diamond lattice of size $N\approx130,000$, we sample a particular lattice configuration by randomly occupying lattice sites with either a $^{13}$C nucleus or a paramagnetic impurity, using a binomial trial consistent with their respective concentrations. Periodic boundary conditions are applied to suppress finite-size effects and more accurately represent bulk transport dynamics. Around each paramagnetic impurity, we impose a spin diffusion barrier of $r_{c}=16$ \r{A}, in agreement with Eq.~5 in the Methods section. Spins that fall within $r_{c}$ are excluded from the simulations, as they do not effectively participate in the transport process. This results in a system of $N_C$ $^{13}$C nuclei and $N_e$ paramagnetic impurities, typically on the order of $N_C\approx$1300 and $N_e\approx$3. While simulations of larger configurations are possible, we find that this minimal system effectively captures the underlying dynamics, which remain largely unchanged with increasing system size.

Within the semiclassical approach, the polarization dynamics for a particular configuration are determined by: 
\begin{equation}
\begin{aligned}
\dot{p}(t) = (W+R)p(t)
\end{aligned}
\end{equation}
\cite{vugmeisterSpatialSpectralSpin1976a, goldmanNuclearSpinDiffusion1982a}, where $p(t)$ contains the time-dependent polarization of each nuclear spin. The matrices $W$ and $R$ account for dipolar polarization transport and electron-induced relaxation, respectively. To simulate the decoherence dynamics under detuned driving, we employ a bimodal high-frequency expansion of the Floquet Hamiltonian~\cite{scholzOperatorbasedFloquetTheory2010, bukovUniversalHighfrequencyBehavior2015, eckardtHighfrequencyApproximationPeriodically2015a, ivanovFloquetTheoryMagnetic2021}. In general, detuned driving necessitates bimodal Floquet theory, since the system’s dynamics are governed by two characteristic frequencies: the driving frequency $\omega_d$ and the effective energy gap $\omega_{\rm eff}$ of the system. Within the semi-classical picture the resulting hopping rates describing the diffusive dynamics can be expressed as follows~\cite{abragamNuclearMagnetismOrder1982, goldmanNuclearSpinDiffusion1982a}
\begin{equation}
\label{eq:coupling_scaling}
\begin{aligned}
W_{ij}=\kappa^{2}d_{ij}^{2}J_{\rm ZQ}(0),
\end{aligned}
\end{equation}
where $d_{ij}$ is the dipolar coupling constant between nuclear spins $i$ and $j$, given by 
\begin{equation}
\begin{aligned}
d_{ij}=-\frac{\mu_{0}}{4\pi}\frac{\hbar\gamma^{2}_C}{r^{3}_{ij}}\frac{1}{2}(3\cos^{2}(\theta_{ij}) - 1).
\end{aligned}
\end{equation}

\begin{figure}[t]
    \centering
    \includegraphics[width=0.9\columnwidth]{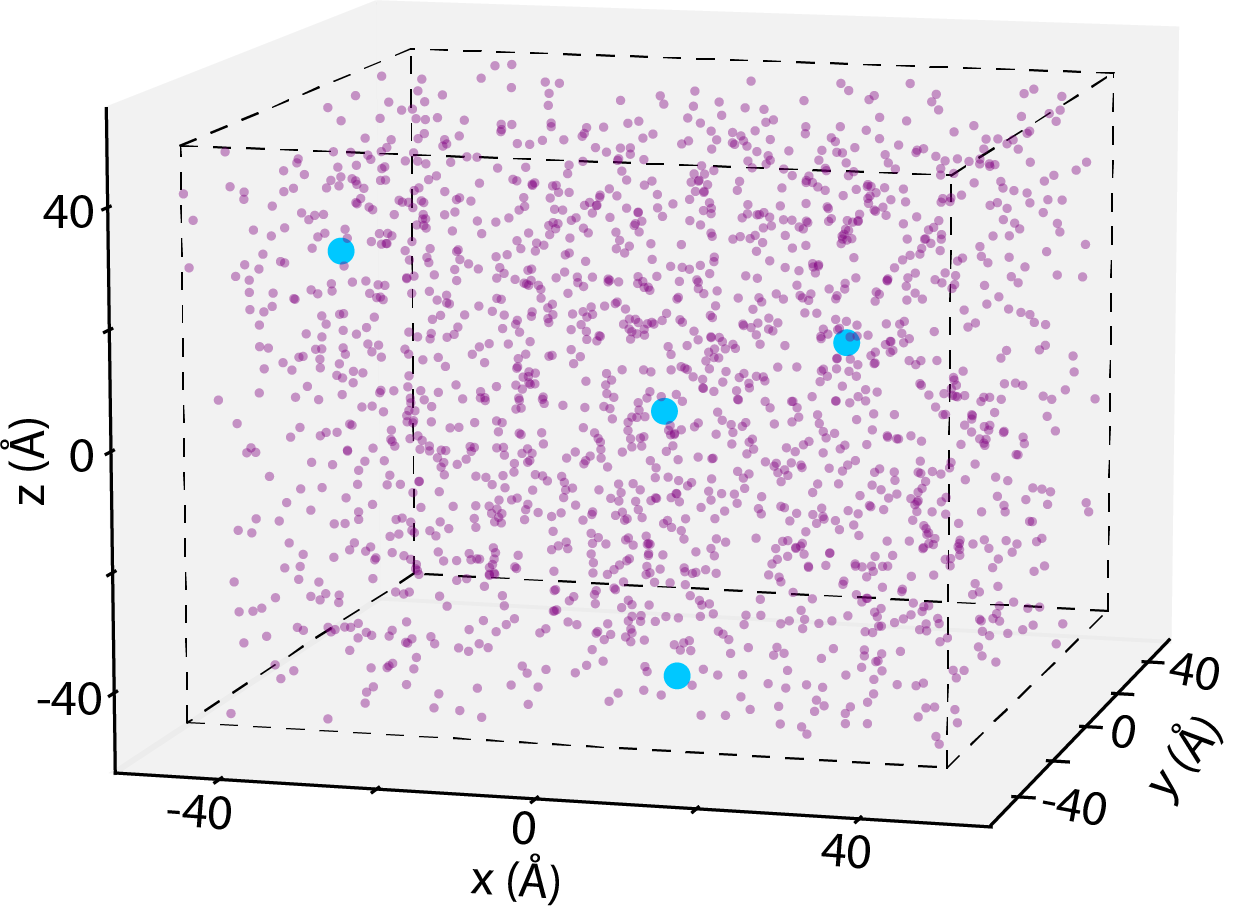}
    \caption{\textbf{Random spin configuration} A typical simulation volume of 80 \r{A}$^{3}$ is employed for the Monte Carlo hopping model. The diamond lattice structure is explicitly modeled, with lattice sites randomly occupied by $^{13}$C nuclear spins (purple) and electron spins (blue). Lattice sites are assigned using binomial sampling, consistent with the species concentrations, preserving the spatial distribution inherent to the NV-diamond platform.}
    \label{SimulationVolume}
\end{figure}

\noi Here, $\mu_{0}$ is the permeability of free space, $\hbar$ is the reduced Planck's constant, $\gamma_{C}$ is the gyromagnetic ratio of $^{13}$C, $r_{ij}$ is the distance between nuclear spins $i$ and $j$, and $\theta_{ij}$ is the angle between the inter-spin vector and the external magnetic field. $J_{\rm ZQ}(\omega)$ represents the zero-quantum
spectrum of the $^{13}$C nuclei quantifying the energy overlap between spins~\cite{abragamNuclearMagnetismOrder1982}. For simplicity, we assume all nuclei have the same resonance frequency, in which case $J_{\rm ZQ}(0)$ is simply related to the intrinsic $T_{2}$ of the nuclei, which we treat as a model parameter. $0<\kappa<1$ is the driving protocol dependent dipolar scaling factor
\begin{equation}
\label{eq:coupling_scaling}
\begin{aligned}
\kappa=\sum_{m=-2}^{+2}T^{-1}\int_{0}^{T}e^{+i m \phi_{\rm eff}}d^{2}_{0m}(\theta_{\rm eff})D^{2}_{m0}{[}P^{\dagger}(t){]}dt.
\end{aligned}
\end{equation}
Here, $T$ is the driving period, $\theta_{\rm eff}$ and $\phi_{\rm eff}$ parameterize the effective quantization axis, $d^{l}_{mn}$ are reduced Wigner matrix elements, and $D^{l}_{m0}{[}P^{\dagger}(t){]}$ are full Wigner matrix elements parametrized by the time-periodic micromotion operator $P^{\dagger}(t)$ as defined within the Floquet framework. The depolarization of the nuclei may be approximated as follows. For a fluctuating spin model~\cite{abragamNuclearMagnetismOrder1982}, the depolarization rates due to hyperfine field fluctuations are approximately given by:
\begin{equation}
\begin{aligned}
R_{ij} = -\eta J^{i}_{\rm env}(\omega_{\rm eff})\delta_{ij}.
\end{aligned}
\end{equation}

\begin{figure}[b]
    \centering
    \includegraphics[width=0.9\columnwidth]{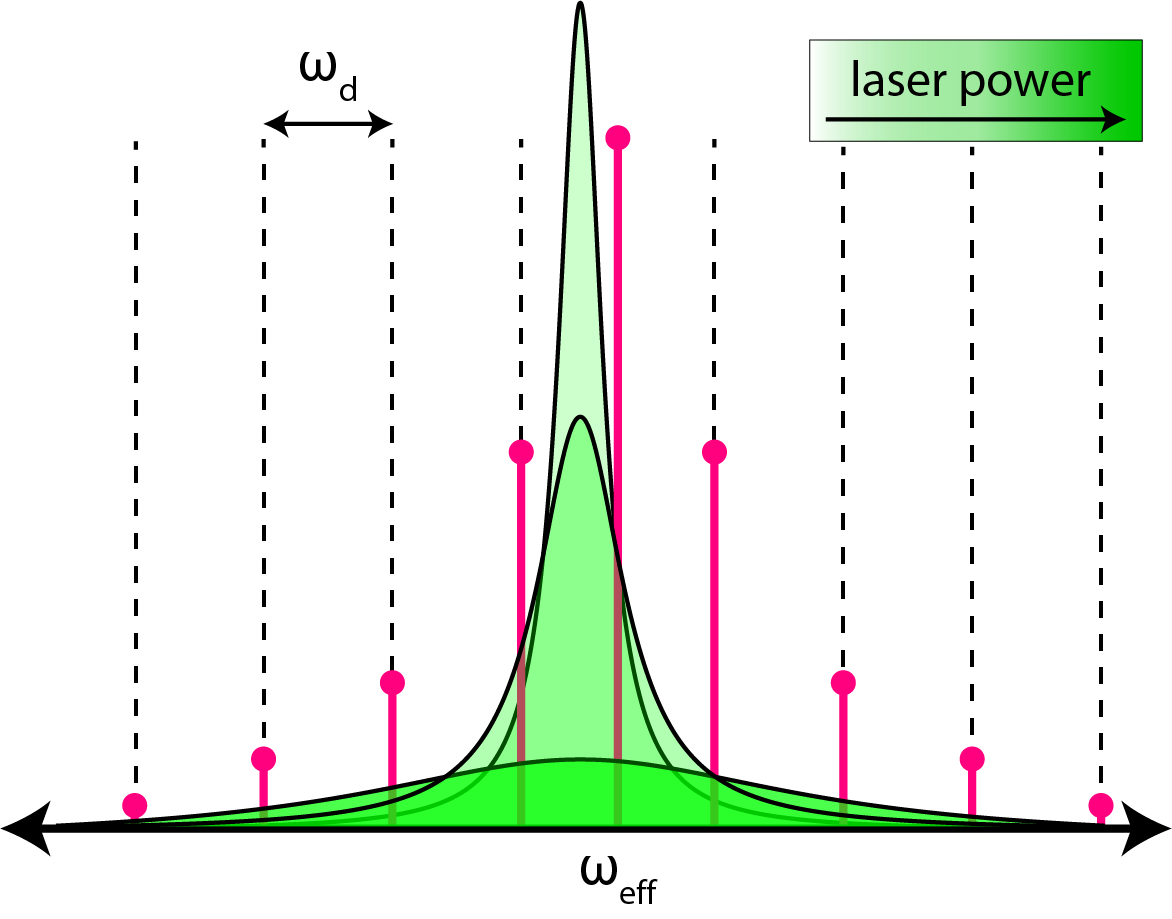}
    \caption{\textbf{Spectral overlap} Interaction between the Floquet filter function and electron noise spectral density under laser illumination. Floquet driving shapes the amplitude and spacing of the spectral comb (pink sticks). For detuned driving, the comb is centered around the effective energy gap $\omega_{\rm eff}$, and spaced at integer intervals of the driving frequency, $\omega_{d}$. Laser illumination broadens the spectral density (green shades) due to a steady decrease in the correlation time, $\tau_{c}$.}
    \label{SpectralOverlap}
\end{figure}

\noi Here, $\delta_{ij}$ is the Kronecker delta and
$\eta$ is a free parameter accounting for the limited number of electrons in the simulation volume and possible deviations in the exact experimental conditions. $J^{i}_{\rm env}(\omega)$ represents the filtered electron noise spectral density
\begin{equation}
\label{J_env}
\begin{aligned}
J^{i}_{\rm env}(\omega)=\sum_{\mu=1}^{N_e} h_{i\mu}^2\int_{-\infty}^{+\infty}J_{e}(\omega-\omega')Y(\omega')d\omega',
\end{aligned}
\end{equation}
where $h_{i\mu}$ is the dipolar coupling constant between nuclear spin $i$ and paramagnetic impurity $\mu$, given by:
\begin{equation}
\begin{aligned}
h_{i\mu}=-\frac{\mu_{0}}{4\pi}\frac{\hbar\gamma_C\gamma_e}{r^{3}_{i\mu}}\frac{1}{2}(3\cos^{2}(\theta_{i\mu}) - 1).
\end{aligned}
\end{equation}
Here, $J_{e}(\omega)$ represents the bare electron noise spectral density, which we assume to be Lorentzian, and $Y(\omega)$ represents the spectral filter function of the Floquet driving protocol~\cite{rhimCalculationSpinLattice1978, furmanSpinDiffusionSpinlattice1999a, harkinsAnomalouslyExtendedFloquet2024}
\begin{equation}
\label{Y}
\begin{aligned}
Y(\omega)=\sum_{k=-\infty}^{+\infty}\vert c_{k}\vert^{2}\delta(k\omega_{d}+\omega),
\end{aligned}
\end{equation}
characterized by the Fourier coefficients $c_{k}$
\begin{equation}
\begin{aligned}
c_{k}=T^{-1}\sum_{m=-1}^{+1}e^{+i m \phi_{\rm eff}}d^{1}_{1m}(\theta_{\rm eff})\int_{0}^{T}D^{1}_{m0}{[}P^{\dagger}(t){]}e^{-i k \omega_{d}}dt.
\end{aligned}
\end{equation}
This is illustrated in Fig.~\ref{SpectralOverlap}, and in close analogy to other dynamical decoupling treatments, such as the CPMG (Carl-Purcell-Meiboom-Gill) train, for example~\cite{cywinskiHowEnhanceDephasing2008, uysOptimizedNoiseFiltration2009, alvarezMeasuringSpectrumColored2011a, ajoyOptimalPulseSpacing2011}. The interplay between the filter function and the noise spectral density is central to how Floquet driving and laser illumination jointly modulate decoherence. While Floquet driving predominantly controls the filter function, laser illumination modulates the spectral density of the environmental noise. 

The total nuclear polarization $P(t)$ is obtained by a configurational average of $p(t)$
\begin{equation}
\begin{aligned}
P(t) = \langle p(t)\rangle_{\rm conf}.
\end{aligned}
\end{equation}
Good convergence is typically achieved after approximately 100 averages. We find that the resulting polarization decay profile is well described by the functional form $e^{-\sqrt{R_p t}}e^{-R_d t}$, displaying both stretched- and mono-exponential character (see~\ref{subsection_optimalstretchingfactor}). 

\subsection{Modeling the Effect of Laser Illumination}
\label{subsection_lasereffects}

\begin{figure}[t]
    \centering
    \includegraphics[width=0.9\columnwidth]{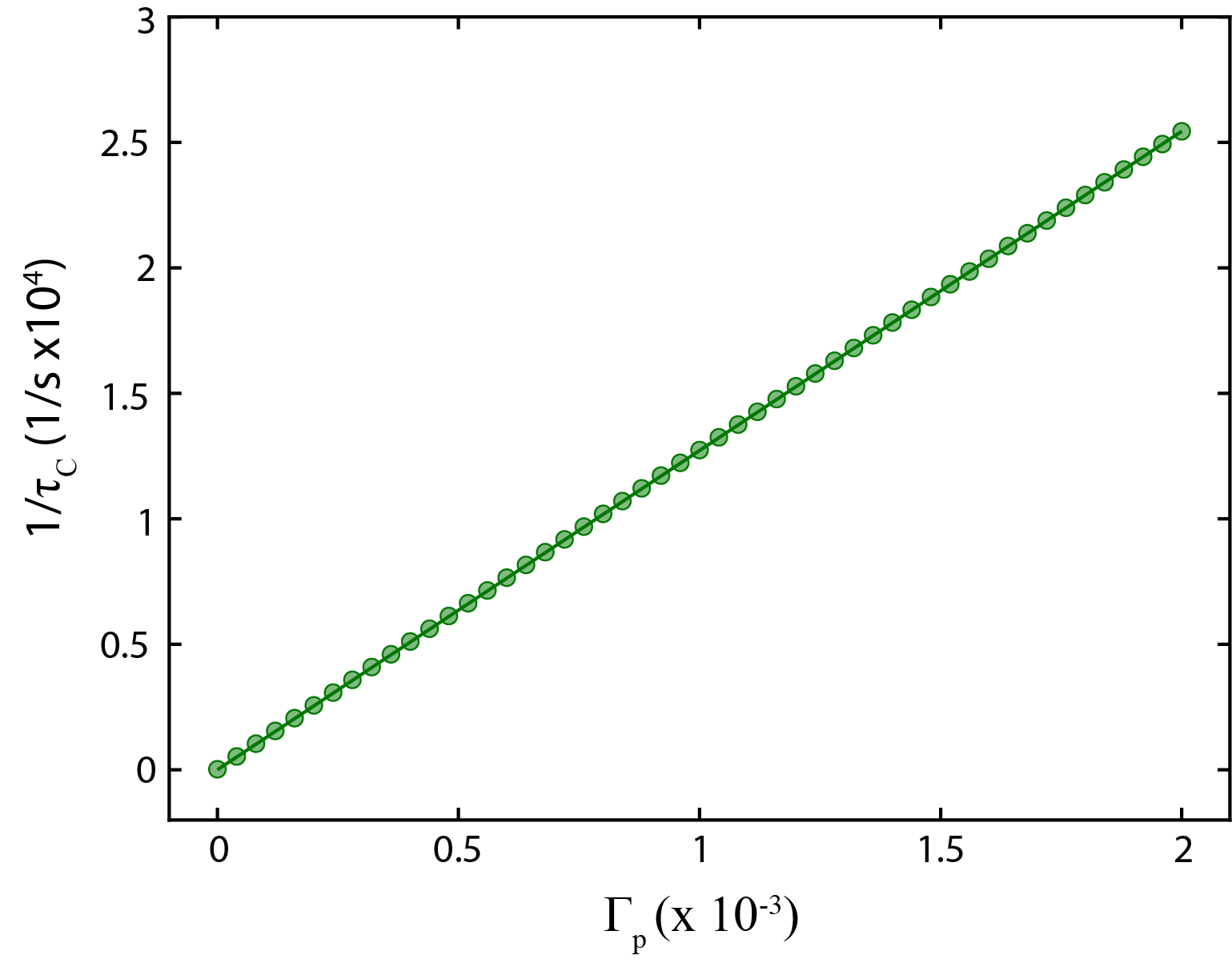}
    \caption{\textbf{Electron correlation time versus laser power} Inverse correlation time $1/\tau_{c}$ plotted against the optical pumping rate $\Gamma_{p}$, illustrating their linear relationship. Simulations as a function of laser power are performed by linearly increasing $1/\tau_{c}$ corresponding to a linear increase in laser power.}
    \label{InvCorrelationTime}
\end{figure}

To investigate the effects of laser illumination, we build upon the NV center optical pumping model described in \cite{hincksStatisticalInferenceQuantum2018}, which approximates the pumping process within an effective Hilbert space spanned by the basis states
\begin{equation}
\begin{aligned}
\mathbbm{B}=\{\vert g,-1\rangle,\vert g,0\rangle,\vert g,+1\rangle,
\vert e,-1\rangle,\vert e,0\rangle,\vert e,+1\rangle,
\vert s\rangle\}
\end{aligned}
\end{equation}
The pumping process itself is described by a set of Lindblad dissipators leading to the following population dynamics
\begin{equation}
\begin{aligned}
{[}\mathcal{D}{]}_{\mathbbm{B}}=\left[\begin{array}{ccccccc}
     -\gamma_{g}^{-1} & 0 & 0 & \gamma_{eg} & \gamma_{01} & 0  &  \gamma_{sg}
     \\
     0 & -\gamma_{g}^{0} & 0  & \gamma_{01} & \gamma_{eg} & \gamma_{01} & \gamma_{sg}
     \\
     0 & 0 &  -\gamma_{g}^{+1} & 0 & \gamma_{01} & \gamma_{eg} &  \gamma_{sg}
     \\
     \Gamma_{p} \gamma_{sg} & 0 & 0 &  -\gamma^{-1}_{e} & 0 & 0 & 0
     \\
     0 & \Gamma_{p} \gamma_{sg} & 0 & 0&  -\gamma^{0}_{e} & 0 & 0
     \\
     0 & 0& \Gamma_{p} \gamma_{sg} & 0 & 0& -\gamma^{+1}_{e} & 0
    \\
      0 & 0& 0 & \gamma_{es} & 0& \gamma_{es} & -\gamma_{s}
\end{array}\right].
\end{aligned}
\end{equation}
The diagonal elements are given by the sum of the respective column, and $\Gamma_{p}$ represents a dimensionless parameter quantifying the pumping efficiency, and may be taken as a measure of the applied laser power~\cite{medinaThermodynamicsOpticalPumping2025}. We augment the optical pumping model by $T_{1}$ relaxation processes as follows
\begin{equation}
\begin{aligned}
&{[}\mathcal{R}{]}_{\mathbbm{B}}=R\oplus R\oplus\mathbbm{1}_{1},
\\
&R=R^{\rm E}_{1}\left[\begin{array}{ccc}
-\Theta(-\omega) & \Theta(\omega) & 0
\\
\Theta(-\omega) & -(\Theta(\omega)+\Theta(-\omega)) & \Theta(\omega)
\\
0 & \Theta(-\omega) & -\Theta(\omega)
\end{array}\right],
\end{aligned}
\end{equation}
with
\begin{equation}
\begin{aligned}
\Theta(\omega)=\exp(-\beta \omega/2).
\end{aligned}
\end{equation}
Although this approach does not fully account for changes in the energy level structure that occur when the sample is moved to high magnetic field regions, it remains sufficient to provide a qualitative understanding. The electron correlation function is computed as follows
\begin{equation}
\begin{aligned}
C(\tau)=\sum_{m,n}\langle m \vert S_{z}\vert m\rangle {[}e^{{[}\mathcal{D}+\mathcal{R}{]}_{\mathbbm{B}}\tau}{]}_{mn}P^{\text{eq}}_{n}\langle n \vert S_{z}\vert n\rangle,
\end{aligned}
\end{equation}
where $P^{\rm eq}$ is the equilibrium distribution of \mbox{${[}\mathcal{D}+\mathcal{R}{]}_{\mathbbm{B}}$}. Utilizing the model parameters described in reference~\cite{hincksStatisticalInferenceQuantum2018}, we find that the electron correlation function decays approximately exponentially
\begin{equation}
\begin{aligned}
C(\tau)\simeq e^{-t/\tau_{c}(\Gamma_{p})}
\end{aligned}
\end{equation}
As shown in Fig.~\ref{InvCorrelationTime}, the inverse correlation time in the augmented model increases approximately linearly with $\Gamma_{p}$ (or laser power), leading to a broadening of the electron spectral density. We incorporate this trend into our Monte Carlo simulations by increasing the inverse electron correlation time entering Eq.~\ref{J_env} linearly with laser power.

\subsection{Data shows Optimal Stretching Factor is 1/2}
\label{subsection_optimalstretchingfactor}

\begin{figure}[t]
    \centering
    \includegraphics[width=0.9\columnwidth]{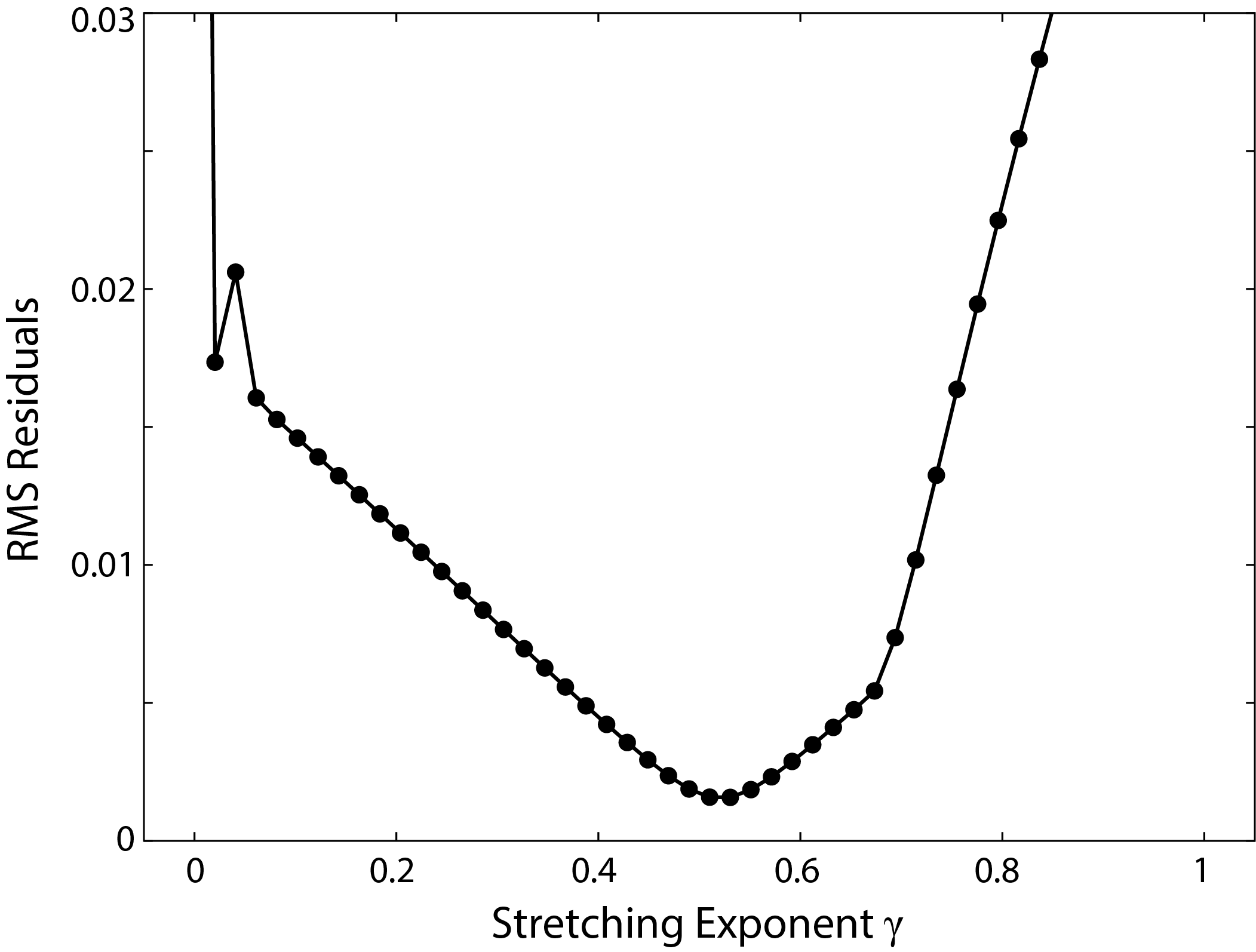}
    \caption{\textbf{Optimal stretching factor} Root mean squared (RMS) residuals as a function of stretching exponent $\gamma$ obtained by fitting Monte Carlo simulation results to the functional form $e^{-(R_{p}t)^{\gamma}}e^{-R_{d}t}$.}
    \label{Optimal_Stretch}
\end{figure}

To quantitatively assess the validity of the emergent decay law, we fit the Monte Carlo simulation results to the function $e^{-(R_{p}t)^{\gamma}}e^{-R_{d}t}$, varying the stretching exponent $\gamma$ over a range of values. The root mean squared (RMS) residuals for each fit are plotted in Fig.~\ref{Optimal_Stretch} as a function of $\gamma$. A clear minimum is observed near $\gamma=0.5$, providing strong numerical support for the predicted form of the decay, which combines a 
$\gamma=\frac{1}{2}$ stretched exponential component with a monoexponential decay. These results highlight the robustness of the universal form and demonstrate that the minimal physical framework successfully reproduces the essential features of the behavior.

\subsection{Modeling Polarization Transport}
\label{subsection_transport}

\begin{figure*}[t]
    \centering
    \includegraphics[width=\textwidth]{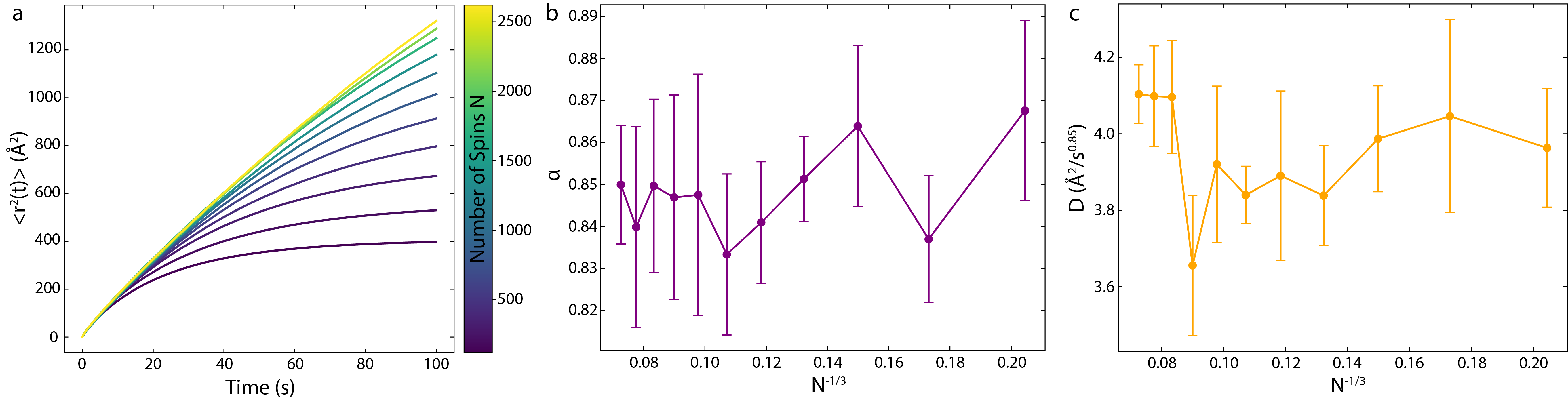}
    \caption{\textbf{Finite-size scaling of polarization transport.} (a) \emph{\textbf{Mean squared displacement (MSD)}} for many different system sizes at 1.1\% $^{13}$C concentration. (b) \emph{\textbf{Diffusion exponent}} $\alpha$ and (c) \emph{\textbf{Diffusion coefficient}} $D$ plotted versus the inverse linear system size $N^{-1/3}$, where $N$ is the number of nuclear spins. Each point is the mean of 5 independent runs of 20 trajectories; error bars denote standard error. Both quantities remain constant within uncertainty across nearly a twenty-fold range in $N$, confirming that the extracted transport parameters are intrinsic and not affected by finite simulation volume.}
    \label{FiniteSizeScaling}
\end{figure*}

To characterize polarization transport in the disordered $^{13}$C spin network, we performed Monte Carlo simulations of the hopping dynamics governed by \zr{hopping_model} in the absence of relaxation ($R=0$). A single lattice realization was first generated at the desired $^{13}$C concentration, with no electrons present, within a cubic simulation volume subject to periodic boundary conditions. The origin was defined by locating the $^{13}$C spin nearest to the center of the box and translating all coordinates such that this spin was positioned exactly at the origin. The initial polarization vector $p(0)$ was set to unity on that site and zero elsewhere, corresponding to a localized excitation.

Time evolution was carried out under the hopping operator $W$, which encodes nuclear–nuclear random hopping dynamics. At each time step, the polarization distribution $p(t)$ was updated, and the mean squared displacement (MSD) was computed as
\begin{equation}
\label{eq:msd}
\langle r^2(t) \rangle = \sum_i \left(x_i^2 + y_i^2 + z_i^2\right)p_i(t),
\end{equation}
where the summation runs over all nuclear sites $i$ in the simulation volume. The MSD quantifies the spreading of polarization away from the origin as a function of time. Each trajectory was evolved until the propagating polarization front reached the boundary of the simulation volume. To avoid finite-size artifacts, the portion of the data beyond this point was excluded from subsequent fitting.

This procedure was repeated for 100 random lattice realizations and averaged to obtain an ensemble-averaged $\langle r^2(t) \rangle$. The ensemble-averaged MSD curve was then fit to the generalized diffusion law,
\begin{equation}
\langle r^2(t) \rangle = 6 D t^{\alpha},
\end{equation}
yielding the diffusion coefficient $D$ and diffusion exponent $\alpha$.

Fig.~\ref{FiniteSizeScaling}a shows representative MSD traces for several system sizes at a fixed $^{13}$C concentration of 1.1\%. Fig.~\ref{FiniteSizeScaling}b–c summarize the extracted parameters $\alpha$ and $D$ as a function of the inverse linear system size $N^{-1/3}$, where $N$ is the number of nuclear spins in the simulation volume. Both quantities remain constant within uncertainty over nearly a twenty-fold range in $N$, confirming that the extracted values of $\alpha$ and $D$ are intrinsic and not limited by finite simulation size. This finite-size scaling analysis therefore verifies that the observed sub-diffusive transport behavior arises from the intrinsic lattice disorder rather than finite-size effects.

\subsection{Possibility of All-Optical Electron Decoupling}
\label{subsection_allopticaldecoupling}

\begin{figure}[t]
    \centering
    \includegraphics[width=0.9\columnwidth]{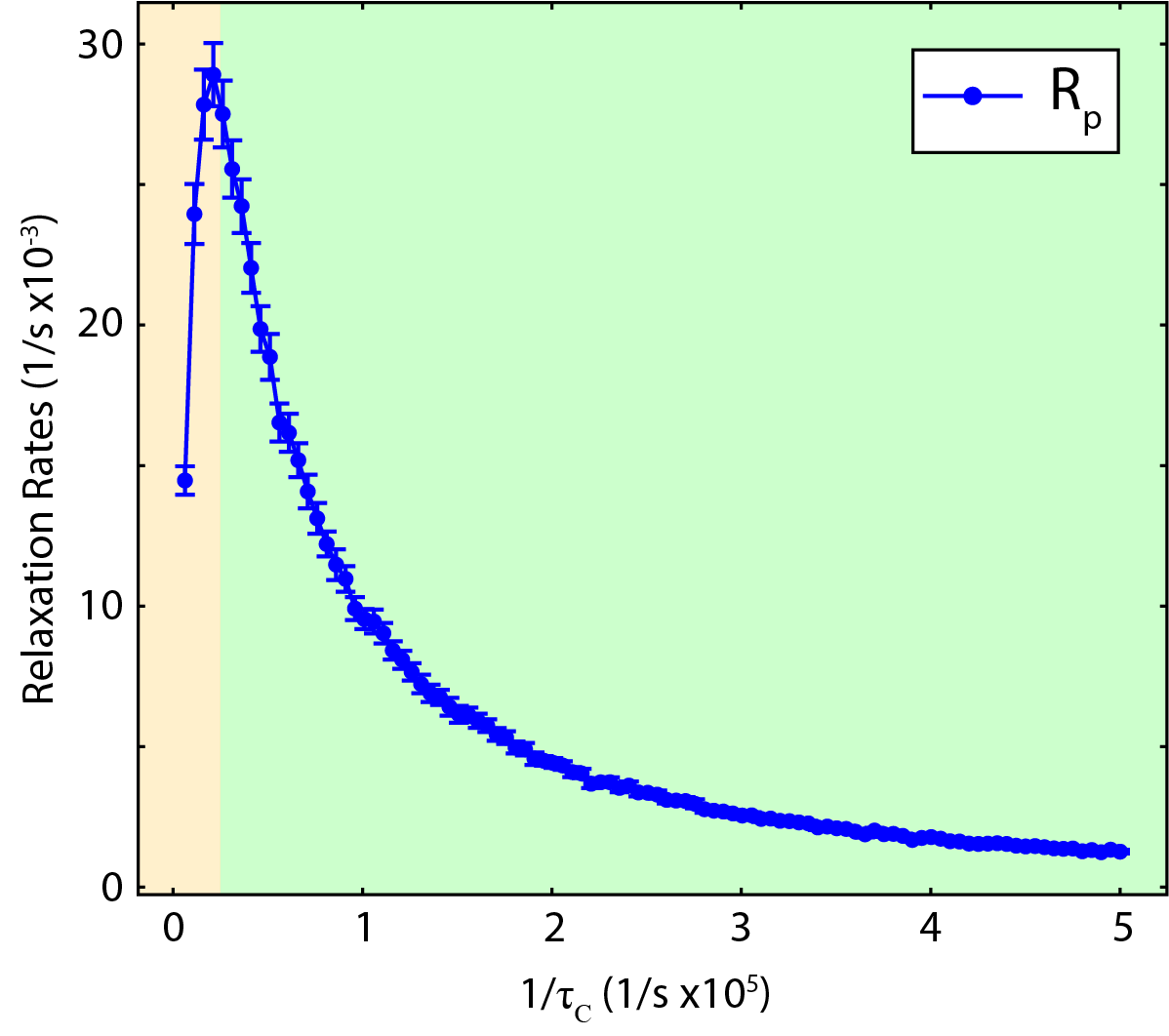}
    \caption{\textbf{All-optical decoupling simulation} Monte Carlo simulation of the paramagnetic relaxation rate $R_{p}$ as a function of $1/\tau_{c}$, illustrating the potential for all-optical decoupling. Values of $1/\tau_{c}$ used to fit the experimental data (Fig.~2 in the main text) are highlighted in orange, while extrapolated values beyond this are shown in green.}
    \label{AllOpticalDecouple}
\end{figure}

As discussed in the main text, the increase in inverse correlation time with laser power is modeled as a collective effect arising from the entire electron spin bath, including both NV and P1 centers. This is motivated by the presence of nonsecular interactions between nearby NV and P1 centers, which allow laser illumination to indirectly modulate the dynamics of the P1 bath—an effect likely facilitated by the known NV-P1 clustering in diamond, reported previously~\cite{bussandriP1CenterElectron2024, nir-aradNitrogenSubstitutionsAggregation2024}. To explore the potential for all-optical decoupling of the electron bath from the $^{13}$C spin network, we extend our Monte Carlo simulations to include inverse correlation times beyond those needed to fit experimental data. The results are shown in Fig.~\ref{AllOpticalDecouple}, where the range of experimentally relevant inverse correlation times is shaded in orange, and the extrapolated regime is shaded in green. Initially, increasing the laser power enhances the relaxation rate due to greater overlap between the filter function and the bath spectral density (see Fig.~\ref{SpectralOverlap}). However, beyond a critical point, further increases in inverse correlation time broaden the spectral density such that its spectral weight becomes nearly uniform across all frequencies. In this regime, the relaxation rate begins to decrease monotonically—indicating a form of all-optical decoupling—highlighted in the green shaded region of Fig.~\ref{AllOpticalDecouple}. We note that extremely high optical powers may introduce additional effects such as charge state conversion or heating which are not considered here.

\subsection{Anticipated Effects of Lowering Temperature}
\label{subsection_temperature}

A crucial parameter in the relaxation model is the electron 
$T_{1}$ time, which determines the correlation time of the fluctuating hyperfine fields responsible for nuclear spin relaxation. Previous experiments~\cite{takahashiQuenchingSpinDecoherence2008, jarmolaTemperatureMagneticFieldDependentLongitudinal2012a} have shown that the $T_{1}$ times of NV and P1 centers can significantly increase at lower temperatures. When the fluctuating field model applies, the paramagnetic relaxation rate $R_{p}$ approximately scales as $(1-P^{2})J(\omega)$~\cite{abragamNuclearMagnetismOrder1982}, where $P$ is the electron polarization and $J(\omega)$ is the spectral density function. As temperature decreases, the factor 
$1-P^{2}$ decreases due to higher electron polarization, and the spectral density $J(\omega)$ narrows and decreases in magnitude at $\omega$ because longer $T_{1}$ times correspond to slower fluctuations. Consequently, the paramagnetic relaxation rate $R_{p}$ can be significantly suppressed. Furthermore, since diffusive relaxation arises from diffusion gradients toward rapidly relaxing nuclei near paramagnetic centers, the diffusion-driven relaxation rate $R_{d}$ is similarly expected to decrease. Taken together, these effects imply that operating at low temperatures may significantly extend nuclear spin lifetimes. Additionally, because electron $T_{1}$ times depend sensitively on electron concentration~\cite{jarmolaTemperatureMagneticFieldDependentLongitudinal2012a}—which varies between samples—using samples with lower defect densities in combination with lower temperature can further reduce electron-mediated relaxation. These combined insights suggest that substantial enhancements in $^{13}$C lifetimes are achievable through a strategic combination of low-temperature operation and careful control of electron defect concentrations.

\section{Relaxation Landscape}
\label{section_relaxationlandscape}

\begin{figure}[t]
    \centering
    \includegraphics[width=0.9\columnwidth]{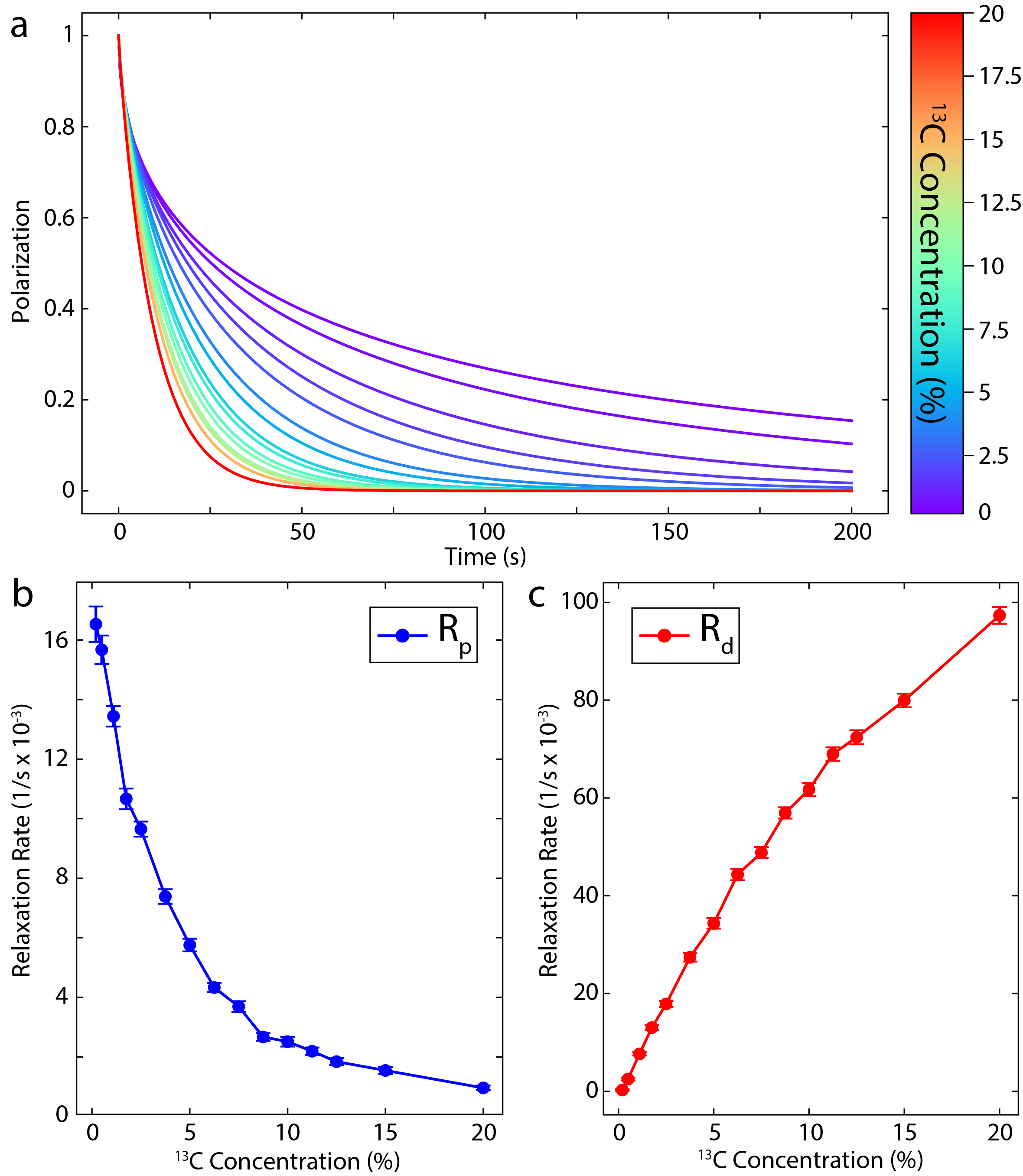}
    \caption{\textbf{Relaxation versus $^{13}$C concentration} (a) \emph{\textbf{Polarization decay curves}} for varying $^{13}$C concentrations (0.2\% to 20\%) at fixed electron concentration (30 ppm) and fixed simulation box size ($\approx$80 \r{A}$^{3}$), averaged over 100 disorder realizations. (b)-(c) \emph{\textbf{Relaxation rates}} $R_p$ and $R_d$ as a function of $^{13}$C concentration. At low concentrations $R_d=0$ and at high concentrations $R_p=0$, reflecting a transition from the diffusion-limited regime to the diffusion-dominated regime.}
    \label{Concentrations}
\end{figure}

To better understand the relaxation landscape presented in Fig.~3b in the main text, where both nuclear and electron spin concentrations are varied simultaneously, here we isolate the effect of nuclear spin concentration alone. Specifically, we fix the electron concentration at 30 ppm and hold the simulation box size constant at approximately 80 \r{A}$^{3}$, while systematically varying the $^{13}$C concentration from 0.2\% to 20\%. This corresponds to a 1D horizontal slice through the relaxation landscape. As shown in Fig.~\ref{Concentrations}a, the relaxation becomes markedly faster with increasing $^{13}$C concentration. This trend reflects the transition between the diffusion-limited and diffusion-dominated regimes discussed in the main text, where at higher concentrations polarization can rapidly diffuse toward electron ``sinkholes,'' resulting in a faster decay. 

This transition is quantitatively captured in Fig~\ref{Concentrations}b-c, which plot the decoherence rates $R_p$ and $R_d$ as functions of $^{13}$C concentration. At low concentrations, diffusion is severely inhibited resulting in a purely stretched-exponential decay (i.e., $R_{d}\rightarrow0$).
At higher $^{13}$C concentrations, polarization rapidly hops between neighboring nuclear spins, effectively averaging out site-specific differences in relaxation resulting in a mono-exponential decay governed by the mean relaxation rate ($R_{p}\rightarrow0$).

\begin{figure*}[t]
    \centering
    \includegraphics[width=0.9\textwidth]{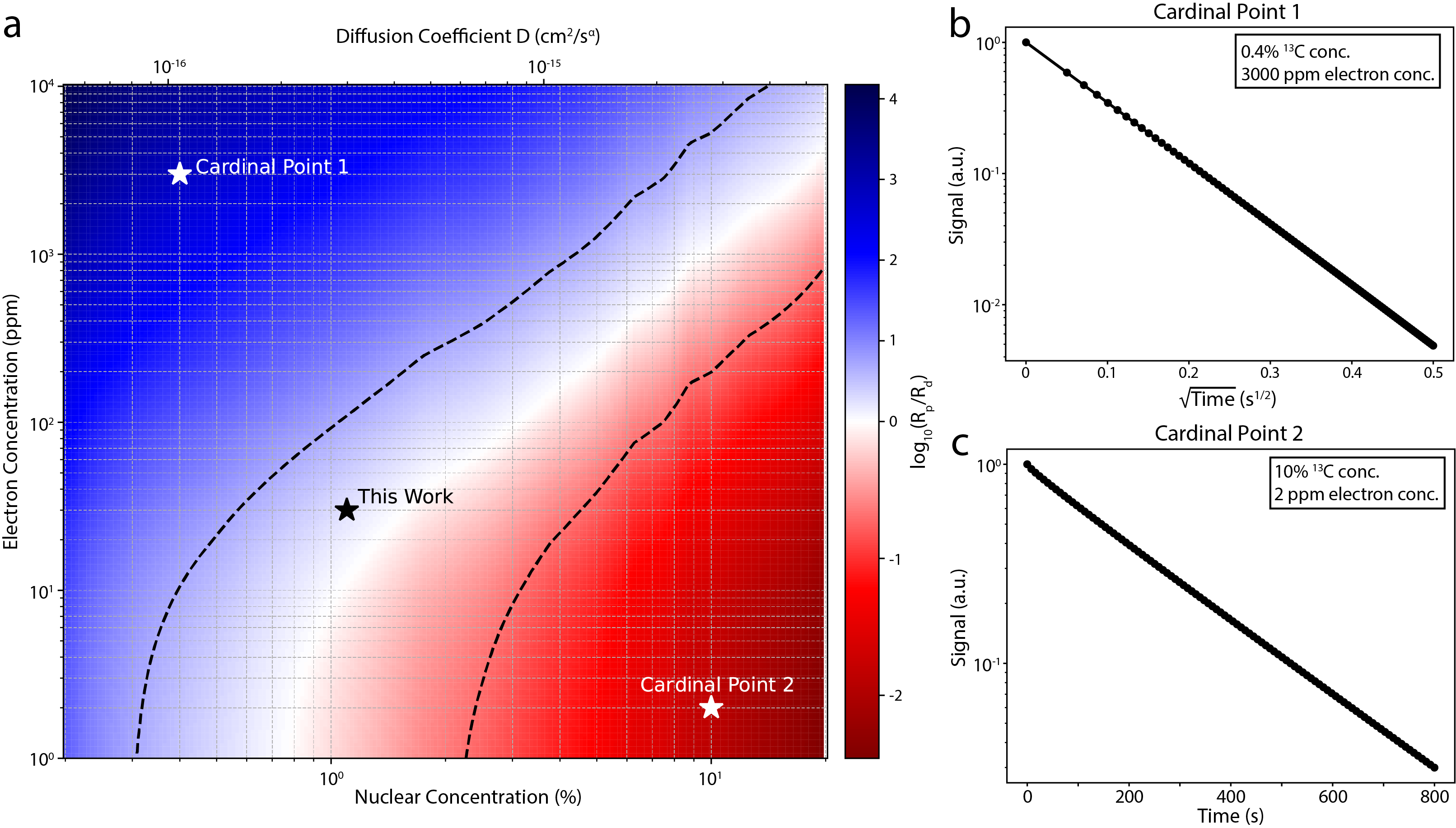}
    \caption{\textbf{Understanding the relaxation landscape} (a) \emph{\textbf{Relaxation landscape}} showing the relative contributions of stretched and monoexponential components as a function of electron and $^{13}$C concentrations. The two ``cardinal points'' are indicated: a diffusion-limited regime (0.4\% $^{13}$C, 3000 ppm electrons; Cardinal Point 1) and a diffusion-dominated regime (10\% $^{13}$C, 2 ppm electrons; Cardinal Point 2). (b) \emph{\textbf{Cardinal point 1}} signal decay plotted on a logarithmic scale versus $\sqrt{t}$, revealing a linear trend consistent with a purely stretched exponential decay of the form $e^{-\sqrt{R_{p}t}}$. The signal decays within $\approx0.25$ s. (c) \emph{\textbf{Cardinal point 2}} signal decay plotted on a logarithmic scale versus linear time, showing a straight line indicative of monoexponential relaxation governed by $e^{-R_{d}t}$. The signal persists for over 600 seconds.}
    \label{Cardinal_Points}
\end{figure*}

To illustrate the distinct dynamical regimes captured by the relaxation landscape presented in Fig.~3b of the main text, we examine two representative ``cardinal'' points corresponding to extreme ends of the electron and $^{13}$C concentration parameter space.

Cardinal Point 1 corresponds to a system with 0.4\% $^{13}$C concentration and 3000 ppm electron concentration, located in the upper-left region of the relaxation landscape (see Fig.~\ref{Cardinal_Points}a). In this regime, the polarization dynamics are diffusion-limited. This behavior is reflected in the relaxation function, which is well-described by a purely stretched exponential decay with 
$R_{p}\gg R_{d}$. To highlight this, we plot the simulated signal decay on a logarithmic scale versus 
$\sqrt{t}$ (Fig.~\ref{Cardinal_Points}b), revealing a linear relationship that confirms the $e^{-\sqrt{R_{p}t}}$ form. The absolute timescale of relaxation in this regime is short, with the signal decaying almost completely within $0.25$ s.

In contrast, Cardinal Point 2 corresponds to a system with 10\% $^{13}$C concentration and 2 ppm electron concentration, situated in the bottom-right region of the relaxation landscape (see Fig.~\ref{Cardinal_Points}a). Here, the dynamics are diffusion-dominated, with polarization decaying primarily via the mono-exponential term 
$e^{-R_{d}t}$. In this case, $R_{d}\gg R_{p}$, and the stretched exponential component plays a negligible role. The decay is plotted on a logarithmic scale versus linear time (Fig.~\ref{Cardinal_Points}c), again yielding a straight line indicative of purely exponential relaxation. Notably, the timescale of decay in this regime is significantly longer: the signal persists for more than 600 seconds.

These two cardinal points exemplify the limiting behaviors of the emergent decay law introduced in Eq.~1 of the main text. The sharp contrast in both functional form and relaxation timescale underscores the importance of considering both nuclear and electronic concentrations in interpreting decoherence behavior. Between these extremes lies the intermediate regime explored experimentally, where both stretched- and mono-exponential components contribute significantly—highlighting the full expressive power of the universal decay law.

\section{Eigenmode Decomposition}
\label{section_eigenmode}

\subsection{Derivation of Asymptotic Eigenmode Decomposition}
\label{subsection_derivation}

To gain further insight into the time dependence of the polarization $p(t)$, we analyze the eigenmode decomposition of the relaxation dynamics. This allows us to express the polarization as a product of a mono-exponential decay originating from the slowest decaying eigenmode and a second term arising from the collective contributions of the faster decaying eigenmodes, which we hypothesize to resemble a stretched exponential with a stretching factor of $1/2$.

We can write the polarization $p(t)$ as a sum over all eigenmodes:
\begin{equation}
\begin{aligned}
p(t)=\sum_{j=0}^{N-1}a_{j}e^{-\lambda_{j}t},
\end{aligned}
\end{equation}
where $\lambda_{0}$ is the slowest eigenvalue (smallest magnitude) and $a_{j}$ are the projections of the initial state onto each eigenmode. Factoring out the slowest decaying eigenmode gives:
\begin{equation}
\begin{aligned}
p(t)
&=a_{0}\exp(-\lambda_{0}t)(1+a^{-1}_{0}\sum_{j=0}^{N-1}a_{j}e^{-(\lambda_{j}-\lambda_{0})t}).
\end{aligned}
\end{equation}
Taking the logarithm:
\begin{equation}
\begin{aligned}
\ln(p(t))&=\ln(a_{0})-\lambda_{0}t+\ln(1+\sum_{j=1}^{N-1}\frac{a_{j}}{a_{0}}e^{-(\lambda_{j}-\lambda_{0})t}).
\end{aligned}
\end{equation}
For $\frac{a_{j}}{a_{0}}\ll1$ or $(\lambda_{j}-\lambda_{0})t\gg1$ for $j>0$, we can expand the logarithm:
\begin{equation}
\begin{aligned}
\ln(p(t))&\sim \ln(a_{0})-\lambda_{0}t+\sum_{j=1}^{N-1}\frac{a_{j}}{a_{0}}e^{-(\lambda_{j}-\lambda_{0})t}.
\end{aligned}
\end{equation}
Exponentiating both sides gives:
\begin{equation}
\begin{aligned}
p(t)
&\sim a_{0}\exp(-\lambda_{0}t)\exp(\sum_{j=1}^{N-1}\frac{a_{j}}{a_{0}}e^{-(\lambda_{j}-\lambda_{0})t}),
\end{aligned}
\end{equation}
which is Eq.~4 in the main text.

\subsection{Polarization Aligned with Slowest Decaying Eigenmode}
\label{subsection_polarization}

\begin{figure*}[t]
    \centering
    \includegraphics[width=\textwidth]{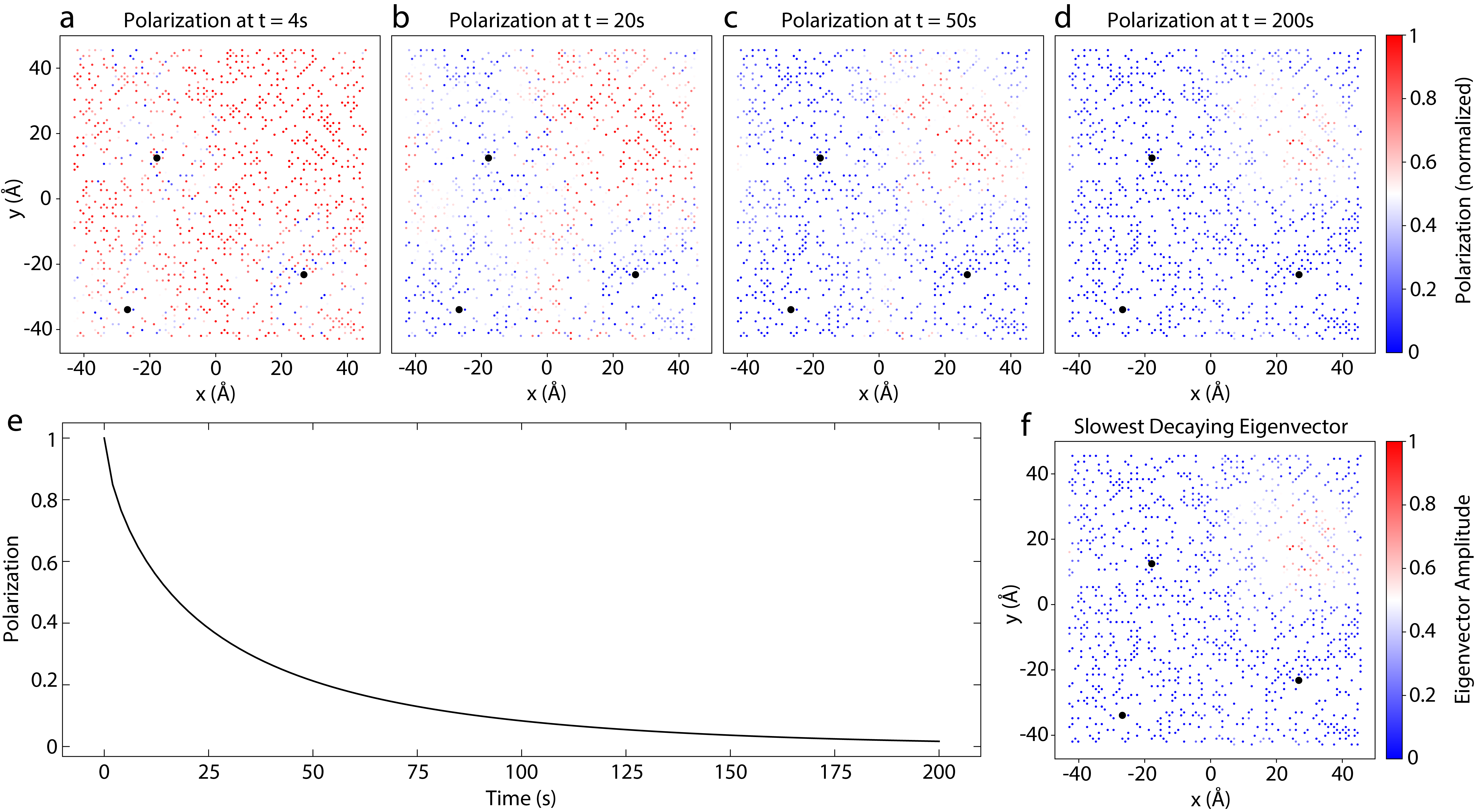}
    \caption{\textbf{Emergence of the slowest relaxation mode} (a-d) \emph{\textbf{Time-resolved polarization heatmaps}} show 2D projections of nuclear spin polarization at times $t=4$s, $20$s, $50$s, and $200$s, respectively, for a single fixed configuration of nuclear and electron spins. Each spin was initialized with uniform polarization at $t=0$. Red indicates polarized spins; blue indicates relaxed (unpolarized) spins. As time progresses, polarization near electron spins rapidly decays, and the polarization appears to gradually concentrate in the spatial region corresponding to the slowest decaying eigenvector. (e) \emph{\textbf{Total polarization versus time}}, showing a stretched-times-mono-exponential decay character. (f) \emph{\textbf{Slowest decaying eigenvector}} illustrated as a heatmap, with red (blue) denoting spin positions with greater (lesser) amplitude in the slowest eigenmode. The near-identical spatial profile of panels d and f confirms that at long times, the system evolves into the slowest relaxation mode.}
    \label{PolarizationVsEigenvector}
\end{figure*} 

To complement the ensemble-averaged heatmaps shown in Fig.~4d-g of the main text—where relaxation dynamics were averaged over 100 $^{13}$C configurations for a fixed electron configuration—we present here the corresponding dynamics for a single fixed realization of both electron and $^{13}$C nuclear spins. This allows an alternative visualization of how polarization evolves over time and eventually concentrations into the spatial region associated with the slowest decaying eigenmode.

We initialized the system with uniform polarization across all $^{13}$C nuclear spins and evolved the system according to the relaxation dynamics described previously. Snapshots of the spatial distribution of polarization projected onto the XY-plane are shown in panels a-d of Fig.~\ref{PolarizationVsEigenvector} at four different times: 4 s, 20 s, 50 s, and 200 s. Each frame represents a 2D heatmap where red indicates spins that remain polarized and blue indicates those that have relaxed.

At early times (panel A, $t=4$s), spins near electrons (black points) have already begun to depolarize, while spins further away remain polarized. As time progresses (panel B, $t=20$s), depolarization continues to spread outward from electron sites. By $t=50$s (panel C), polarization has largely decayed everywhere except in a spatial region roughly matching where the slowest decaying eigenmode is known to be concentrated. Finally, at $t=200$s (panel D), the polarization heatmap becomes indistinguishable from the eigenvector heatmap itself (panel F), indicating that the system has fully relaxed into the slowest mode. 

\begin{figure*}[t]
    \centering
    \includegraphics[width=0.9\textwidth]{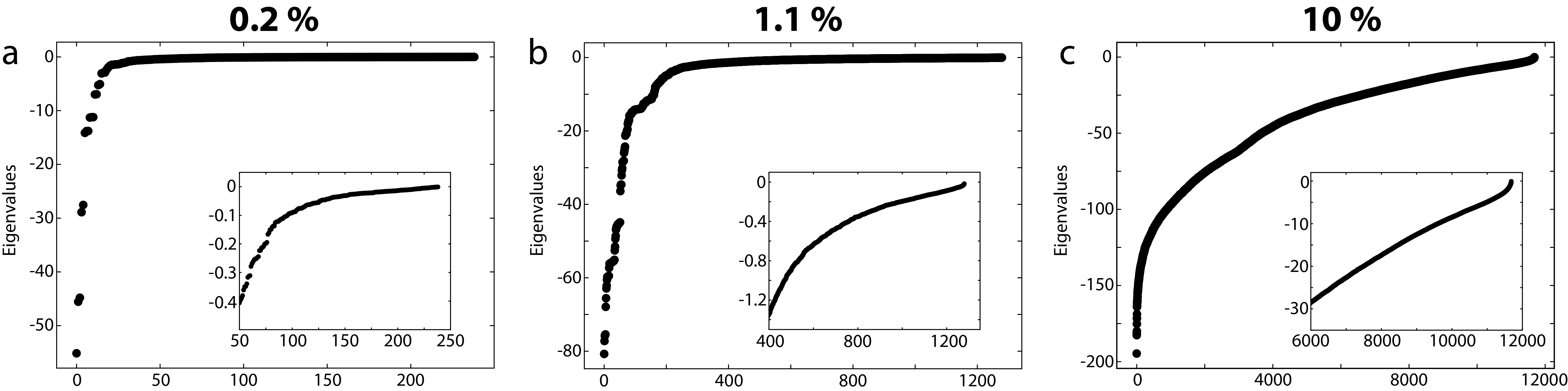}
    \caption{\textbf{Eigenvalue spectra} Eigenvalues of the relaxation matrix $M$ for $^{13}$C concentrations of 0.2\%, 1.1\%, and 10\%, respectively, at a fixed electron concentration of 30 ppm. Each plot shows all sorted eigenvalues, with insets highlighting the dense cluster of slowly decaying modes near zero. At low concentration (a), the spectrum exhibits a clear separation between a few fast-relaxing modes and many slow ones, indicating the presence of isolated, long-lived regions. As the concentration increases (b-c), the spectral gap fills in, reflecting enhanced network connectivity and fewer isolated modes.}
    \label{RelaxationEigenvalues}
\end{figure*}

\subsection{Relaxation Matrix Eigenvalues}
\label{subsection_eigenvalues}

To further understand the eigen-decomposition of the relaxation matrix, we examine the full eigenvalue spectrum across varying nuclear spin concentrations. Fig.~\ref{RelaxationEigenvalues} shows the eigenvalue spectra of the relaxation matrix $M$ for three different $^{13}$C concentrations: 0.2\%, 1.1\%, and 10\%, with a fixed electron concentration of 30 ppm (consistent with Fig.~4h-j in the main text). At all concentrations, we observe a subset of large-magnitude (i.e., fast-decaying) eigenvalues, corresponding to nuclear spins in close proximity to electrons, where relaxation occurs rapidly. In contrast, a large number of small-magnitude eigenvalues represent slowly relaxing modes associated with spins located farther from electrons. Each panel includes an inset that zooms in on this slow regime to better visualize the dense cluster of small eigenvalues.

At low $^{13}$C concentration (e.g., 0.2\%), the spectrum exhibits a sharp separation between the few fast modes and the many slow ones, indicating the presence of isolated regions where polarization can remain ``trapped'' and avoid rapid relaxation (diffusion-limited regime). As the concentration increases, this spectral gap fills in: more eigenvalues appear in the intermediate regime, and the number of very slowly decaying modes is significantly reduced. This trend reflects the increased connectivity of the spin network at higher concentrations, which allows polarization to spread more efficiently and reduces the likelihood of forming isolated ``traps'' or slow-relaxing regions (diffusion-dominated regime).

\section{Disorder-Induced Lifetime Extension}
\label{section_disorder}

\begin{figure}[b]
    \centering
    \includegraphics[width=0.9\columnwidth]{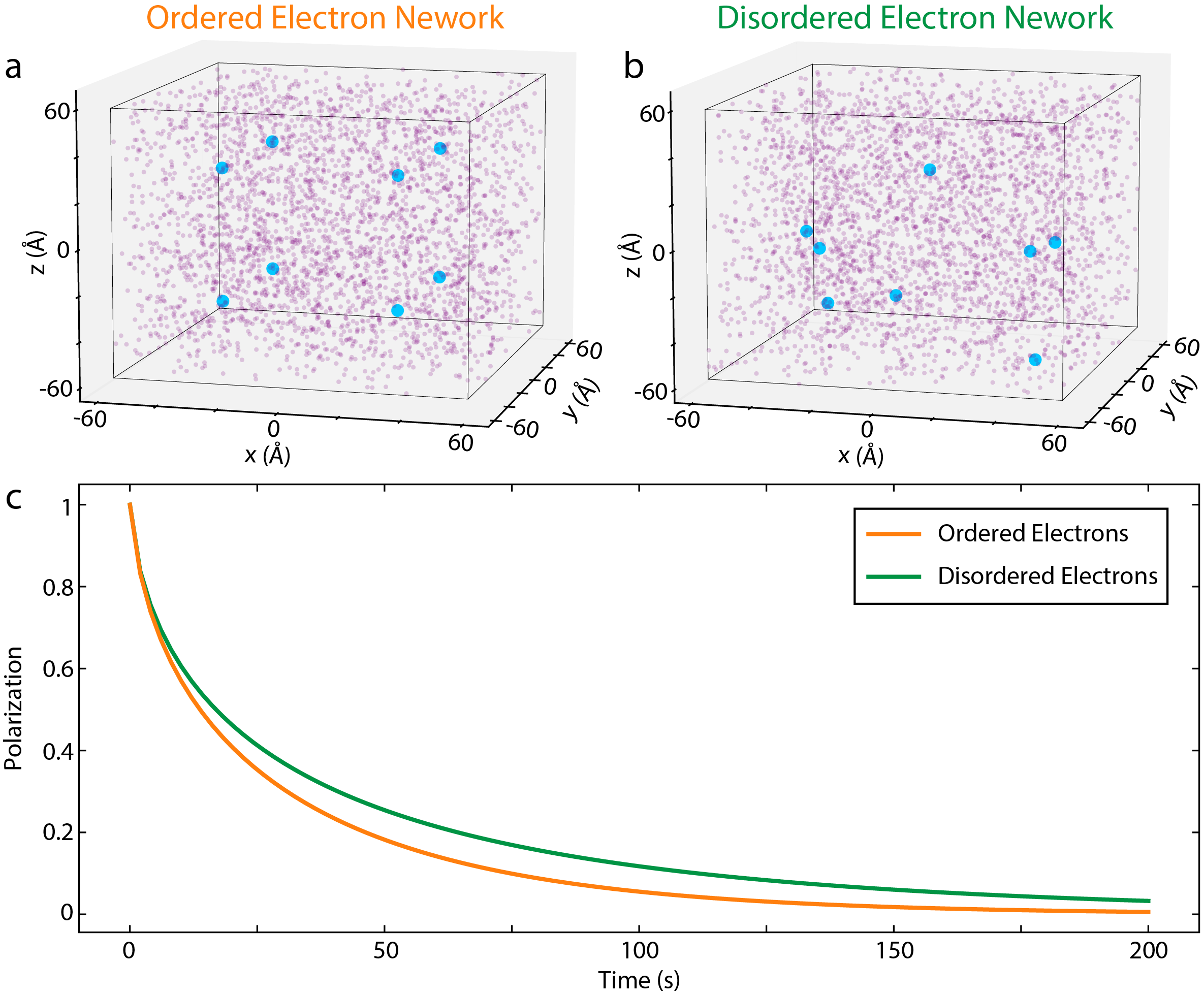}
    \caption{\textbf{Effect of electron disorder on polarization lifetime} (a) \emph{\textbf{Ordered electron network}} with eight electrons placed at the centers of the eight octants of the simulation cube. $^{13}$C nuclei are randomly placed on the diamond lattice. (b) \emph{\textbf{Random electron network}}, with both electrons and $^{13}$C nuclei randomly positioned on the diamond lattice. (c) \emph{\textbf{Polarization decay curves}} comparing the ordered (orange) and random (green) electron configurations, averaged over 100 independent configurations in each case. The ordered configuration leads to consistently faster decay, reflecting a shorter polarization lifetime due to the suppression of spatially isolated, trap-free regions. In contrast, the random electron networks support the formation of electron-free ``pockets'', which act as reservoirs for long-lived polarization and give rise to the slower decay observed.}
    \label{OrderedElectrons}
\end{figure} 

To probe the role of electron disorder in sustaining long-lived polarization dynamics, we performed comparative simulations using both ordered and random electron configurations. In the ordered case, eight electrons were positioned at the centers of the eight octants of the simulation cube, and the simulation cube size was chosen such that eight electrons would be present for a concentration of 30ppm. This arrangement was chosen to ensure maximal coverage of the simulation volume with periodic boundary conditions and serves as a useful reference for a minimally disordered electron configuration. For each configurational average, the electrons were held fixed while the $^{13}$C were randomly placed on the diamond lattice as in all previous simulations. In contrast, the random case consisted of using the same simulation size and randomly placing electrons and $^{13}$C on the diamond lattice at their respective concentrations, as done in all previous simulations. Fig.~\ref{OrderedElectrons}a illustrates the ordered electron network within the random $^{13}$C lattice, while Fig.~\ref{OrderedElectrons}b shows a representative configuration with the randomly placed electrons. Fig.~\ref{OrderedElectrons}c compares the polarization decay curves for the two cases, with the ordered-electron result plotted in orange and the random-electron result in green. Notably, the decay in the ordered case is consistently faster, indicating a shorter polarization lifetime. This difference can be attributed to the suppression of spatially isolated trap-free (electron-free) regions in the ordered configuration. In the configurations with random-electron networks, certain regions of the lattice remain relatively far from any electron spin, effectively forming trap-free domains that serve as reservoirs for long-lived polarization. By contrast, in the ordered electron network, the uniform placement of electrons minimizes the formation of such regions, leading to more homogeneous relaxation and a more rapid overall decay.

\section{Goodness of Fit Analysis}
\label{section_fitting}

\begin{figure}[t]
    \centering
    \includegraphics[width=0.9\columnwidth]{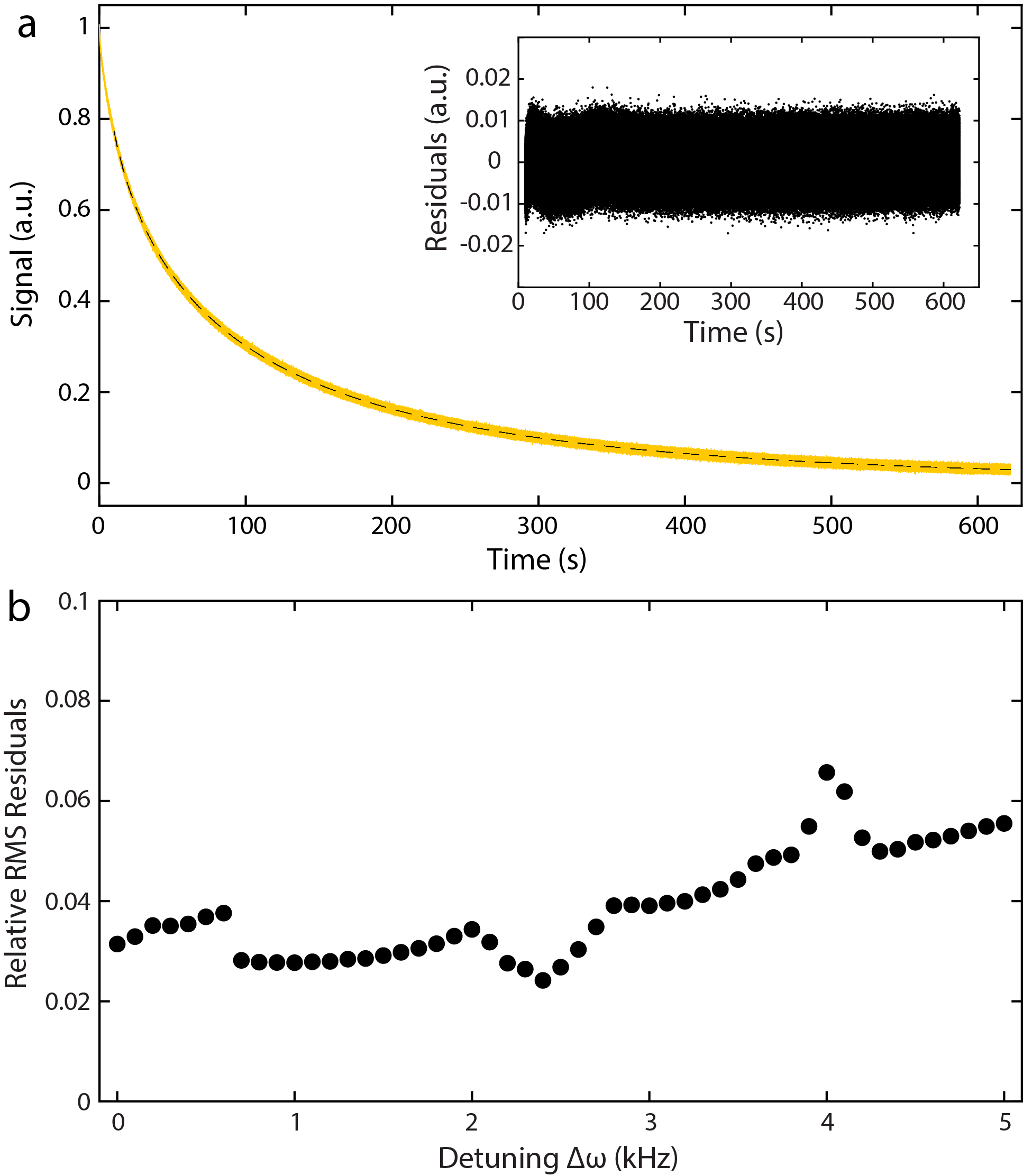}
    \caption{\textbf{Goodness of fit analysis} (a) \emph{\textbf{Fit and residuals}} for $\Delta\omega=2$ kHz. Yellow trace shows raw experimental data acquired over 600 s ($\sim8$M data points sampled every $\approx$80 $\mu$s); black dashed line shows best fit to emergent law $e^{-\sqrt{R_p t}} e^{-R_d t}$ (b) \emph{\textbf{Relative RMS residuals}} for 50 decay curves with detuning $\Delta\omega$ spanning 0-5 kHz, demonstrating uniformly high fit quality and confirming the robustness of the emergent law across the parameter space.}
    \label{Fitting}
\end{figure}

Each experimental decay trace was independently fit to the emergent decoherence law (\zr{emerge}), $M(t)=e^{-\sqrt{R_p t}} e^{-R_d t}$, using the raw data obtained over the 600 s measurement window containing $\sim$8 million data points sampled every $\approx$80 $\mu$s. To exclude short-time transients associated with spins in the frozen core (see Methods), the first 10 s of each trace were omitted from the fit. Fig.~\ref{Fitting}a shows a representative decay curve ($\Delta\omega=2$ kHz, yellow) together with the best fit (black dashed line). The inset displays the residuals, which are structureless and remain within the noise floor, demonstrating that the functional form captures the full temporal behavior of the signal. To quantify the goodness of fit across all data, we evaluated the relative root-mean-square (rRMS) residual,
\begin{equation}
\mathrm{rRMS} = 
\frac{\sqrt{\langle (S_\text{data}-S_\text{fit})^2\rangle}}
     {\sqrt{\langle S_\text{data}^2\rangle}},
\end{equation}
for 50 traces with detuning $\Delta\omega$ spanning 0–5 kHz.  
As shown in Fig.~\ref{Fitting}b, the rRMS values are relatively constant, indicating excellent agreement between data and model and confirming the robustness of \zr{emerge} across the full parameter space. Residuals and overall fit quality exhibit the same behavior for data acquired as a function of laser power (\zfr{fig2}, main text), further confirming the robustness of the emergent law across distinct experimental conditions.

\section{Heating Effects} 
\label{section_heating}

\begin{figure}[b]
    \centering
    \includegraphics[width=0.9\columnwidth]{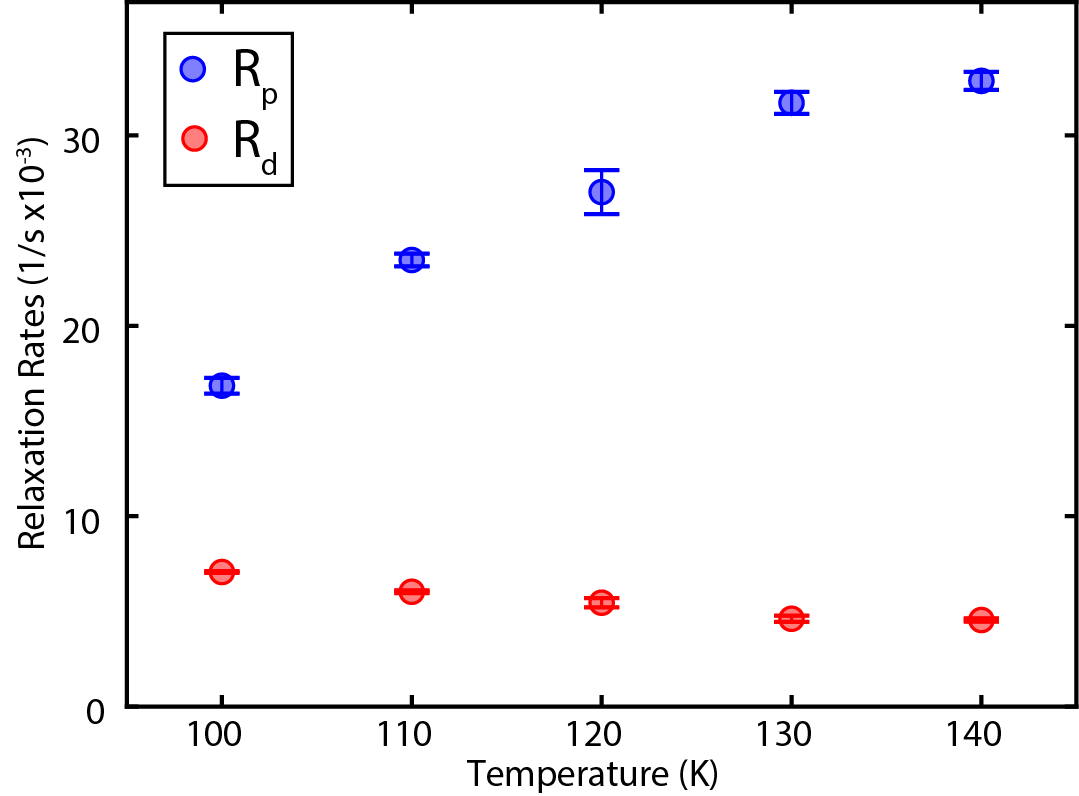}
    \caption{\textbf{Relaxation rates versus temperature} Experimentally measured relaxation rates $R_{p}$ and $R_{d}$ as a function of temperature for resonant ($\Delta\omega=0$) $\theta=90^{\circ}$ pulses (Regime \T{I}). Each measurement was repeated three times and error bars represent the standard error of the mean.}
    \label{TempRates}
\end{figure}

To determine whether sample heating contributes to the trends observed in Fig.~2 of the main text, we experimentally measured the relaxation rates $R_{p}$ and $R_{d}$ as a function of temperature by systematically heating the cryostat, under on-resonance ($\Delta\omega = 0$) and $\xt = 90^\circ$ pulses (Regime \T{I}), as shown in Fig.~\ref{TempRates}. The resulting temperature-dependent behavior is qualitatively distinct from that observed under laser illumination, indicating that heating alone cannot account for the laser-induced effects. Notably, under laser illumination, the paramagnetic relaxation rate $R_{p}$ initially increases but then decreases beyond a certain rate. In contrast, $R_{p}$ increases monotonically with temperature and ultimately exceeds the values observed under illumination. These distinct behaviors reflect the fact that different correlation times are being modulated in each case. Furthermore, the diffusive relaxation rate $R_{d}$ remains largely insensitive to temperature changes, in contrast to its pronounced increase under laser illumination. This disparity further supports the conclusion that the observed changes in $R_{d}$ are not thermal in origin but instead arise from active modulation of the electron environment. Specifically, laser illumination preferentially populates the $m_{s}=0$ state of the NV center and enhances fluctuations in the surrounding electron spin bath. These combined effects reduce the effective hyperfine field experienced by the nuclear spins, thereby enhancing diffusive relaxation leading to an increased value of $R_{d}$. The absence of similar changes under thermal modulation confirms that the behavior observed under laser illumination cannot be attributed to heating. 

\begin{figure}[t]
    \centering
    \includegraphics[width=0.9\columnwidth]{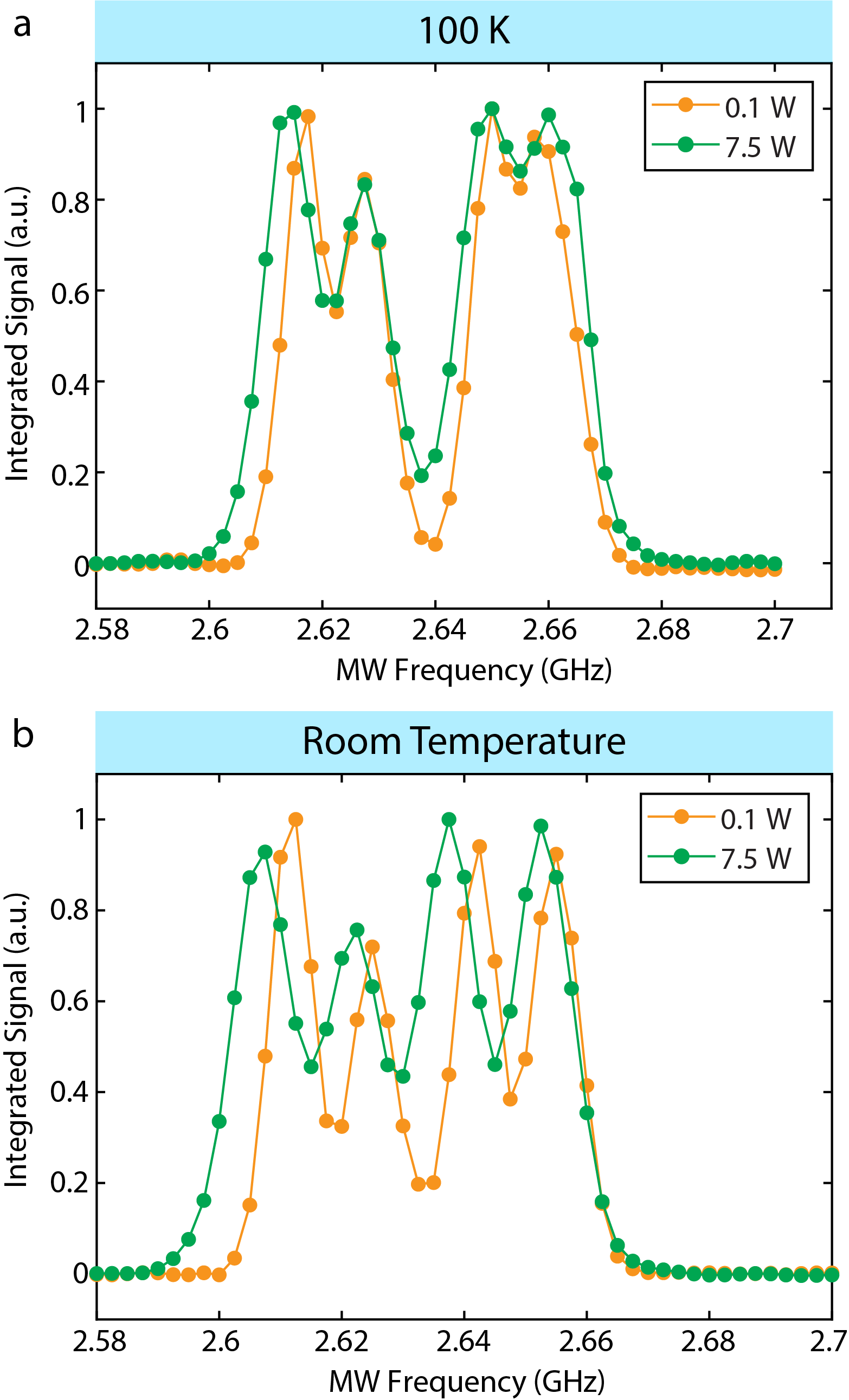}
    \caption{\textbf{NV-center EPR spectra as temperature probe.} (a) \emph{\textbf{NV-center EPR spectra at 100 K}} under laser powers of 0.1 W (orange) and 7.5 W (green). The spectra are obtained by integrating the $^{13}$C NMR signal as a function of the applied microwave (MW) center frequency during optical hyperpolarization. Sample heating would modify the NV center's zero field splitting (ZFS) and thus shift the EPR spectrum; however, no measurable shift is observed between the two laser powers. Because the ZFS is only weakly temperature dependent near 100 K, this indicates that any heating is smaller than our experimental resolution. (b) \emph{\textbf{Equivalent measurement performed at room temperature}}. The EPR spectrum obtained under 7.5 W illumination (green) exhibits a small shift of one frequency increment (2.5 MHz) relative to the 0.1 W spectrum (orange), corresponding to a temperature rise of approximately 40 K. This places an upper-bound on the possible temperature increase from laser illumination at 100K.}
    \label{EPR_Data}
\end{figure}

To further assess whether laser-induced heating contributes to the effects reported in the main text, we measured the electron paramagnetic resonance (EPR) spectrum of the NV centers indirectly through the $^{13}$C nuclei under varying laser powers. This approach, established in previous works~\cite{pillaiElectrontonuclearSpectralMapping2023a}, exploits the transfer of polarization from optically pumped NV centers to surrounding $^{13}$C spins. In brief, laser illumination and chirped microwave (MW) fields are applied to hyperpolarize the $^{13}$C nuclei, and the resulting NMR signal amplitude is measured as a function of the MW center frequency. Repeating this process across a range of MW center frequencies reconstructs the NV center's EPR spectrum as indirectly detected by the nuclear spins.

The NV center’s zero-field splitting (ZFS) is known to depend sensitively on temperature~\cite{acostaTemperatureDependenceNitrogenVacancy2010, cambriaPhysicallyMotivatedAnalytical2023}; heating of the sample would therefore manifest as a measurable shift of the EPR spectrum. We performed this measurement at two laser powers, 0.1 W and 7.5 W, using the same conditions as those used for experiments in the main text. At each MW center frequency, after applying laser illumination and chirped MW excitation with a 5 MHz bandwidth for 120 s, we recorded the full $^{13}$C decay curve under on-resonance ($\Delta\omega = 0$) Floquet driving with $\xt = 90^\circ$ pulses (Regime \T{I}), integrated the first two seconds of the signal decay over time, and normalized the result such that the maximum integrated signal equals one. The normalized signal amplitude as a function of MW center frequency yields the NV center's EPR spectrum shown in Fig.~\ref{EPR_Data}a for $T=100$ K, where the sample is actively cooled with liquid nitrogen.

At 100 K, the EPR spectra obtained under 0.1 W and 7.5 W illumination are indistinguishable within the experimental resolution (2.5 MHz frequency steps). The absence of any measurable shift indicates that illumination at 7.5 W does not significantly heat the sample under cryogenic conditions. We note, however, that the temperature dependence of the NV ZFS flattens near 100 K~\cite{cambriaPhysicallyMotivatedAnalytical2023}, limiting the sensitivity of this measurement. Consequently, these data constrain the possible temperature increase to less than approximately 100 K, but cannot rule out smaller temperature changes below this range.

To provide a more quantitative upper bound, we repeated the same measurement at room temperature, where the NV ZFS varies more strongly with temperature. As shown in Fig.~\ref{EPR_Data}b, the EPR spectrum acquired under 7.5 W illumination shifts by about two frequency increments (5 MHz) relative to that obtained at 0.1 W. Given the known temperature dependence of the NV ZFS at room temperature, this corresponds to a temperature increase of approximately 60 K, within the uncertainty set by our frequency resolution. Because the 100 K measurements are performed under active cryogenic cooling, we anticipate the temperature rise in those experiments to be substantially smaller than this upper bound.

\clearpage
\bibliography{DiffusionPaper}  

@article{abaninColloquiumManybodyLocalization2019,
  title = {{\emph{Colloquium}} : {{Many-body}} Localization, Thermalization, and Entanglement},
  shorttitle = {{\emph{Colloquium}}},
  author = {Abanin, Dmitry A. and Altman, Ehud and Bloch, Immanuel and Serbyn, Maksym},
  year = 2019,
  month = may,
  journal = {Reviews of Modern Physics},
  volume = {91},
  number = {2},
  pages = {021001},
  issn = {0034-6861, 1539-0756},
  doi = {10.1103/RevModPhys.91.021001},
  urldate = {2025-11-06},
  langid = {english}
}

@article{abobeihAtomicscaleImaging27nuclearspin2019,
  title = {Atomic-Scale Imaging of a 27-Nuclear-Spin Cluster Using a Quantum Sensor},
  author = {Abobeih, M. H. and Randall, J. and Bradley, C. E. and Bartling, H. P. and Bakker, M. A. and Degen, M. J. and Markham, M. and Twitchen, D. J. and Taminiau, T. H.},
  year = 2019,
  month = dec,
  journal = {Nature},
  volume = {576},
  number = {7787},
  pages = {411--415},
  publisher = {Nature Publishing Group},
  issn = {1476-4687},
  doi = {10.1038/s41586-019-1834-7},
  urldate = {2025-11-06},
  copyright = {2019 The Author(s), under exclusive licence to Springer Nature Limited},
  langid = {english}
}

@book{abragamNuclearMagnetismOrder1982,
  title = {Nuclear Magnetism : Order and Disorder},
  shorttitle = {Nuclear Magnetism},
  author = {Abragam, A.},
  year = 1982,
  publisher = {Oxford : Clarendon Press ; New York : Oxford University Press},
  urldate = {2025-07-11},
  collaborator = {{Internet Archive}},
  isbn = {978-0-19-851294-3},
  langid = {english}
}

@book{abragamPrinciplesNuclearMagnetism1962,
  title = {The Principles of Nuclear Magnetism},
  author = {Abragam, A.},
  year = 1962,
  publisher = {Oxford, Clarendon Press [1962]},
  urldate = {2025-07-11},
  collaborator = {{Internet Archive}},
  langid = {english}
}

@article{acostaTemperatureDependenceNitrogenVacancy2010,
  title = {Temperature {{Dependence}} of the {{Nitrogen-Vacancy Magnetic Resonance}} in {{Diamond}}},
  author = {Acosta, V. M. and Bauch, E. and Ledbetter, M. P. and Waxman, A. and Bouchard, L.-S. and Budker, D.},
  year = 2010,
  month = feb,
  journal = {Physical Review Letters},
  volume = {104},
  number = {7},
  pages = {070801},
  issn = {0031-9007, 1079-7114},
  doi = {10.1103/PhysRevLett.104.070801},
  urldate = {2025-11-03},
  copyright = {http://link.aps.org/licenses/aps-default-license},
  langid = {english}
}

@article{ajoyOptimalPulseSpacing2011,
  title = {Optimal Pulse Spacing for Dynamical Decoupling in the Presence of a Purely Dephasing Spin Bath},
  author = {Ajoy, Ashok and {\'A}lvarez, Gonzalo A. and Suter, Dieter},
  year = 2011,
  month = mar,
  journal = {Physical Review A},
  volume = {83},
  number = {3},
  pages = {032303},
  publisher = {American Physical Society},
  doi = {10.1103/PhysRevA.83.032303},
  urldate = {2025-07-21}
}

@article{ajoyOrientationindependentRoomTemperature2018a,
  title = {Orientation-Independent Room Temperature Optical {{13C}} Hyperpolarization in Powdered Diamond},
  author = {Ajoy, Ashok and Liu, Kristina and Nazaryan, Raffi and Lv, Xudong and Zangara, Pablo R. and Safvati, Benjamin and Wang, Guoqing and Arnold, Daniel and Li, Grace and Lin, Arthur and Raghavan, Priyanka and Druga, Emanuel and Dhomkar, Siddharth and Pagliero, Daniela and Reimer, Jeffrey A. and Suter, Dieter and Meriles, Carlos A. and Pines, Alexander},
  year = 2018,
  month = may,
  journal = {Science Advances},
  volume = {4},
  number = {5},
  pages = {eaar5492},
  publisher = {American Association for the Advancement of Science},
  doi = {10.1126/sciadv.aar5492},
  urldate = {2025-05-19}
}

@article{alvarezMeasuringSpectrumColored2011a,
  title = {Measuring the {{Spectrum}} of {{Colored Noise}} by {{Dynamical Decoupling}}},
  author = {{\'A}lvarez, Gonzalo A. and Suter, Dieter},
  year = 2011,
  month = nov,
  journal = {Physical Review Letters},
  volume = {107},
  number = {23},
  pages = {230501},
  publisher = {American Physical Society},
  doi = {10.1103/PhysRevLett.107.230501},
  urldate = {2025-07-11}
}

@article{aslamPhotoinducedIonizationDynamics2013,
  title = {Photo-Induced Ionization Dynamics of the Nitrogen Vacancy Defect in Diamond Investigated by Single-Shot Charge State Detection},
  author = {Aslam, N and Waldherr, G and Neumann, P and Jelezko, F and Wrachtrup, J},
  year = 2013,
  month = jan,
  journal = {New Journal of Physics},
  volume = {15},
  number = {1},
  pages = {013064},
  publisher = {IOP Publishing},
  issn = {1367-2630},
  doi = {10.1088/1367-2630/15/1/013064},
  urldate = {2025-10-29},
  langid = {english}
}

@article{beatrezFloquetPrethermalizationLifetime2021a,
  title = {Floquet {{Prethermalization}} with {{Lifetime Exceeding}} 90 s in a {{Bulk Hyperpolarized Solid}}},
  author = {Beatrez, William and Janes, Otto and Akkiraju, Amala and Pillai, Arjun and Oddo, Alexander and Reshetikhin, Paul and Druga, Emanuel and McAllister, Maxwell and Elo, Mark and Gilbert, Benjamin and Suter, Dieter and Ajoy, Ashok},
  year = 2021,
  month = oct,
  journal = {Physical Review Letters},
  volume = {127},
  number = {17},
  pages = {170603},
  publisher = {American Physical Society},
  doi = {10.1103/PhysRevLett.127.170603},
  urldate = {2025-05-19}
}

@article{bernienProbingManybodyDynamics2017,
  title = {Probing Many-Body Dynamics on a 51-Atom Quantum Simulator},
  author = {Bernien, Hannes and Schwartz, Sylvain and Keesling, Alexander and Levine, Harry and Omran, Ahmed and Pichler, Hannes and Choi, Soonwon and Zibrov, Alexander S. and Endres, Manuel and Greiner, Markus and Vuleti{\'c}, Vladan and Lukin, Mikhail D.},
  year = 2017,
  month = nov,
  journal = {Nature},
  volume = {551},
  number = {7682},
  pages = {579--584},
  publisher = {Nature Publishing Group},
  issn = {1476-4687},
  doi = {10.1038/nature24622},
  urldate = {2025-11-06},
  copyright = {2017 Macmillan Publishers Limited, part of Springer Nature. All rights reserved.},
  langid = {english}
}

@article{bloembergenInteractionNuclearSpins1949a,
  title = {On the Interaction of Nuclear Spins in a Crystalline Lattice},
  author = {Bloembergen, N.},
  year = 1949,
  month = may,
  journal = {Physica},
  volume = {15},
  number = {3},
  pages = {386--426},
  issn = {0031-8914},
  doi = {10.1016/0031-8914(49)90114-7},
  urldate = {2025-07-11}
}

@article{blumbergNuclearSpinLatticeRelaxation1960b,
  title = {Nuclear {{Spin-Lattice Relaxation Caused}} by {{Paramagnetic Impurities}}},
  author = {Blumberg, W. E.},
  year = 1960,
  month = jul,
  journal = {Physical Review},
  volume = {119},
  number = {1},
  pages = {79--84},
  publisher = {American Physical Society},
  doi = {10.1103/PhysRev.119.79},
  urldate = {2025-07-11}
}

@article{boutisSpinDiffusionCorrelated2004,
  title = {Spin {{Diffusion}} of {{Correlated Two-Spin States}} in a {{Dielectric Crystal}}},
  author = {Boutis, G. and Greenbaum, D. and Cho, H. and Cory, D. and Ramanathan, C.},
  year = 2004,
  month = mar,
  journal = {Physical Review Letters},
  volume = {92},
  number = {13},
  pages = {137201},
  issn = {0031-9007, 1079-7114},
  doi = {10.1103/PhysRevLett.92.137201},
  urldate = {2025-10-16},
  langid = {english}
}

@article{bukovUniversalHighfrequencyBehavior2015,
  title = {Universal High-Frequency Behavior of Periodically Driven Systems: From Dynamical Stabilization to {{Floquet}} Engineering},
  shorttitle = {Universal High-Frequency Behavior of Periodically Driven Systems},
  author = {Bukov, Marin and D'Alessio, Luca and Polkovnikov, Anatoli},
  year = 2015,
  month = mar,
  journal = {Advances in Physics},
  volume = {64},
  number = {2},
  pages = {139--226},
  publisher = {Taylor \& Francis},
  issn = {0001-8732},
  doi = {10.1080/00018732.2015.1055918},
  urldate = {2025-07-11}
}

@article{bussandriP1CenterElectron2024,
  title = {P1 {{Center Electron Spin Clusters Are Prevalent}} in {{Type Ib Diamonds}}},
  author = {Bussandri, Santiago and Shimon, Daphna and Equbal, Asif and Ren, Yuhang and Takahashi, Susumu and Ramanathan, Chandrasekhar and Han, Songi},
  year = 2024,
  month = feb,
  journal = {Journal of the American Chemical Society},
  volume = {146},
  number = {8},
  pages = {5088--5099},
  publisher = {American Chemical Society},
  issn = {0002-7863},
  doi = {10.1021/jacs.3c06705},
  urldate = {2025-07-11}
}

@article{cambriaPhysicallyMotivatedAnalytical2023,
  title = {Physically Motivated Analytical Expression for the Temperature Dependence of the Zero-Field Splitting of the Nitrogen-Vacancy Center in Diamond},
  author = {Cambria, M. C. and Thiering, G. and Norambuena, A. and Dinani, H. T. and Gardill, A. and Kemeny, I. and Lordi, V. and Gali, {\'A}. and Maze, J. R. and Kolkowitz, S.},
  year = 2023,
  month = nov,
  journal = {Physical Review B},
  volume = {108},
  number = {18},
  pages = {L180102},
  issn = {2469-9950, 2469-9969},
  doi = {10.1103/PhysRevB.108.L180102},
  urldate = {2025-11-03},
  langid = {english}
}

@article{capozziThermalAnnihilationPhotoinduced2017,
  title = {Thermal Annihilation of Photo-Induced Radicals Following Dynamic Nuclear Polarization to Produce Transportable Frozen Hyperpolarized {{13C-substrates}}},
  author = {Capozzi, Andrea and Cheng, Tian and Boero, Giovanni and Roussel, Christophe and Comment, Arnaud},
  year = 2017,
  month = jun,
  journal = {Nature Communications},
  volume = {8},
  number = {1},
  pages = {15757},
  publisher = {Nature Publishing Group},
  issn = {2041-1723},
  doi = {10.1038/ncomms15757},
  urldate = {2025-11-03},
  copyright = {2017 The Author(s)},
  langid = {english}
}

@book{chakrabartiQuantumSemiclassicalPercolation2009,
  title = {Quantum and {{Semi-classical Percolation}} and {{Breakdown}} in {{Disordered Solids}}},
  editor = {Chakrabarti, Bikas K. and Bardhan, Kamal K. and Sen, Asok K.},
  year = 2009,
  series = {Lecture {{Notes}} in {{Physics}}},
  volume = {762},
  publisher = {Springer},
  address = {Berlin, Heidelberg},
  doi = {10.1007/978-3-540-85428-9},
  urldate = {2025-11-03},
  copyright = {https://www.springernature.com/gp/researchers/text-and-data-mining},
  isbn = {978-3-540-85427-2 978-3-540-85428-9},
  langid = {english}
}

@article{choiExploringManybodyLocalization2016a,
  title = {Exploring the Many-Body Localization Transition in Two Dimensions},
  author = {Choi, Jae-yoon and Hild, Sebastian and Zeiher, Johannes and Schau{\ss}, Peter and {Rubio-Abadal}, Antonio and Yefsah, Tarik and Khemani, Vedika and Huse, David A. and Bloch, Immanuel and Gross, Christian},
  year = 2016,
  month = jun,
  journal = {Science},
  volume = {352},
  number = {6293},
  pages = {1547--1552},
  publisher = {American Association for the Advancement of Science},
  doi = {10.1126/science.aaf8834},
  urldate = {2025-11-06}
}

@article{choiRobustDynamicHamiltonian2020,
  title = {Robust {{Dynamic Hamiltonian Engineering}} of {{Many-Body Spin Systems}}},
  author = {Choi, Joonhee and Zhou, Hengyun and Knowles, Helena S. and Landig, Renate and Choi, Soonwon and Lukin, Mikhail D.},
  year = 2020,
  month = jul,
  journal = {Physical Review X},
  volume = {10},
  number = {3},
  pages = {031002},
  issn = {2160-3308},
  doi = {10.1103/PhysRevX.10.031002},
  urldate = {2025-11-06},
  langid = {english}
}

@article{cywinskiHowEnhanceDephasing2008,
  title = {How to Enhance Dephasing Time in Superconducting Qubits},
  author = {Cywi{\'n}ski, {\L}ukasz and Lutchyn, Roman M. and Nave, Cody P. and Das Sarma, S.},
  year = 2008,
  month = may,
  journal = {Physical Review B},
  volume = {77},
  number = {17},
  pages = {174509},
  publisher = {American Physical Society},
  doi = {10.1103/PhysRevB.77.174509},
  urldate = {2025-07-11}
}

@article{davisProbingManybodyDynamics2023,
  title = {Probing Many-Body Dynamics in a Two-Dimensional Dipolar Spin Ensemble},
  author = {Davis, E. J. and Ye, B. and Machado, F. and Meynell, S. A. and Wu, W. and Mittiga, T. and Schenken, W. and Joos, M. and Kobrin, B. and Lyu, Y. and Wang, Z. and Bluvstein, D. and Choi, S. and Zu, C. and Jayich, A. C. Bleszynski and Yao, N. Y.},
  year = 2023,
  month = jun,
  journal = {Nature Physics},
  volume = {19},
  number = {6},
  pages = {836--844},
  publisher = {Nature Publishing Group},
  issn = {1745-2481},
  doi = {10.1038/s41567-023-01944-5},
  urldate = {2025-10-22},
  copyright = {2023 The Author(s)},
  langid = {english}
}

@article{dohertyTheoryGroundstateSpin2012,
  title = {Theory of the Ground-State Spin of the {{NV}} - Center in Diamond},
  author = {Doherty, M. W. and Dolde, F. and Fedder, H. and Jelezko, F. and Wrachtrup, J. and Manson, N. B. and Hollenberg, L. C. L.},
  year = 2012,
  month = may,
  journal = {Physical Review B},
  volume = {85},
  number = {20},
  pages = {205203},
  issn = {1098-0121, 1550-235X},
  doi = {10.1103/PhysRevB.85.205203},
  urldate = {2025-10-29},
  copyright = {http://link.aps.org/licenses/aps-default-license},
  langid = {english}
}

@article{dsouzaCryogenicFieldcyclingInstrument2025,
  title = {Cryogenic Field-Cycling Instrument for Optical {{NMR}} Hyperpolarization Studies},
  author = {D'Souza, Noella and Harkins, Kieren A. and Selco, Cooper and Basumallick, Ushoshi and Breuer, Samantha and Zhang, Zhuorui and Reshetikhin, Paul and Ho, Marcus and Nayak, Aniruddha and McAllister, Maxwell and Druga, Emanuel and Marchiori, David and Ajoy, Ashok},
  year = 2025,
  month = jun,
  journal = {Journal of Magnetic Resonance},
  volume = {375},
  pages = {107874},
  issn = {1090-7807},
  doi = {10.1016/j.jmr.2025.107874},
  urldate = {2025-07-11}
}

@article{eckardtHighfrequencyApproximationPeriodically2015a,
  title = {High-Frequency Approximation for Periodically Driven Quantum Systems from a {{Floquet-space}} Perspective},
  author = {Eckardt, Andr{\'e} and Anisimovas, Egidijus},
  year = 2015,
  month = sep,
  journal = {New Journal of Physics},
  volume = {17},
  number = {9},
  pages = {093039},
  publisher = {IOP Publishing},
  issn = {1367-2630},
  doi = {10.1088/1367-2630/17/9/093039},
  urldate = {2025-07-11},
  langid = {english}
}

@article{eisertQuantumManybodySystems2015,
  title = {Quantum Many-Body Systems out of Equilibrium},
  author = {Eisert, J. and Friesdorf, M. and Gogolin, C.},
  year = 2015,
  month = feb,
  journal = {Nature Physics},
  volume = {11},
  number = {2},
  pages = {124--130},
  publisher = {Nature Publishing Group},
  issn = {1745-2481},
  doi = {10.1038/nphys3215},
  urldate = {2025-11-06},
  copyright = {2014 Springer Nature Limited},
  langid = {english}
}

@article{ernstNuclearMagneticDouble1966,
  title = {Nuclear {{Magnetic Double Resonance}} with an {{Incoherent Radio}}-{{Frequency Field}}},
  author = {Ernst, R. R.},
  year = 1966,
  month = nov,
  journal = {The Journal of Chemical Physics},
  volume = {45},
  number = {10},
  pages = {3845--3861},
  issn = {0021-9606},
  doi = {10.1063/1.1727409},
  urldate = {2025-07-11}
}

@article{friedmanDiffusiveHydrodynamicsIntegrability2020,
  title = {Diffusive Hydrodynamics from Integrability Breaking},
  author = {Friedman, Aaron J. and Gopalakrishnan, Sarang and Vasseur, Romain},
  year = 2020,
  month = may,
  journal = {Physical Review B},
  volume = {101},
  number = {18},
  pages = {180302},
  issn = {2469-9950, 2469-9969},
  doi = {10.1103/PhysRevB.101.180302},
  urldate = {2025-10-31},
  langid = {english}
}

@article{furmanNuclearSpinlatticeRelaxation1995a,
  title = {Nuclear Spin-Lattice Relaxation via Paramagnetic Impurities in Solids with Arbitrary Space Dimension},
  author = {Furman, G. B. and Kunoff, E. M. and Goren, S. D. and Pasquier, V. and Tinet, D.},
  year = 1995,
  month = oct,
  journal = {Physical Review B},
  volume = {52},
  number = {14},
  pages = {10182--10187},
  publisher = {American Physical Society},
  doi = {10.1103/PhysRevB.52.10182},
  urldate = {2025-07-11}
}

@article{furmanNuclearSpinlatticeRelaxation1997a,
  title = {Nuclear Spin-Lattice Relaxation of Dipolar Order Caused by Paramagnetic Impurities},
  author = {Furman, G. B. and Panich, A. M. and Yochelis, A. and Kunoff, E. M. and Goren, S. D.},
  year = 1997,
  month = jan,
  journal = {Physical Review B},
  volume = {55},
  number = {1},
  pages = {439--444},
  publisher = {American Physical Society},
  doi = {10.1103/PhysRevB.55.439},
  urldate = {2025-05-20}
}

@article{furmanSpinDiffusionSpinlattice1999a,
  title = {Spin Diffusion and Spin-Lattice Relaxation in Pulse Spin-Locking in Solids Containing Paramagnetic Impurities},
  author = {Furman, Gregory B. and Goren, Shaul D.},
  year = 1999,
  month = oct,
  journal = {Journal of Physics: Condensed Matter},
  volume = {11},
  number = {42},
  pages = {8275},
  issn = {0953-8984},
  doi = {10.1088/0953-8984/11/42/308},
  urldate = {2025-05-20},
  langid = {english}
}

@article{galiInitioSupercellCalculations2008,
  title = {Ab Initio Supercell Calculations on Nitrogen-Vacancy Center in Diamond: {{Electronic}} Structure and Hyperfine Tensors},
  shorttitle = {Ab Initio Supercell Calculations on Nitrogen-Vacancy Center in Diamond},
  author = {Gali, Adam and Fyta, Maria and Kaxiras, Efthimios},
  year = 2008,
  month = apr,
  journal = {Physical Review B},
  volume = {77},
  number = {15},
  pages = {155206},
  publisher = {American Physical Society},
  doi = {10.1103/PhysRevB.77.155206},
  urldate = {2025-07-21}
}

@article{gaoEffectsDisorderSuperconducting2025,
  title = {The Effects of Disorder in Superconducting Materials on Qubit Coherence},
  author = {Gao, Ran and Wu, Feng and Sun, Hantao and Chen, Jianjun and Deng, Hao and Ma, Xizheng and Miao, Xiaohe and Song, Zhijun and Wan, Xin and Wang, Fei and Xia, Tian and Ying, Make and Zhang, Chao and Shi, Yaoyun and Zhao, Hui-Hai and Deng, Chunqing},
  year = 2025,
  month = apr,
  journal = {Nature Communications},
  volume = {16},
  number = {1},
  pages = {3620},
  publisher = {Nature Publishing Group},
  issn = {2041-1723},
  doi = {10.1038/s41467-025-58745-y},
  urldate = {2025-11-03},
  copyright = {2025 The Author(s)},
  langid = {english}
}

@article{goldmanNuclearSpinDiffusion1982a,
  title = {Nuclear Spin Diffusion in a Rare Spin Species},
  author = {Goldman, M. and Jacquinot, J. F.},
  year = 1982,
  month = jul,
  journal = {Journal de Physique},
  volume = {43},
  number = {7},
  pages = {1049--1058},
  publisher = {Soci\'et\'e Fran\c caise de Physique},
  issn = {0302-0738, 2777-3396},
  doi = {10.1051/jphys:019820043070104900},
  urldate = {2025-05-20},
  langid = {english}
}

@article{golubevQuantumDecoherenceDisordered1998,
  title = {Quantum {{Decoherence}} in {{Disordered Mesoscopic Systems}}},
  author = {Golubev, Dmitrii and Zaikin, Andrei},
  year = 1998,
  month = aug,
  journal = {Physical Review Letters},
  volume = {81},
  number = {5},
  pages = {1074--1077},
  issn = {0031-9007, 1079-7114},
  doi = {10.1103/PhysRevLett.81.1074},
  urldate = {2025-11-06},
  langid = {english}
}

@article{gopalakrishnanHydrodynamicsOperatorSpreading2018,
  title = {Hydrodynamics of Operator Spreading and Quasiparticle Diffusion in Interacting Integrable Systems},
  author = {Gopalakrishnan, Sarang and Huse, David A. and Khemani, Vedika and Vasseur, Romain},
  year = 2018,
  month = dec,
  journal = {Physical Review B},
  volume = {98},
  number = {22},
  pages = {220303},
  issn = {2469-9950, 2469-9969},
  doi = {10.1103/PhysRevB.98.220303},
  urldate = {2025-11-06},
  langid = {english}
}

@article{grassbergerLongTimeProperties1982,
  title = {The Long Time Properties of Diffusion in a Medium with Static Traps},
  author = {Grassberger, Peter and Procaccia, Itamar},
  year = 1982,
  month = dec,
  journal = {The Journal of Chemical Physics},
  volume = {77},
  number = {12},
  pages = {6281--6284},
  issn = {0021-9606},
  doi = {10.1063/1.443832},
  urldate = {2025-07-21}
}

@misc{harkinsAnomalouslyExtendedFloquet2024,
  title = {Anomalously Extended {{Floquet}} Prethermal Lifetimes and Applications to Long-Time Quantum Sensing},
  author = {Harkins, Kieren A. and Selco, Cooper and Bengs, Christian and Marchiori, David and Moon, Leo Joon Il and Zhang, Zhuo-Rui and Yang, Aristotle and Singh, Angad and Druga, Emanuel and Song, Yi-Qiao and Ajoy, Ashok},
  year = 2024,
  month = oct,
  number = {arXiv:2410.09028},
  eprint = {2410.09028},
  primaryclass = {quant-ph},
  publisher = {arXiv},
  doi = {10.48550/arXiv.2410.09028},
  urldate = {2025-05-20},
  archiveprefix = {arXiv}
}

@article{hincksStatisticalInferenceQuantum2018,
  title = {Statistical Inference with Quantum Measurements: Methodologies for Nitrogen Vacancy Centers in Diamond},
  shorttitle = {Statistical Inference with Quantum Measurements},
  author = {Hincks, Ian and Granade, Christopher and Cory, David G},
  year = 2018,
  month = jan,
  journal = {New Journal of Physics},
  volume = {20},
  number = {1},
  pages = {013022},
  publisher = {IOP Publishing},
  issn = {1367-2630},
  doi = {10.1088/1367-2630/aa9c9f},
  urldate = {2025-04-30},
  langid = {english}
}

@article{ivanovFloquetTheoryMagnetic2021,
  title = {Floquet Theory in Magnetic Resonance: {{Formalism}} and Applications},
  shorttitle = {Floquet Theory in Magnetic Resonance},
  author = {Ivanov, Konstantin L. and Mote, Kaustubh R. and Ernst, Matthias and Equbal, Asif and Madhu, Perunthiruthy K.},
  year = 2021,
  month = oct,
  journal = {Progress in Nuclear Magnetic Resonance Spectroscopy},
  volume = {126--127},
  pages = {17--58},
  issn = {0079-6565},
  doi = {10.1016/j.pnmrs.2021.05.002},
  urldate = {2025-07-11}
}

@article{jarmolaTemperatureMagneticFieldDependentLongitudinal2012a,
  title = {Temperature- and {{Magnetic-Field-Dependent Longitudinal Spin Relaxation}} in {{Nitrogen-Vacancy Ensembles}} in {{Diamond}}},
  author = {Jarmola, A. and Acosta, V. M. and Jensen, K. and Chemerisov, S. and Budker, D.},
  year = 2012,
  month = may,
  journal = {Physical Review Letters},
  volume = {108},
  number = {19},
  pages = {197601},
  publisher = {American Physical Society},
  doi = {10.1103/PhysRevLett.108.197601},
  urldate = {2025-07-21}
}

@article{jelezkoSingleDefectCentres2006,
  title = {Single Defect Centres in Diamond: {{A}} Review},
  shorttitle = {Single Defect Centres in Diamond},
  author = {Jelezko, F. and Wrachtrup, J.},
  year = 2006,
  journal = {physica status solidi (a)},
  volume = {203},
  number = {13},
  pages = {3207--3225},
  issn = {1862-6319},
  doi = {10.1002/pssa.200671403},
  urldate = {2025-10-29},
  copyright = {Copyright \copyright{} 2006 WILEY-VCH Verlag GmbH \& Co. KGaA, Weinheim},
  langid = {english}
}

@article{joosProtectingQubitCoherence2022a,
  title = {Protecting Qubit Coherence by Spectrally Engineered Driving of the Spin Environment},
  author = {Joos, Maxime and Bluvstein, Dolev and Lyu, Yuanqi and Weld, David and Bleszynski Jayich, Ania},
  year = 2022,
  month = apr,
  journal = {npj Quantum Information},
  volume = {8},
  number = {1},
  pages = {47},
  publisher = {Nature Publishing Group},
  issn = {2056-6387},
  doi = {10.1038/s41534-022-00560-0},
  urldate = {2025-10-29},
  copyright = {2022 The Author(s)},
  langid = {english}
}

@article{joshiObservingEmergentHydrodynamics2022a,
  title = {Observing Emergent Hydrodynamics in a Long-Range Quantum Magnet},
  author = {Joshi, M. K. and Kranzl, F. and Schuckert, A. and Lovas, I. and Maier, C. and Blatt, R. and Knap, M. and Roos, C. F.},
  year = 2022,
  month = may,
  journal = {Science},
  volume = {376},
  number = {6594},
  pages = {720--724},
  publisher = {American Association for the Advancement of Science},
  doi = {10.1126/science.abk2400},
  urldate = {2025-11-06}
}

@article{khemaniOperatorSpreadingEmergence2018a,
  title = {Operator {{Spreading}} and the {{Emergence}} of {{Dissipative Hydrodynamics}} under {{Unitary Evolution}} with {{Conservation Laws}}},
  author = {Khemani, Vedika and Vishwanath, Ashvin and Huse, David A.},
  year = 2018,
  month = sep,
  journal = {Physical Review X},
  volume = {8},
  number = {3},
  pages = {031057},
  issn = {2160-3308},
  doi = {10.1103/PhysRevX.8.031057},
  urldate = {2025-11-06},
  langid = {english}
}

@article{khutsishviliSPINDIFFUSION1966,
  title = {{{SPIN DIFFUSION}}},
  author = {Khutsishvili, G. R.},
  year = 1966,
  month = may,
  journal = {Soviet Physics Uspekhi},
  volume = {8},
  number = {5},
  pages = {743},
  publisher = {IOP Publishing},
  issn = {0038-5670},
  doi = {10.1070/PU1966v008n05ABEH003035},
  urldate = {2025-10-29},
  langid = {english}
}

@article{khutsishviliSPINDIFFUSIONNUCLEAR1969,
  title = {{{SPIN DIFFUSION AND NUCLEAR MAGNETIC RELAXATION IN A CRYSTAL CONTAININGA MAGNETIC IMPURITY}}},
  author = {Khutsishvili, G. R.},
  year = 1969,
  month = jun,
  journal = {Soviet Physics Uspekhi},
  volume = {11},
  number = {6},
  pages = {802},
  publisher = {IOP Publishing},
  issn = {0038-5670},
  doi = {10.1070/PU1969v011n06ABEH003776},
  urldate = {2025-05-20},
  langid = {english}
}

@article{kirkpatrickTimeDependentTransport1982,
  title = {Time Dependent Transport in a Fluid with Static Traps},
  author = {Kirkpatrick, T. R.},
  year = 1982,
  month = apr,
  journal = {The Journal of Chemical Physics},
  volume = {76},
  number = {8},
  pages = {4255--4259},
  issn = {0021-9606},
  doi = {10.1063/1.443503},
  urldate = {2025-07-21}
}

@article{levittBroadbandHeteronuclearDecoupling1982,
  title = {Broadband Heteronuclear Decoupling},
  author = {Levitt, Malcolm H and Freeman, Ray and Frenkiel, Thomas},
  year = 1982,
  month = apr,
  journal = {Journal of Magnetic Resonance (1969)},
  volume = {47},
  number = {2},
  pages = {328--330},
  issn = {0022-2364},
  doi = {10.1016/0022-2364(82)90124-X},
  urldate = {2025-05-19}
}

@article{lunghiHowPhononsRelax2019,
  title = {How Do Phonons Relax Molecular Spins?},
  author = {Lunghi, Alessandro and Sanvito, Stefano},
  year = 2019,
  month = sep,
  journal = {Science Advances},
  volume = {5},
  number = {9},
  pages = {eaax7163},
  publisher = {American Association for the Advancement of Science},
  doi = {10.1126/sciadv.aax7163},
  urldate = {2025-11-03}
}

@misc{medinaThermodynamicsOpticalPumping2025,
  title = {Thermodynamics of the Optical Pumping Process in {{Nitrogen-Vacancy}} Centers},
  author = {Medina, Ivan and Muniz, S{\'e}rgio R. and Goettems, Elisa I. and {Soares-Pinto}, Diogo O.},
  year = 2025,
  month = mar,
  number = {arXiv:2503.08769},
  eprint = {2503.08769},
  primaryclass = {quant-ph},
  publisher = {arXiv},
  doi = {10.48550/arXiv.2503.08769},
  urldate = {2025-04-30},
  archiveprefix = {arXiv}
}

@article{miInformationScramblingQuantum2021,
  title = {Information Scrambling in Quantum Circuits},
  author = {Mi, Xiao and Roushan, Pedram and Quintana, Chris and Mandr{\`a}, Salvatore and Marshall, Jeffrey and Neill, Charles and Arute, Frank and Arya, Kunal and Atalaya, Juan and Babbush, Ryan and Bardin, Joseph C. and Barends, Rami and Basso, Joao and Bengtsson, Andreas and Boixo, Sergio and Bourassa, Alexandre and Broughton, Michael and Buckley, Bob B. and Buell, David A. and Burkett, Brian and Bushnell, Nicholas and Chen, Zijun and Chiaro, Benjamin and Collins, Roberto and Courtney, William and Demura, Sean and Derk, Alan R. and Dunsworth, Andrew and Eppens, Daniel and Erickson, Catherine and Farhi, Edward and Fowler, Austin G. and Foxen, Brooks and Gidney, Craig and Giustina, Marissa and Gross, Jonathan A. and Harrigan, Matthew P. and Harrington, Sean D. and Hilton, Jeremy and Ho, Alan and Hong, Sabrina and Huang, Trent and Huggins, William J. and Ioffe, L. B. and Isakov, Sergei V. and Jeffrey, Evan and Jiang, Zhang and Jones, Cody and Kafri, Dvir and Kelly, Julian and Kim, Seon and Kitaev, Alexei and Klimov, Paul V. and Korotkov, Alexander N. and Kostritsa, Fedor and Landhuis, David and Laptev, Pavel and Lucero, Erik and Martin, Orion and McClean, Jarrod R. and McCourt, Trevor and McEwen, Matt and Megrant, Anthony and Miao, Kevin C. and Mohseni, Masoud and Montazeri, Shirin and Mruczkiewicz, Wojciech and Mutus, Josh and Naaman, Ofer and Neeley, Matthew and Newman, Michael and Niu, Murphy Yuezhen and O'Brien, Thomas E. and Opremcak, Alex and Ostby, Eric and Pato, Balint and Petukhov, Andre and Redd, Nicholas and Rubin, Nicholas C. and Sank, Daniel and Satzinger, Kevin J. and Shvarts, Vladimir and Strain, Doug and Szalay, Marco and Trevithick, Matthew D. and Villalonga, Benjamin and White, Theodore and Yao, Z. Jamie and Yeh, Ping and Zalcman, Adam and Neven, Hartmut and Aleiner, Igor and Kechedzhi, Kostyantyn and Smelyanskiy, Vadim and Chen, Yu},
  year = 2021,
  month = dec,
  journal = {Science},
  volume = {374},
  number = {6574},
  pages = {1479--1483},
  publisher = {American Association for the Advancement of Science},
  doi = {10.1126/science.abg5029},
  urldate = {2025-11-06}
}

@article{millington-hotzeNuclearSpinDiffusion2023,
  title = {Nuclear Spin Diffusion in the Central Spin System of a {{GaAs}}/{{AlGaAs}} Quantum Dot},
  author = {{Millington-Hotze}, Peter and Manna, Santanu and {Covre da Silva}, Saimon F. and Rastelli, Armando and Chekhovich, Evgeny A.},
  year = 2023,
  month = may,
  journal = {Nature Communications},
  volume = {14},
  number = {1},
  pages = {2677},
  publisher = {Nature Publishing Group},
  issn = {2041-1723},
  doi = {10.1038/s41467-023-38349-0},
  urldate = {2025-11-06},
  copyright = {2023 The Author(s)},
  langid = {english}
}

@article{moonHighspeedHighmemoryNMR2025,
  title = {High-Speed, High-Memory {{NMR}} Spectrometer and Hyperpolarizer},
  author = {Moon, Leo Joon Il and Beatrez, William and Ball, Jason and Mercade, Joan and Elo, Mark and Singh, Angad and Druga, Emanuel and Ajoy, Ashok},
  year = 2025,
  month = nov,
  journal = {Journal of Magnetic Resonance},
  volume = {380},
  pages = {107952},
  issn = {1090-7807},
  doi = {10.1016/j.jmr.2025.107952},
  urldate = {2025-10-23}
}

@article{nir-aradNitrogenSubstitutionsAggregation2024,
  title = {Nitrogen {{Substitutions Aggregation}} and {{Clustering}} in {{Diamonds}} as {{Revealed}} by {{High-Field Electron Paramagnetic Resonance}}},
  author = {{Nir-Arad}, Orit and Shlomi, David H. and Manukovsky, Nurit and Laster, Eyal and Kaminker, Ilia},
  year = 2024,
  month = feb,
  journal = {Journal of the American Chemical Society},
  volume = {146},
  number = {8},
  pages = {5100--5107},
  publisher = {American Chemical Society},
  issn = {0002-7863},
  doi = {10.1021/jacs.3c06739},
  urldate = {2025-07-11}
}

@article{onizhukColloquiumDecoherenceSolidstate2025,
  title = {{\emph{Colloquium}} : {{Decoherence}} of Solid-State Spin Qubits: {{A}} Computational Perspective},
  shorttitle = {{\emph{Colloquium}}},
  author = {Onizhuk, Mykyta and Galli, Giulia},
  year = 2025,
  month = apr,
  journal = {Reviews of Modern Physics},
  volume = {97},
  number = {2},
  pages = {021001},
  issn = {0034-6861, 1539-0756},
  doi = {10.1103/RevModPhys.97.021001},
  urldate = {2025-11-06},
  langid = {english}
}

@article{pillaiElectrontonuclearSpectralMapping2023a,
  title = {Electron-to-Nuclear Spectral Mapping via Dynamic Nuclear Polarization},
  author = {Pillai, Arjun and Elanchezhian, Moniish and Virtanen, Teemu and Conti, Sophie and Ajoy, Ashok},
  year = 2023,
  month = oct,
  journal = {The Journal of Chemical Physics},
  volume = {159},
  number = {15},
  pages = {154201},
  issn = {0021-9606},
  doi = {10.1063/5.0157954},
  urldate = {2025-11-03}
}

@article{priscoScalingAnalysesHyperpolarization2021,
  title = {Scaling Analyses for Hyperpolarization Transfer across a Spin-Diffusion Barrier and into Bulk Solid Media},
  author = {Prisco, Nathan A. and Pinon, Arthur C. and Emsley, Lyndon and Chmelka, Bradley F.},
  year = 2021,
  journal = {Physical Chemistry Chemical Physics},
  volume = {23},
  number = {2},
  pages = {1006--1020},
  issn = {1463-9076, 1463-9084},
  doi = {10.1039/D0CP03195J},
  urldate = {2025-11-04},
  langid = {english}
}

@article{ramanathanDynamicNuclearPolarization2008a,
  title = {Dynamic {{Nuclear Polarization}} and {{Spin Diffusion}} in {{Nonconducting Solids}}},
  author = {Ramanathan, C.},
  year = 2008,
  month = aug,
  journal = {Applied Magnetic Resonance},
  volume = {34},
  number = {3},
  pages = {409--421},
  issn = {1613-7507},
  doi = {10.1007/s00723-008-0123-7},
  urldate = {2025-10-29},
  langid = {english}
}

@article{rauReversibleQuantumMicrodynamics1996,
  title = {From Reversible Quantum Microdynamics to Irreversible Quantum Transport},
  author = {Rau, J. and M{\"u}ller, B.},
  year = 1996,
  month = jul,
  journal = {Physics Reports},
  volume = {272},
  number = {1},
  pages = {1--59},
  issn = {0370-1573},
  doi = {10.1016/0370-1573(95)00077-1},
  urldate = {2025-11-06}
}

@book{resnickAdventuresStochasticProcesses2002,
  title = {Adventures in {{Stochastic Processes}}},
  author = {Resnick, Sidney I.},
  year = 2002,
  publisher = {Birkh\"auser},
  address = {Boston, MA},
  doi = {10.1007/978-1-4612-0387-2},
  urldate = {2025-10-16},
  copyright = {http://www.springer.com/tdm},
  isbn = {978-1-4612-6738-6 978-1-4612-0387-2},
  langid = {english}
}

@article{rhimCalculationSpinLattice1978,
  title = {Calculation of Spin--Lattice Relaxation during Pulsed Spin Locking in Solids},
  author = {Rhim, W.-K. and Burum, D. P. and Elleman, D. D.},
  year = 1978,
  month = jan,
  journal = {The Journal of Chemical Physics},
  volume = {68},
  number = {2},
  pages = {692--695},
  issn = {0021-9606},
  doi = {10.1063/1.435742},
  urldate = {2025-05-20}
}

@article{ryanRobustDecouplingTechniques2010,
  title = {Robust {{Decoupling Techniques}} to {{Extend Quantum Coherence}} in {{Diamond}}},
  author = {Ryan, C. A. and Hodges, J. S. and Cory, D. G.},
  year = 2010,
  month = nov,
  journal = {Physical Review Letters},
  volume = {105},
  number = {20},
  pages = {200402},
  issn = {0031-9007, 1079-7114},
  doi = {10.1103/PhysRevLett.105.200402},
  urldate = {2025-10-31},
  copyright = {http://link.aps.org/licenses/aps-default-license},
  langid = {english}
}

@article{sahinHighFieldMagnetometry2022c,
  title = {High Field Magnetometry with Hyperpolarized Nuclear Spins},
  author = {Sahin, Ozgur and {de Leon Sanchez}, Erica and Conti, Sophie and Akkiraju, Amala and Reshetikhin, Paul and Druga, Emanuel and Aggarwal, Aakriti and Gilbert, Benjamin and Bhave, Sunil and Ajoy, Ashok},
  year = 2022,
  month = sep,
  journal = {Nature Communications},
  volume = {13},
  number = {1},
  pages = {5486},
  publisher = {Nature Publishing Group},
  issn = {2041-1723},
  doi = {10.1038/s41467-022-32907-8},
  urldate = {2025-11-06},
  copyright = {2022 The Author(s)},
  langid = {english}
}

@article{sangtawesinOriginsDiamondSurface2019,
  title = {Origins of {{Diamond Surface Noise Probed}} by {{Correlating Single-Spin Measurements}} with {{Surface Spectroscopy}}},
  author = {Sangtawesin, Sorawis and Dwyer, Bo L. and Srinivasan, Srikanth and Allred, James J. and Rodgers, Lila V. H. and De Greve, Kristiaan and Stacey, Alastair and Dontschuk, Nikolai and O'Donnell, Kane M. and Hu, Di and Evans, D. Andrew and Jaye, Cherno and Fischer, Daniel A. and Markham, Matthew L. and Twitchen, Daniel J. and Park, Hongkun and Lukin, Mikhail D. and De Leon, Nathalie P.},
  year = 2019,
  month = sep,
  journal = {Physical Review X},
  volume = {9},
  number = {3},
  pages = {031052},
  issn = {2160-3308},
  doi = {10.1103/PhysRevX.9.031052},
  urldate = {2025-11-03},
  langid = {english}
}

@article{sarkarRapidlyEnhancedSpinPolarization2022a,
  title = {Rapidly {{Enhanced Spin-Polarization Injection}} in an {{Optically Pumped Spin Ratchet}}},
  author = {Sarkar, Adrisha and Blankenship, Brian and Druga, Emanuel and Pillai, Arjun and Nirodi, Ruhee and Singh, Siddharth and Oddo, Alexander and Reshetikhin, Paul and Ajoy, Ashok},
  year = 2022,
  month = sep,
  journal = {Physical Review Applied},
  volume = {18},
  number = {3},
  pages = {034079},
  publisher = {American Physical Society},
  doi = {10.1103/PhysRevApplied.18.034079},
  urldate = {2025-05-19}
}

@book{schlosshauerDecoherenceQuantumToClassicalTransition2007,
  title = {Decoherence and the {{Quantum-To-Classical Transition}}},
  author = {Schlosshauer, Maximilian},
  year = 2007,
  series = {Frontiers {{Collection}}},
  publisher = {Springer},
  address = {Berlin, Heidelberg},
  issn = {1612-3018},
  doi = {10.1007/978-3-540-35775-9},
  urldate = {2025-11-06},
  copyright = {http://www.springer.com/tdm},
  isbn = {978-3-540-35773-5 978-3-540-35775-9},
  langid = {english}
}

@article{schlosshauerQuantumDecoherence2019,
  title = {Quantum Decoherence},
  author = {Schlosshauer, Maximilian},
  year = 2019,
  month = oct,
  journal = {Physics Reports},
  series = {Quantum Decoherence},
  volume = {831},
  pages = {1--57},
  issn = {0370-1573},
  doi = {10.1016/j.physrep.2019.10.001},
  urldate = {2025-11-06}
}

@article{scholzOperatorbasedFloquetTheory2010,
  title = {Operator-Based {{Floquet}} Theory in Solid-State {{NMR}}},
  author = {Scholz, Ingo and {van Beek}, Jacco D. and Ernst, Matthias},
  year = 2010,
  month = may,
  journal = {Solid State Nuclear Magnetic Resonance},
  volume = {37},
  number = {3},
  pages = {39--59},
  issn = {0926-2040},
  doi = {10.1016/j.ssnmr.2010.04.003},
  urldate = {2025-07-11}
}

@article{siddiqiEngineeringHighcoherenceSuperconducting2021,
  title = {Engineering High-Coherence Superconducting Qubits},
  author = {Siddiqi, Irfan},
  year = 2021,
  month = oct,
  journal = {Nature Reviews Materials},
  volume = {6},
  number = {10},
  pages = {875--891},
  publisher = {Nature Publishing Group},
  issn = {2058-8437},
  doi = {10.1038/s41578-021-00370-4},
  urldate = {2025-11-03},
  copyright = {2021 Springer Nature Limited},
  langid = {english}
}

@article{simmonsNuclearSpinLatticeRelaxation1962,
  title = {Nuclear {{Spin-Lattice Relaxation}} in {{Dilute Paramagnetic Sapphire}}},
  author = {Simmons, W. W. and O'Sullivan, W. J. and Robinson, W. A.},
  year = 1962,
  month = aug,
  journal = {Physical Review},
  volume = {127},
  number = {4},
  pages = {1168--1178},
  issn = {0031-899X},
  doi = {10.1103/PhysRev.127.1168},
  urldate = {2025-10-16},
  copyright = {http://link.aps.org/licenses/aps-default-license},
  langid = {english}
}

@article{smithManybodyLocalizationQuantum2016b,
  title = {Many-Body Localization in a Quantum Simulator with Programmable Random Disorder},
  author = {Smith, J. and Lee, A. and Richerme, P. and Neyenhuis, B. and Hess, P. W. and Hauke, P. and Heyl, M. and Huse, D. A. and Monroe, C.},
  year = 2016,
  month = oct,
  journal = {Nature Physics},
  volume = {12},
  number = {10},
  pages = {907--911},
  publisher = {Nature Publishing Group},
  issn = {1745-2481},
  doi = {10.1038/nphys3783},
  urldate = {2025-11-06},
  copyright = {2016 Springer Nature Limited},
  langid = {english}
}

@article{sternDirectObservationHyperpolarization2021a,
  title = {Direct Observation of Hyperpolarization Breaking through the Spin Diffusion Barrier},
  author = {Stern, Quentin and Cousin, Samuel Fran{\c c}ois and {Mentink-Vigier}, Fr{\'e}d{\'e}ric and Pinon, Arthur C{\'e}sar and Elliott, Stuart James and Cala, Olivier and Jannin, Sami},
  year = 2021,
  month = apr,
  journal = {Science Advances},
  volume = {7},
  number = {18},
  pages = {eabf5735},
  publisher = {American Association for the Advancement of Science},
  doi = {10.1126/sciadv.abf5735},
  urldate = {2025-10-29}
}

@article{takahashiQuenchingSpinDecoherence2008,
  title = {Quenching {{Spin Decoherence}} in {{Diamond}} through {{Spin Bath Polarization}}},
  author = {Takahashi, Susumu and Hanson, Ronald and {van Tol}, Johan and Sherwin, Mark S. and Awschalom, David D.},
  year = 2008,
  month = jul,
  journal = {Physical Review Letters},
  volume = {101},
  number = {4},
  pages = {047601},
  publisher = {American Physical Society},
  doi = {10.1103/PhysRevLett.101.047601},
  urldate = {2025-07-11}
}

@article{trotzkyProbingRelaxationEquilibrium2012,
  title = {Probing the Relaxation towards Equilibrium in an Isolated Strongly Correlated One-Dimensional {{Bose}} Gas},
  author = {Trotzky, S. and Chen, Y.-A. and Flesch, A. and McCulloch, I. P. and Schollw{\"o}ck, U. and Eisert, J. and Bloch, I.},
  year = 2012,
  month = apr,
  journal = {Nature Physics},
  volume = {8},
  number = {4},
  pages = {325--330},
  publisher = {Nature Publishing Group},
  issn = {1745-2481},
  doi = {10.1038/nphys2232},
  urldate = {2025-11-06},
  copyright = {2012 Springer Nature Limited},
  langid = {english}
}

@article{tseNuclearSpinLatticeRelaxation1968a,
  title = {Nuclear {{Spin-Lattice Relaxation Via Paramagnetic Centers Without Spin Diffusion}}},
  author = {Tse, D. and Hartmann, S. R.},
  year = 1968,
  month = aug,
  journal = {Physical Review Letters},
  volume = {21},
  number = {8},
  pages = {511--514},
  publisher = {American Physical Society},
  doi = {10.1103/PhysRevLett.21.511},
  urldate = {2025-05-20}
}

@article{uysOptimizedNoiseFiltration2009,
  title = {Optimized {{Noise Filtration}} through {{Dynamical Decoupling}}},
  author = {Uys, Hermann and Biercuk, Michael J. and Bollinger, John J.},
  year = 2009,
  month = jul,
  journal = {Physical Review Letters},
  volume = {103},
  number = {4},
  pages = {040501},
  publisher = {American Physical Society},
  doi = {10.1103/PhysRevLett.103.040501},
  urldate = {2025-07-11}
}

@article{vugmeisterSpatialSpectralSpin1976a,
  title = {Spatial and {{Spectral Spin Diffusion}} in {{Dilute Spin Systems}}},
  author = {Vugmeister, B. E.},
  year = 1976,
  journal = {physica status solidi (b)},
  volume = {76},
  number = {1},
  pages = {161--170},
  issn = {1521-3951},
  doi = {10.1002/pssb.2220760116},
  urldate = {2025-05-20},
  copyright = {Copyright \copyright{} 1976 WILEY-VCH Verlag GmbH \& Co. KGaA},
  langid = {english}
}

@article{wiersmaRandomQuantumNetworks2010,
  title = {Random {{Quantum Networks}}},
  author = {Wiersma, Diederik S.},
  year = 2010,
  month = mar,
  journal = {Science},
  volume = {327},
  number = {5971},
  pages = {1333--1334},
  publisher = {American Association for the Advancement of Science},
  doi = {10.1126/science.1187084},
  urldate = {2025-11-03}
}

@article{wolfeDirectObservationNuclear1973,
  title = {Direct {{Observation}} of a {{Nuclear Spin Diffusion Barrier}}},
  author = {Wolfe, J. P.},
  year = 1973,
  month = oct,
  journal = {Physical Review Letters},
  volume = {31},
  number = {15},
  pages = {907--910},
  issn = {0031-9007},
  doi = {10.1103/PhysRevLett.31.907},
  urldate = {2025-11-04},
  copyright = {http://link.aps.org/licenses/aps-default-license},
  langid = {english}
}

@article{wuIntrinsicDecoherenceIsolated2017,
  title = {Intrinsic Decoherence in Isolated Quantum Systems},
  author = {Wu, Yang-Le and Deng, Dong-Ling and Li, Xiaopeng and Das Sarma, S.},
  year = 2017,
  month = jan,
  journal = {Physical Review B},
  volume = {95},
  number = {1},
  pages = {014202},
  issn = {2469-9950, 2469-9969},
  doi = {10.1103/PhysRevB.95.014202},
  urldate = {2025-11-06},
  copyright = {http://link.aps.org/licenses/aps-default-license},
  langid = {english}
}

@article{yeEmergentHydrodynamicsNonequilibrium2020,
  title = {Emergent {{Hydrodynamics}} in {{Nonequilibrium Quantum Systems}}},
  author = {Ye, Bingtian and Machado, Francisco and White, Christopher David and Mong, Roger S. K. and Yao, Norman Y.},
  year = 2020,
  month = jul,
  journal = {Physical Review Letters},
  volume = {125},
  number = {3},
  pages = {030601},
  issn = {0031-9007, 1079-7114},
  doi = {10.1103/PhysRevLett.125.030601},
  urldate = {2025-11-06},
  langid = {english}
}

@article{zuEmergentHydrodynamicsStrongly2021,
  title = {Emergent Hydrodynamics in a Strongly Interacting Dipolar Spin Ensemble},
  author = {Zu, C. and Machado, F. and Ye, B. and Choi, S. and Kobrin, B. and Mittiga, T. and Hsieh, S. and Bhattacharyya, P. and Markham, M. and Twitchen, D. and Jarmola, A. and Budker, D. and Laumann, C. R. and Moore, J. E. and Yao, N. Y.},
  year = 2021,
  month = sep,
  journal = {Nature},
  volume = {597},
  number = {7874},
  pages = {45--50},
  publisher = {Nature Publishing Group},
  issn = {1476-4687},
  doi = {10.1038/s41586-021-03763-1},
  urldate = {2025-05-20},
  copyright = {2021 The Author(s), under exclusive licence to Springer Nature Limited},
  langid = {english}
}

@article{zurekDecoherenceEinselectionQuantum2003,
  title = {Decoherence, Einselection, and the Quantum Origins of the Classical},
  author = {Zurek, Wojciech Hubert},
  year = 2003,
  journal = {Reviews of Modern Physics},
  volume = {75},
  number = {3},
  pages = {715--775},
  doi = {10.1103/RevModPhys.75.715}
}

\end{document}